\documentclass{aa}
\usepackage{epsfig,graphics}

\usepackage[usenames]{color}

\newcommand{\teff}{$T_{\rm eff}$}

\begin{document}

\title{Searching for a link between the magnetic nature and other observed properties of Herbig Ae/Be
stars and stars with debris disks\thanks{Based on observations obtained at the European Southern Observatory,
Paranal, Chile (ESO programmes 077.C-0521(A) and 081.C-0410(A)).}}

\author{
S.~Hubrig\inst{1,2}
\and
B.~Stelzer\inst{3}
\and
M.~Sch\"oller\inst{4}
\and
C.~Grady\inst{5}
\and
O.~Sch\"utz\inst{2}
\and
M.~A.~Pogodin\inst{6,7}
\and
M.~Cur\'e\inst{8}
\and
K.~Hamaguchi\inst{9}
\and
R.~V.~Yudin\inst{6,7}
}

\institute{
Astrophysikalisches Institut Potsdam, An der Sternwarte 16, 14482 Potsdam, Germany
\and
European Southern Observatory, Casilla 19001, Santiago 19, Chile
\and
INAF-Osservatorio Astronomico di Palermo, Piazza del Parlamento 1, 90134 Palermo, Italy
\and
European Southern Observatory, Karl-Schwarzschild-Str.\ 2, 85748 Garching, Germany
\and
Eureka Scientific, 2452 Delmer, Suite 100, Oakland, CA 96002, USA
\and
Pulkovo Observatory, Saint-Petersburg, 196140, Russia
\and
Isaac Newton Institute of Chile, Saint-Petersburg Branch, Russia
\and
Departamento de F\'isica y Astronom\'ia, Facultad de Ciencias, Universidad de Valpara\'iso, Chile
\and
Astrophysics Science Division, NASA's Goddard Space Flight Center, Greenbelt, MD 20771, USA
}

%\email{shubrig@eso.org}}
\date{Received date / Accepted date}

\abstract
{Recently, evidence for the presence of weak magnetic fields in Herbig Ae/Be stars has been found 
in several studies.
}
{We seek to expand the sample of intermediate-mass pre-main sequence stars with
circular polarization data to measure their magnetic fields,
and to determine whether magnetic field properties in these stars are correlated
with mass-accretion rate,  disk inclination,
companions, silicates, PAHs, or show a
correlation with age and X-ray emission as expected for the decay of a remnant dynamo.
}
{
Spectropolarimetric observations of 21 Herbig Ae/Be stars and
six debris disk stars have been obtained at the European Southern Observatory with FORS\,1 
mounted on the 8\,m Kueyen telescope of the VLT. With  
the GRISM\,600B in the wavelength range 3250--6215\,\AA{} we were able
to cover all hydrogen Balmer lines from H$\beta$ to the Balmer jump.
In all observations a slit width of 0$\farcs$4 was used to obtain a spectral resolving power 
of $R\approx2000$.
}
{
Among the 21 Herbig Ae/Be stars studied, new detections of a magnetic field were achieved in six stars.
For three Herbig Ae/Be stars, we confirm previous magnetic field detections. 
The largest longitudinal magnetic field, $\left<B_{\rm z}\right>$\,=\,$-$454$\pm$42\,G, was 
detected in the Herbig Ae/Be star HD\,101412 using hydrogen lines.
No field detection   
at a significance level of 3$\sigma$ was achieved in stars with debris disks.
%Only for HD\,181327 the field detection is close to a 3$\sigma$ level.
Our study does not indicate any correlation of the strength of the longitudinal magnetic field 
with disk orientation, disk geometry, or the presence 
of a companion.
We also do not see any simple dependence on the mass-accretion rate.
However, it is likely that the range of observed field values qualitatively supports
%is in agreement with
the expectations from magnetospheric accretion models giving support for dipole-like field geometries.
Both the magnetic field strength and the X-ray emission show hints of a decline with age
in the range of $\sim$2--14\,Myrs probed by our sample, supporting a dynamo mechanism 
that decays with age. 
However, our study of rotation does not show any obvious trend of the strength of the longitudinal
magnetic field with rotation period.
Furthermore, the stars seem to obey the
universal power-law relation between magnetic flux and X-ray luminosity established for the Sun
and main-sequence active dwarf stars.
}
{}
%{
%}


\titlerunning{Magnetic field measurements of Herbig Ae/Be stars with FORS\,1/VLT}
\authorrunning{S.\ Hubrig et al.}

\maketitle

%________________________________________________________________

%\linenumbers

\section{Introduction}
\label{sect:intro}

Magnetic fields are important ingredients of the star formation
process (McKee \& Ostriker \cite{McKee07.1}).
Models of magnetically driven accretion and outflows
(e.g., Shu et al.\ \cite{Shu2000}; Shu et al.\ \cite{Shu1995}) 
successfully reproduce many observational properties of low-mass
pre-main sequence stars, the classical T\,Tauri stars (cTTS). 
Indirect observational evidence for the presence
of magnetic fields in these stars is seen in strong X-ray, FUV, and UV emission 
(e.g., Feigelson \& Montmerle \cite{Feigelson99.1}; Brown et al.\ \cite{Brown1985}).
The high-energy radiation of young stars related to magnetic processes may be critical 
for the annealing and melting of some refractory grain species (Shu et al.\ \cite{Shu2001}),
for the heating intermediate-depth portions of disks around young stars (Najita \cite{Najita2004}) and 
for mass loss from planets forming in those disks (Penz et al.\ \cite{Penz08.1}). 
The irradiation-induced heating mediated by the magnetic
activity enlarges the zone in circumstellar disks where chemistry
may proceed to assemble pre-biotic materials which could produce life-bearing worlds. 
Therefore, understanding the interaction between central stars, their
magnetic fields and protoplanetary disks is crucial to reconstruct the Solar
System's history, and to account for the diversity of exo-planetary systems. 

Direct magnetic field measurements of cTTS (e.g.\ Johns-Krull \cite{JohnsKrull2007}) corroborate the 
scenario outlined above for the formation of low-mass stars, but the picture is
less clear for higher-mass stars. The presence of protoplanetary disks around
intermediate-mass pre-main sequence (PMS) stars, the Herbig Ae/Be stars 
(e.g., Herbig \cite{Herbig1960};
Finkenzeller \& Mundt \cite{FinkenzellerMundt1984}; Th\'e et al.\ \cite{The1994}),
has been established
through their thermal emission in the IR, and more recently by direct imaging 
with polarimetric and interferometric techniques (e.g., Perrin et al.\ \cite{Perrin04.1}; Monnier et al.\ \cite{Monnier05.1}). 
Dusty disk models explain the IR excess emission 
(e.g.\ Dullemond et al.\ \cite{Dullemond01.1}) and magnetospheric accretion models describe the line 
profiles of Herbig Ae/Be stars (Muzerolle et al.\ \cite{Muzerolle04.1}), in close analogy to cTTS.  
However, the disk lifetime appears to be shorter for Herbig Ae/Be stars
%Primordial disks around HAeBe stars dissipate within 4 Myrs, while the disk lifetimes of 
%cTTS are closer to 10 Myr 
(Uzpen et al.\ \cite{Uzpen09.1}; Carpenter et al.\ \cite{Carpenter06.1}) and within the Herbig Ae/Be class there is a 
trend of more massive stars dispersing their disks more rapidly (Alonso-Albi et al.\ \cite{AlonsoAlbi09.1}).
Herbig Ae stars 
%($0.7$ to $2.4\,M_\odot$) 
are closer analogs to TTS than Herbig Be stars.
%that may be affected by ???.
Bipolar outflows or associated Herbig-Haro knots are now known for six optically visible 
Herbig Ae stars drawn from coronographic imaging
surveys (e.g.\ Melnikov et al.\ \cite{Melnikov2008}).
Due to the probable role of magnetic fields 
in launching and collimating jets, this implies 
%that magnetic fields are thought to be essential in launching and collimating jets: the implication is 
that a significant fraction of Herbig Ae
stars should have measurable magnetic fields, which could persist 
through much, if not all, of the star's PMS lifetime. 
X-ray activity is known for a number of Herbig Ae stars
(Hamaguchi et al.\ \cite{Hamaguchi2005}; 
Feigelson et al.\ \cite{Feigelson2003};
Swartz et al.\ \cite{Swartz2005};
Stelzer et al.\ \cite{Stelzer2006}, \cite{Stelzer2008};
Skinner et al.\ \cite{Skinner2004}),
suggesting the presence of magnetic fields.
Recent advances in instrumentation have resulted in the first magnetic field measurements
for Herbig Ae stars, derived from spectropolarimetry. A few stars have magnetic field strengths
derived from circular polarization measurements of a few hundred Gauss or near 100\,G 
(Hubrig et al.\ \cite{Hubrig2004a};
Wade et al.\ \cite{Wade2005}, \cite{Wade2007}; Hubrig et al.\ \cite{Hubrig2006a}, \cite{Hubrig2007a}). 
Other objects seem to have either smaller average field strengths,
or exhibit significant variability. 

In our previous studies, we reported 
detections at a level higher than 3$\sigma$ for three out of seven
Herbig Ae stars observed with FORS\,1 (Hubrig et al.\ \cite{Hubrig2004a}, \cite{Hubrig2006a}, \cite{Hubrig2007a}). 
The results of the spectropolarimetric
observations of the two Herbig Ae stars HD\,31648 and HD\,190073
with FORS\,1 at a resolution of $R\approx4000$ turned out to be 
especially remarkable (Hubrig et al.\ \cite{Hubrig2006a}; Hubrig et al.\ \cite{Hubrig2007a}). 
These two stars
possess almost the same effective temperature, but are different in
luminosity and projected rotation velocity. In the Stokes~$V$ spectra of HD\,190073, the profiles 
exhibit a number of blueshifted local
absorption components, whereas in HD\,31648, only one blueshifted and one
redshifted feature are observed in both H\&K lines of the \ion{Ca}{ii} doublet.
It is noteworthy that recent Goddard Fabry-Perot narrow band images revealed
the counterjet and bright HH knots in HD\,31648 (Stecklum et al.\ 2009, in preparation).
This star demonstrates notable emission in the H$\beta$, H$\gamma$, and H$\delta$ lines, which
indicates the presence of a significant stellar wind. The profiles of
these lines are of P\,Cyg-type with very deep
blueshifted absorptions.

In this work we seek to expand the sample of intermediate-mass PMS stars with circular polarization
data used to derive stellar magnetic fields.
We then investigate 
whether the magnetic field properties of these stars are correlated with 
mass-accretion rate,
companions, silicates, or PAHs.
We also search for the first time for a correlation between field strength and  
age or X-ray emission, such as expected for a decaying fossil field. Further, we examine
the inclination dependence of the polarization data, and explore the dependence of the 
strength of the longitudinal magnetic field on stellar rotation.

\section{Observations and data reduction}
\label{sect:observations}

The observations reported here were carried out
on May 22 and 23, 2008 in visitor mode at the European Southern Observatory with FORS\,1 
mounted on the 8\,m Kueyen telescope of the VLT. This multi-mode instrument is equipped with
polarisation analyzing optics comprising super-achromatic half-wave and quarter-wave 
phase retarder plates, and a Wollaston prism with a beam divergence of 22$\arcsec$  in 
standard resolution mode. 
In 2007, a new mosaic detector
with blue optimized E2V chips was installed in FORS\,1.
It has a  pixel size of 15\,$\mu$m (compared to 24\,$\mu$m for the
previous Tektronix chip) and  a higher efficiency in the wavelength range below 6000\,\AA{}.
To achieve the highest possible signal-to-noise (S/N) ratio -- as is 
required for accurate measurements of stellar magnetic fields -- we used
the (200kHz, low, 1$\times$1) readout mode, which makes it possible to 
achieve a S/N  ratio of about 1000--1200 with only one single exposure. With  
the GRISM\,600B in the wavelength range 3250--6215\,\AA{} we were able
to cover all hydrogen Balmer lines from H$\beta$ to the Balmer jump.
In all observations, a slit width of 0$\farcs$4 was used to obtain a spectral resolving power 
of $R\approx2000$.
Usually, we took three to five continuous series of two exposures for each object,
leading to signal-to-noise ratios of typically 3000 to 4000.
For the faintest Herbig Ae star in our sample, VV\,Ser,
only two series were 
taken due to the rather long exposure time for each sub-exposure. 
More details on the observing technique with FORS\,1 can be 
found elsewhere (e.g., 
Hubrig et al.\ \cite{Hubrig2004a}, \cite{Hubrig2004c}, and references therein).

The mean longitudinal magnetic field is the average over the stellar hemisphere
visible at the time of observation of the component of the magnetic field
parallel to the line of sight, weighted by the local emergent spectral line
intensity.
Its determination is based on the use of the equation

\begin{equation}
\frac{V}{I} = -\frac{g_{\rm eff} e \lambda^2}{4\pi{}m_ec^2}\ \frac{1}{I}\ \frac{{\rm d}I}{{\rm d}\lambda} \left<B_z\right>,
\label{eqn:one}
\end{equation}

%It is diagnosed from the slope of a linear regression of $V/I$ versus the quantity
%$-\frac{g_{\rm eff}e}{4\pi{}m_ec^2} \lambda^2 \frac{1}{I} \frac{{\mathrm d}I}{{\mathrm d}\lambda} \left<B_z\right> + V_%0/I_0$,
where $V$ is the Stokes parameter which measures the circular polarization,
$I$ is the intensity observed in unpolarized light,
$g_{\rm eff}$ is the effective Land\'e factor,
$e$ is the electron charge,
%$\lambda$  is the wavelength expressed in \AA{},
$\lambda$  is the wavelength,
$m_e$ the electron mass,
$c$ the speed of light,
${{\rm d}I/{\rm d}\lambda}$ is the derivative of Stokes~$I$,
and $\left<B_z\right>$ is the mean longitudinal magnetic field.
To minimize the cross-talk effect, we executed the sequence
+45$-$45, +45$-$45, +45$-$45 etc. and
calculated the values $V/I$ using:

\begin{equation}
\frac{V}{I} =
\frac{1}{2} \left\{ \left( \frac{f^{\rm o} - f^{\rm e}}{f^{\rm o} + f^{\rm e}} \right)_{\alpha=-45^{\circ}}
- \left( \frac{f^{\rm o} - f^{\rm e}}{f^{\rm o} + f^{\rm e}} \right)_{\alpha=+45^{\circ}} \right\},
\label{eqn:two}  
\end{equation}

\noindent
where $\alpha$ denotes the position angle of the retarder waveplate
and $f^{\rm o}$ and $f^{\rm e}$ are ordinary and
extraordinary beams, respectively. Stokes $I$ values were obtained from the sum of the 
ordinary and extraordinary beams, which are recorded simultaneously by the detector.
To derive $\left<B_z \right>$, a least-squares technique was
used to minimize the expression

%\begin{displaymath}
\begin{equation}
\chi^2 = \sum_i \frac{(y_i - \left<B_z \right> x_i - b)^2}{\sigma_i^2}
\label{eqn:three}  
\end{equation}
%\end{displaymath}

\noindent
where, for each spectral point $i$, $y_i = (V/I)_i$,
$x_i = -\frac{g_{\rm eff} e \lambda_i^2}{4\pi{}m_ec^2}\ (1/I\ \times\ {\rm d}I/{\rm d}\lambda)_i$,
and $b$ is a constant term that, assuming that Eq.~\ref{eqn:one} is
correct, approximates the fraction of instrumental polarization not removed
after the application of Eq.~\ref{eqn:two} to the observations.
During the commissioning of FORS\,1, this 
instrumental polarization term was found to be wavelength independent.
We assume that the only
  source of uncertainty in our field measurements is from the photon count
  statistics of the observations. Therefore, the longitudinal
  field uncertainty is obtained from the formal uncertainty of the linear 
  regression.
%The longitudinal field uncertainty is obtained from the formal uncertainty of the linear regression.
%$\sigma_i$ associated with each reduced spectral pixel from the formal 
%uncertainty in the linear regression.
%A wavelength-dependent instrumental polarization  would also be visible 
%in the $V$/$I$  spectra, but we do not see anything like this in the data.
For each spectral point $i$, the derivative of Stokes $I$ with respect
to the wavelength was evaluated following

%\begin{displaymath}
\begin{equation}
\left( \frac{{\rm d}I}{{\rm d}\lambda} \right)_{\lambda=\lambda_i}
= \frac{N_{i+1}-N_{i-1}}{\lambda_{i+1}-\lambda_{i-1}},
\label{eqn:four}  
\end{equation}
%\end{displaymath}

\noindent
where $N_i$ is the photon count at wavelength $\lambda_i$.
Since noise strongly influences the derivative, 
we interpolate the data after spectrum extraction with splines.
Our experience from a study of a large sample of magnetic and non-magnetic
Ap and Bp stars (Hubrig et al.\ \cite{Hubrig2006b}) revealed that this regression technique is very robust
and that detections with $\left<B_z\right>$$\geq$3$\sigma$ result only for stars possessing magnetic fields.

%{\bf  Our experience from a study of a sample of seven Vega-like and normal A
%  stars (Hubrig et al.\ \cite{Hubrig2006a}; \cite{Hubrig2006b}) revealed that this regression technique is
%  fairly robust and generally yields detections with $\left<B_z\right>$$\geq$3$\sigma$ only for stars
%possessing magnetic fields.
%<Bz> > 3sigma;
%However, as we show in Sect.~\ref{sect:results}, the distribution of the detection 
%significance $\left<B_{\rm z}\right>$/$\sigma_{\left<B_{\rm z}\right>}$ for
%our programme Herbig Ae/Be stars and stars with debris disks does not appear Gaussian as would be expected.}

%Our experience from a study of a large sample of magnetic and non-magnetic
%Ap and Bp stars (Hubrig et al.\ \cite{Hubrig2006a}) revealed that this regression technique is very robust
%and that detections with $\left<B_z\right>$$\geq$3$\sigma$ result only for stars
%possessing magnetic fields.

\begin{table}
\centering
\caption{
Target stars for which spectropolarimetric data were obtained during our observing run.
}
\label{tab:targetlist}
\centering
\begin{tabular}{llrl}
\hline
\hline
\multicolumn{1}{c}{Object \rule{0pt}{2.6ex}} &
\multicolumn{1}{c}{Other} &
\multicolumn{1}{c}{V} &
\multicolumn{1}{c}{Spectral} \\
\multicolumn{1}{c}{name} &
\multicolumn{1}{c}{identifier} &
 &
\multicolumn{1}{c}{type} \\
\hline
PDS\,2       & CPD$-$53 295  & 10.7 & F2        \\
HD\,47839    & 15\,Mon       & 4.7  & O7 Ve     \\
HD\,95881    & Hen\,3-554    & 8.3  & A1 III    \\
HD\,97048    & CU\,Cha       & 8.5  & A0pshe    \\
HD\,97300    & CED 112 IRS 3 & 9.0  & B9 V      \\
HD\,100453   & CD-53 4102    & 7.8  & A9 Ve     \\
HD\,100546   & KR\,Mus       & 6.7  & B9 Vne    \\
HD\,101412   & V1052\,Cen    & 9.0  & B9.5 V    \\
HD\,135344B  & SAO\,206462   & 7.8  & F4--F8    \\
HD\,139614   & Hen 3-1086    & 8.3  & A7 Ve     \\
HD\,144432   & Hen 3-1141    & 8.2  & A9/F0 V   \\
HD\,144668   & HR\,5999      & 7.0  & A7 IVe    \\
HD\,150193   & V2307\,Oph    & 8.9  & A1 Ve     \\
HD\,152404   & AK\,Sco       & 9.1  & F5 Ve     \\
HD\,158643   & 51\,Oph       & 4.8  & A0 V      \\
HD\,163296   & MWC\,275      & 6.9  & A1 Ve     \\
HD\,169142   & MWC\,925      & 8.2  & B9 Ve     \\
VV\,Ser      & FMC\,39       & 11.6 & A2e       \\
HD\,176386   & CD$-$37 13023 & 7.3  & B9 IVe    \\
HD\,179218   & BD$+$15 1991  & 7.2  & B9e       \\
HD\,190073   & MWC\,325      & 7.8  & A2 IVpe   \\
\hline
\multicolumn{4}{c}{Stars with debris disks}\\
\hline
HD\,9672     & 49\,Cet       & 5.6  & A1 V      \\
HD\,39060    & $\beta$\,Pic  & 3.9  & A6V       \\
HD\,109573   & HR\,4796      & 5.8  & A0 V      \\
HD\,164249   & CD-51 11312   & 7.3  & F5 V      \\
HD\,172555   & HR\,7012      & 4.8  & A7 V      \\
HD\,181327   & CD-54 8270    & 7.0  & F5/F6V    \\
\hline
\multicolumn{4}{c}{Magnetic Herbig Ae/Be stars from previous studies}\\
\hline
HD\,31648  & MWC\,480    & 7.7 & A3pshe     \\
HD\,104237 & Hen 3-741   & 6.6 & A:pe       \\
HD\,200775 & MWC\,361    & 7.4 & B2 Ve      \\
V380\,Ori  & MWC\,765    & 10.7& A0         \\
BF\,Ori    & BD$-$061259 & 10.3& A5II/IIIev \\
\hline
\multicolumn{4}{c}{Standard Ap star with a weak magnetic field}\\
\hline
HD\,162725   & V951\,Sco     & 6.4  & Ap        \\
\hline
\end{tabular}
\begin{flushleft}
Notes:
Spectral types and visual magnitudes are taken from SIMBAD.
\end{flushleft}
\end{table}

\begin{table*}
\centering
\caption{
%Target stars for which spectropolarimetric data were obtained during our observing run.
Targets discussed in this paper.
%The mass-accretion rates are from Garcia Lopez et al.\ (\cite{GarciaLopez2006}).
}
\label{tab:targetvals}
\centering
\begin{tabular}{llllllllll}
\hline
\hline
\multicolumn{1}{c}{Object \rule{0pt}{2.6ex}} &
\multicolumn{1}{c}{$v\sin i$} &
\multicolumn{1}{c}{log \.M$_{\rm acc}$} &
\multicolumn{1}{c}{IR } &
\multicolumn{1}{c}{Comp.} & 
\multicolumn{1}{c}{log $T_{\rm eff}$} & 
\multicolumn{1}{c}{log $L$} & 
\multicolumn{1}{c}{$M$} & 
\multicolumn{1}{c}{age} \\
\multicolumn{1}{c}{name} &
\multicolumn{1}{c}{[km s$^{-1}$]} &
\multicolumn{1}{c}{[M$_{\sun}$/yr]} &
\multicolumn{1}{c}{feature} &
 &
\multicolumn{1}{c}{[K]} & 
\multicolumn{1}{c}{[L$_{\sun}$]} & 
\multicolumn{1}{c}{[M$_{\sun}$]} & 
\multicolumn{1}{c}{[Myr]} \\
\hline
PDS\,2       & 175$^a$       &           &                & no           & 3.98$^{d}$  & 1.68$^{d}$  & 2.5 & \\
HD\,47839$^*$& 120$^b$       &           &                & yes$^{ee}$   & 4.57$^{uu}$ & 5.23$^{uu}$ &     & $\sim$0.3$^{eee}$   \\
HD\,95881    & 50$^c$        &$-$8.04    & Si, PAH$^r$    & yes$^{ff}$   & 3.86$^{vv}$ & 1.84$^{vv}$ & 3.1 & $>$3$^{x}$       \\
HD\,97048    & 140$^d$       &$-$7.17    & PAH$^s$        & no$^{gg}$    & 4.00$^{s}$  & 1.64$^{s}$  & 2.5 & $>$2$^{d}$       \\
HD\,97300    &               &$<$$-$8.18 & PAH$^t$        & yes$^{hh}$   & 4.02$^{ww}$ & 1.54$^{ww}$ & 2.5 & $>$3$^{ww}$       \\
HD\,100453   & 39$^{e}$      &$-$8.04    & PAH$^{u}$      & yes$^{ii}$   & 3.87$^{s}$  & 0.90$^{s}$  & 1.7 & 14$\pm$4$^{x}$   \\
HD\,100546   & 65$^f$        &           & Si, PAH$^{u}$  & no$^{gg}$    & 4.02$^{s}$  & 1.51$^{s}$  & 2.5 & $>$10$^{d}$      \\
HD\,101412   & 5$^g$         &           & Si, PAH$^{v}$  & yes$^{gg}$   & 3.98$^{xx}$ & 1.74$^{xx}$ & 2.5 & $\sim$2$^{fff}$    \\
HD\,135344B  & 69$^{h}$      &$-$8.27    & Si$^s$, PAH$^{w}$ & yes$^{jj}$& 3.82$^{s}$  & 0.91$^{s}$  & 1.6& 8$\pm$4$^{x}$    \\
HD\,139614   & 15$^i$        &$-$7.99    & Si$^{u}$       & yes$^{ee}$   & 3.90$^{s}$  & 0.91$^{s}$  & 1.8 & $>$10$^{x}$      \\
HD\,144432   & 70$^i$        &$-$7.07    & Si$^{u}$       & yes$^{kk}$   & 3.87$^{s}$  & 1.01$^{s}$  & 1.8 & 10$\pm$5$^{x}$   \\
HD\,144668   & 100$^i$       &$-$6.37    & Si$^{x}$       & yes$^{ll}$   & 3.90$^{s}$  & 1.94$^{s}$  & 3.1 & 0.6$\pm$0.4$^{ww}$\\
HD\,150193   & 100$^f$       &$-$7.29    & Si$^s$         & yes$^{mm}$   & 3.95$^{s}$  & 1.38$^{s}$  & 2.2 & $>$2$^{d}$       \\
HD\,152404   & 18.5$^{j}$    &           & Si$^{y}$       & yes$^{j}$    & 3.81$^{s}$  & 0.95$^{s}$  & 1.7 & $\sim$8$^{vv}$    \\
HD\,158643   & 267$^{h}$/228$^k$ &$-$6.87 & Si$^{u}$      & no$^{nn}$    & 4.00$^{d}$  & 2.39$^{d}$  & 3.9 & 0.3$^{d}$        \\
HD\,163296   & 130$^i$       &$-$7.12    & Si$^{u}$       & no$^{gg}$    & 3.94$^{s}$  & 1.38$^{s}$  & 2.2 & 5$\pm$2$^{x}$    \\
HD\,169142   & 66$^{h}$      &$-$7.40    & PAH$^{w}$      & yes$^{oo}$   & 3.91$^{s}$  & 1.16$^{s}$  & 2.0 & 6$\pm$3$^{x}$    \\
VV\,Ser      & 142$^a$/229$^{l}$ &$-$6.34 & Si$^{z}$      & no$^{gg}$    & 3.95$^{s}$  & 1.27$^{s}$  & 2.0 & 3.5$\pm$0.5$^{ggg}$\\
HD\,176386   &               &$-$8.11    & PAH$^{aa}$     & yes$^{ee}$   & 4.03$^{ww}$ & 1.69$^{ww}$ & 2.7  & $>$2$^{ww}$       \\
HD\,179218   & 60$^b$        &$-$6.59    & Si, PAH$^{u}$  & no$^{gg}$    & 4.02$^{s}$  & 2.00$^{s}$  & 3.0 & 1.3$\pm$0.5$^{x}$\\
HD\,190073   & 12$^i$        &           & Si, PAH$^r$    & yes$^{ff}$   & 3.95$^{d}$  & 1.80$^{d}$  & 2.7 & 1.2$\pm$0.6$^{hhh}$\\
\hline
\multicolumn{6}{c}{Stars with debris disks}\\
\hline
HD\,9672     & 196$^k$       &           & --$^{bb}$      & no          & 4.00$^{yy}$  & 1.42$^{yy}$  & 2.3 & $\sim$8$^{iii}$    \\
HD\,39060    & 130$^{m}$     &           & Si$^{cc}$      & no         & 3.91$^{zz}$   & 0.95$^{zz}$  & 1.8 & $\sim$12$^{jjj}$   \\
HD\,109573   & 152$^k$       &$<$$-$8.53 & --$^{bb}$      & yes$^{pp}$  & 3.97$^{aaa}$ & 1.32$^{aaa}$ & 2.2 & 8$\pm$2$^{pp}$    \\
HD\,164249   &               &           &                & yes$^{qq}$ & 3.81$^{bbb}$  & 0.48$^{n}$   & 1.4 & $\sim$12$^{jjj}$   \\
HD\,172555   &               &           & Si$^{dd}$      & no         & 3.89$^{bbb}$  & 0.96$^{n}$   & 1.7 & $\sim$12$^{jjj}$   \\
HD\,181327   & 16$^{n}$      &           &                & no         & 3.81$^{bbb}$  & 0.54$^{n}$   & 1.4 & $\sim$12$^{jjj}$   \\
\hline
\multicolumn{6}{c}{Magnetic Herbig Ae/Be stars from previous studies}\\
\hline
HD\,31648$^*$  & 90$^{o}$  &          & Si$^{s}$      & no$^{rr}$  & 3.94$^{s}$  & 1.14$^{s}$ & 2.0 & $\sim$7$^{kkk}$      \\     
HD\,104237$^*$ & 10$^{h}$  &  $-$7.45 & Si$^{s}$      & yes$^{ss}$ & 3.93$^{s}$  & 1.54$^{s}$ & 2.2 & 2.0$\pm$0.5$^{ww}$  \\
HD\,200775$^*$ & 26$^{p}$  &          &               & yes$^{tt}$ & 4.29$^{s}$  & 3.82$^{s}$ &     & 0.1$\pm$0.05$^{ww}$ \\
V380\,Ori      & 200$^{f}$ &          & Si$^{x}$      & yes$^{ee}$ & 3.97$^{d}$  & 2.04$^{ddd}$ & 3.0 & 1.5$\pm$0.5$^{x}$  \\
BF\,Ori        & 37$^{l}$  &          & Si$^{s}$      & no         & 3.95$^{s}$  & 1.53$^{s}$ & 2.3 & $\sim$2$^{lll}$      \\
\hline
\multicolumn{6}{c}{Standard Ap star with a weak magnetic field}\\
\hline
HD\,162725   & 32$^q$   &          &                &            & 3.99$^{ccc}$ & 2.33$^{ccc}$ &     &  \\
\hline
\end{tabular}
\begin{flushleft}
Notes:
The information on $v\sin i$ values, mass-accretion rate, IR features, binarity, effective temperature,
luminosity, and age was collected from the literature. 
The mass was deduced from the position of the star in the H-R diagram (see Sect.~\ref{sect:link}).
Objects marked with $^*$ are excluded from the correlation analysis
(see Sects.~\ref{sect:observations}, \ref{sect:prev_detect}, and \ref{subsect:accr_rate}).\\
References:
$^a$Vieira et al.\ (\cite{Vieira2003}),
$^b$Bernacca \& Perinotto (\cite{BernaccaPerinotto1970}),
$^c$Grady et al.\ (\cite{Grady1996}),
$^d$van den Ancker et al.\ (\cite{vandenAncker1998}),
$^{e}$Acke \& Waelkens (\cite{AckeWaelkens2004}),
$^f$Hamidouche et al.\ (\cite{Hamidouche2008}),
$^{g}$this work,
$^{h}$Dunkin et al.\ (\cite{Dunkin1997}),
$^i$Hubrig et al.\ (\cite{Hubrig2007b}),
$^{j}$Alencar et al.\ (\cite{Alencar2003}),
$^k$Royer et al.\ (\cite{Royer2007}),
$^{l}$Mora et al.\ (\cite{Mora2001}),
$^{m}$Slettebak (\cite{Slettebak1982}),
$^{n}$de la Reza \& Pinz\'on (\cite{delaRezaPinzon2004}),
$^{o}$Beskrovnaya et al.\ (\cite{Beskrovnaya2004}), 
$^{p}$Alecian et al.\ (\cite{Alecian2008}),
$^q$Landstreet et al.\ (\cite{Landstreet2008}),
$^r$Boersma et al.\ (\cite{Boersma2008}),
$^s$Acke \& van den Ancker (\cite{Acke2004}),
$^t$Siebenmorgen et al.\ (\cite{Siebenmorgen1998}),
$^{u}$Meeus et al.\ (\cite{Meeus2001}),
$^{v}$Geers et al.\ (\cite{Geers2007}),
$^{w}$Sloan et al.\ (\cite{Sloan2005}),
$^{x}$van Boekel et al.\ (\cite{vanBoekel2005}),
$^{y}$Przygodda et al.\ (\cite{Przygodda2003}),
$^{z}$Pontoppidan et al.\ (\cite{Pontoppidan2007b}),
$^{aa}$Siebenmorgen et al.\ (\cite{Siebenmorgen2000}),
$^{bb}$Kessler-Silacci et al.\ (\cite{KesslerSilacci2005}),
$^{cc}$Okamoto et al.\ (\cite{Okamoto2004}),
$^{dd}$Sch\"utz et al.\ (\cite{Schuetz2005b}),
$^{ee}$Dommanget \& Nys (\cite{Dommanget1994}),
$^{ff}$Baines et al.\ (\cite{Baines2006}),
$^{gg}$Corporon \& Lagrange (\cite{Corporon1999}),
$^{hh}$Stelzer et al.\ (\cite{Stelzer2006}),
$^{ii}$Chen et al.\ (\cite{Chen2006a}),
$^{jj}$Augereau et al.\ (\cite{Augereau2001}),
$^{kk}$Carmona et al.\ (\cite{Carmona2007}),
$^{ll}$Stecklum et al.\ (\cite{Stecklum1995}),
$^{mm}$Reipurth \& Zinnecker (\cite{Reipurth1993}),
$^{nn}$Roberge et al.\ (\cite{Roberge2002}),
$^{oo}$Grady et al.\ (\cite{Grady2007}),
$^{pp}$Stauffer et al.\ (\cite{Stauffer1995}),
$^{qq}$SIMBAD,
$^{rr}$Eisner et al.\ (\cite{Eisner2004}),
$^{ss}$Grady et al.\ (\cite{Grady2004}),
$^{tt}$Pirzkal et al.\ (\cite{Pirzkal1997}),
$^{uu}$Schnerr et al.\ (\cite{Schnerr2007}),
$^{vv}$Blondel et al.\ (\cite{Blondel2006}),
$^{ww}$van den Ancker et al.\ (\cite{vandenAncker1997}),
$^{xx}$Wade et al.\ (\cite{Wade2007}),
$^{yy}$Hughes et al.\ (\cite{Hughes2008}),
$^{zz}$di Folco et al.\ (\cite{diFolco2004}),
$^{aaa}$Debes et al.\ (\cite{Debes2008}),
$^{bbb}$Zuckerman \& Song (\cite{ZuckermanSong2004}),
$^{ccc}$Folsom et al.\ (\cite{Folsom2007}),
$^{ddd}$Corcoran \& Ray (\cite{CorcoranRay1998}),
$^{eee}$Walsh (\cite{Walsh1980}),
$^{fff}$Wade et al.\ (\cite{Wade2005}),
$^{ggg}$Pontoppidan et al.\ (\cite{Pontoppidan2007a}),
$^{hhh}$Catala et al.\ (\cite{Catala2007}),
$^{iii}$Thi et al.\ (\cite{Thi2001}),
$^{jjj}$Zuckerman et al.\ (\cite{Zuckerman2001}), 
$^{kkk}$Simon et al.\ (\cite{Simon2000}), and
$^{lll}$Natta et al.\ (\cite{Natta1997}).
%$^{ee}$van Boekel et al.\ (\cite{vanBoekel2005}),
%$^{vv}$van den Ancker et al.\ (\cite{vandenAncker1998}),
%$^{fff}$de la Reza \& Pinz\'on (\cite{delaRezaPinzon2004}),
\end{flushleft}
\end{table*}

The list of the studied stars is presented in Table~\ref{tab:targetlist}.
In the four columns we give the object name,
another identifier from SIMBAD, the visual magnitude and the spectral type. 
The star HD\,47839 of spectral type O7~V was included in our sample since it is classified 
as a pre-main sequence star in the SIMBAD database,
probably due to its close proximity to the Cone Nebula. Markova et al.\ (\cite{Markova2004})
consider this star as a Galactic O-type star with a mass of 32\,M$_\odot$ and 
$T_{\rm eff}$=37\,500\,K. 
In Table~\ref{tab:targetvals} we list 
various properties of the studied stars compiled from the literature, such as  
$v$\,sin\,$i$,
mass-accretion rate,
features in the mid-IR spectra,
the presence of a companion,
effective temperature,
luminosity, and age.
The mass was deduced from the position of the star in the H-R diagram (see Sect.~\ref{sect:link}).
The mass-accretion rates were determined by Garcia Lopez et al.\ (\cite{GarciaLopez2006}) from the 
luminosity of the Br$\gamma $ line seen in emission in Herbig Ae/Be stars.
Mid-IR features such as 
crystalline silicate grains indicate thermal dust processing subsequent to the formation of the 
star-disk system, while PAHs (polycyclic aromatic hydrocarbons)
are abundant in interstellar and circumstellar environments and can be 
important constituents in the energy balance of those environments as sources of 
photoelectrons that heat the gas component (Kamp \& Dullemond \cite{Kamp2004}).
%\changea{SH:  Carol, do we need to put here more explanations on the meaning of these features for the 
%properties of the disks?}
Since the existence of a link between the presence of magnetic fields 
and the disk characteristics remains unproven, we decided to include in this study also the information on mid-IR features.
Additionally, in Tables~\ref{tab:targetlist} and \ref{tab:targetvals}
we added five Herbig Ae/Be stars for which magnetic fields were 
detected in previous studies by various authors.
The magnetic field for HD\,31648 was detected by Hubrig et al.\ (\cite{Hubrig2006b}),
while V380\,Ori and BF\,Ori were studied by Wade et al.\ (\cite{Wade2005}).
The detection of a weak magnetic field in HD\,104237 was reported by
Donati et al.\ (\cite{Donati1997}) and the study of 
the magnetic field of HD\,200775 was carried out by Alecian et al.\ (\cite{Alecian2008}). 
We note, however, that, as was shown by  Hubrig et al.\ (\cite{Hubrig2007b}), the detected 
magnetic field of HD\,31648
is predominantly circumstellar (CS) and that the strength of the photospheric 
magnetic field remains unknown. Also, the field detection in HD\,104237 of the order of 
$\sim$50\,G was just marginal,
and was not confirmed in the follow-up study by Wade et al.\ (\cite{Wade2007}).

Since caution is called for in the detection of magnetic fields with low resolution spectropolarimeters,
from time to time we observed magnetic Ap/Bp stars with well-known variation curves.
% known from the literature. 
These observations confirm that our 
measurements are usually in good agreement with measurements obtained with other 
spectropolarimeters (see e.g.\ Hubrig et al.\ \cite{Hubrig2004b}). 
During our observing run, we also observed the A-type star HD\,162725, for which previous measurements exist
%which has a weak magnetic field measured previously 
both with FORS\,1 and later with ESPaDOnS at 
the Canada-France-Hawaii Telescope at much higher spectral resolution ($R$=65\,000). 
Due to lack of time we obtained only three series of two exposures, resulting in a 
magnetic field measurement $\left<B_{\rm z}\right>$\,=\,128$\pm$30\,G using the full spectrum 
and $\left<B_{\rm z}\right>$\,=\,158$\pm$38\,G using hydrogen lines.
This star was previously observed by Bagnulo et al.\ (\cite{Bagnulo2006}), who failed 
to detect a magnetic field in this star. The non-detection by these authors is probably due to the much 
lower spectral resolution in their spectropolarimetric observations with FORS\,1 (with a slit width 
of 0$\farcs$5--1$\farcs$0). 
%probably due to much lower resolution of their
%spectropolarimetric observations with FORS\,1 (with a slit width of 0$\farcs$5--1$\farcs$0).
%and later by Landstreet et al.\ (\cite{Landstreet2008}) using ESPaDOnS. 
Landstreet et al.\ (\cite{Landstreet2008}) detected a weak magnetic field of the order of $-$100\,G with ESPaDOnS,
and could also show that the field is variable. We confirm their finding with our result, revealing
a positive magnetic field of the order of 130--160\,G, measured in metal and hydrogen 
lines (see Table~\ref{tab:fields}). We note, however, that an accurate ephemeris for HD\,162725 is 
not known and presently it is not possible to prove whether our observations were carried out in 
the opposite rotational phase compared to the observations of Landstreet et al.\ (\cite{Landstreet2008}). 

Certainly, the advantage of using high resolution spectropolarimeters such as  ESPaDOnS at the Canada-France-
Hawaii Telescope,  and NARVAL at the Bernard Lyot
Telescope at Pic du Midi Observatory (France) is  indubitable, since higher spectral resolution 
observations provide more detailed
%to extract 
information about the behaviour of different elements in the presence 
of a magnetic field and on the magnetic field topology.
On the other hand, the use of hydrogen lines, which are usually not used for magnetic field 
measurements with  ESPaDOnS and NARVAL,
%in addition to metal lines 
offers the unique opportunity 
to study the global structure of the detected magnetic field. 
%using the element which is homogeneously distributed over the stellar surface. 
The profiles of metal lines frequently exhibit conspicuous
variations,  which are 
signatures of the circumstellar environment, stellar winds, of a non-uniform distribution of metals 
over the stellar surface, or of the presence 
of temperature spots (which are typical for late type stars).
In our studies with FORS\,1, the longitudinal magnetic fields are measured in two ways: 
using only the absorption hydrogen Balmer 
lines or using the whole spectrum including all available absorption lines, i.e.
we use all lines, variable and non-variable, together. 
As hydrogen is expected to be homogeneously distributed over the 
stellar surface, the longitudinal magnetic field measurements
sample the magnetic field fairly uniformly over the observed hemisphere.

\section{Results}
\label{sect:results}

\begin{table*}
\caption{
The mean longitudinal magnetic field measurements for our sample
of Herbig Ae/Be and debris disk stars observed
with FORS\,1.
%In the first two columns we give the object name
%and the modified Julian date of the middle of the exposures.
%The measured mean longitudinal magnetic fields $\left<B_{\mathrm z}\right>$
%using all lines and hydrogen lines are presented in Cols.~3 and 4. 
}
\label{tab:fields}
\centering
\begin{tabular}{rcrr @{$\pm$} lr @{$\pm$} lc}
\hline
\hline
\multicolumn{1}{c}{Object} &
\multicolumn{1}{c}{MJD} &
\multicolumn{1}{c}{SNR} &
\multicolumn{2}{c}{$\left< B_z\right>_{\rm all}$} &
\multicolumn{2}{c}{$\left< B_z\right>_{\rm hydr}$} &
\multicolumn{1}{c}{Comm.} \\
\multicolumn{1}{c}{name} &
 &
 &
\multicolumn{2}{c}{[G]} &
\multicolumn{2}{c}{[G]} &
 \\
\hline
PDS\,2  & 54610.399      & 1790 & {\bf 103}    & {\bf 29} & $-$4          & 70       & ND \\
47839  & 54609.968       & 3375 & 134          & 52       & 141           & 62       & \\
%53179  & 54608.977       & 2190 & {\bf 73}     & {\bf 11} & {\bf 93}     & {\bf 14}  & ND \\
%85567  & 54609.012       & 3560 & {\bf 83}     & {\bf 16} & {\bf 119}    & {\bf 19}  & ND \\
95881  & 54609.993       & 3745 & 47           & 25      & 24            & 20      & \\
97048  & 54609.137       & 2980 & {\bf 164}    & {\bf 42} & {\bf 188}    &{\bf 47} & ND \\
97300  & 54609.047       & 2495 & 109          & 50       & 96           & 53       & \\
100453 & 54610.022       & 3820 & 1            & 14       & $-$30        & 28       & \\
100546 & 54610.046       & 4195 & {\bf 89}     & {\bf 26} & {\bf 87}     & {\bf 28} & ND \\
101412 & 54609.190       & 3280 & {\bf $-$312} & {\bf 32} & {\bf $-$454} & {\bf 42} & CD \\
       & 54610.081       & 3525 & {\bf $-$207} & {\bf 28} & {\bf $-$317} & {\bf 35} & CD \\
135344B& 54609.243       & 4020 & 32          & 15       & 41           & 30       & \\
       & 54610.145       & 3970 & {\bf $-$38}  &{\bf 11}  & {\bf $-$37}  & {\bf 12} & ND \\
139614 & 54610.201       & 4010 & $-$40        & 25       & $-$58        & 32       & \\
144432 & 54609.383       & 2595 & 36           & 22       & 32           & 42       &  \\
144668 & 54610.238       & 3710 & {\bf $-$62}  & {\bf 18} & {\bf $-$92}  & {\bf 27} & CD\\
150193 & 54609.092       & 3300 & {\bf $-$144} & {\bf 32} & {\bf $-$252} & {\bf 48} & ND \\
152404 & 54609.275       & 2815 & $-$107       & 37       & $-$53        & 23      & \\
158643 & 54609.275       & 4165 & 32           & 20       & 14           & 23       & \\
163296 & 54610.255       & 3535 & 40           & 30       & 105          & 40       & \\
169142 & 54610.175       & 3875 & 50           & 25       & 23           & 32       & \\
VV\,Ser & 54610.332      & 1050 & 200          & 75       & 376          & 141       & \\
176386 & 54610.272       & 3230 & {\bf $-$119} & {\bf 33} & {\bf $-$121} & {\bf 35} & ND\\
179218 & 54609.360       & 3855 & 51           & 30       & 28           & 34       & \\
190073 & 54609.410       & 3900 & {\bf 104}    & {\bf 19} & {\bf 120}    & {\bf 30} & CD\\
\hline
\multicolumn{8}{c}{Stars with debris disks}\\
\hline
9672   & 54345.389       & 4365 & 50           & 22       & 42           & 32       &  \\
39060  & 53455.193       & 4205 & $-$16        & 14       & $-$36        & 22       & \\
109573 & 54610.438       & 3815 & 56           & 30       & 41           & 32       & \\
164249 & 54610.301       & 2970 & 19           & 14       & 1            & 25       & \\
172555 & 54610.287       & 3875 & $-$35        & 20       & $-$31        & 28       & \\
181327 & 54610.364       & 3495 & $-$38        & 16       & $-$75        & 26       & \\
%strange features in spectrum of hd181327
\hline
\multicolumn{8}{c}{ Standard Ap star with a weak magnetic field}\\
\hline
162725 & 54549.413       & 2435 & {\bf 128}  & {\bf 30} & {\bf 158}   & {\bf38}       & \\
\hline
\end{tabular}
\begin{flushleft}
Notes:
All quoted errors are 1$\sigma$ uncertainties.
In Col.~6 we identify new detections by ND and confirmed detections by CD.
We note that all claimed detections have a significance of at
least 3$\sigma$, determined from the formal uncertainties we derive. 
These measurements are indicated in bold face.
\end{flushleft}
\end{table*}

\begin{figure}
\begin{center}
\includegraphics[width=0.45\textwidth,angle=0,clip=]{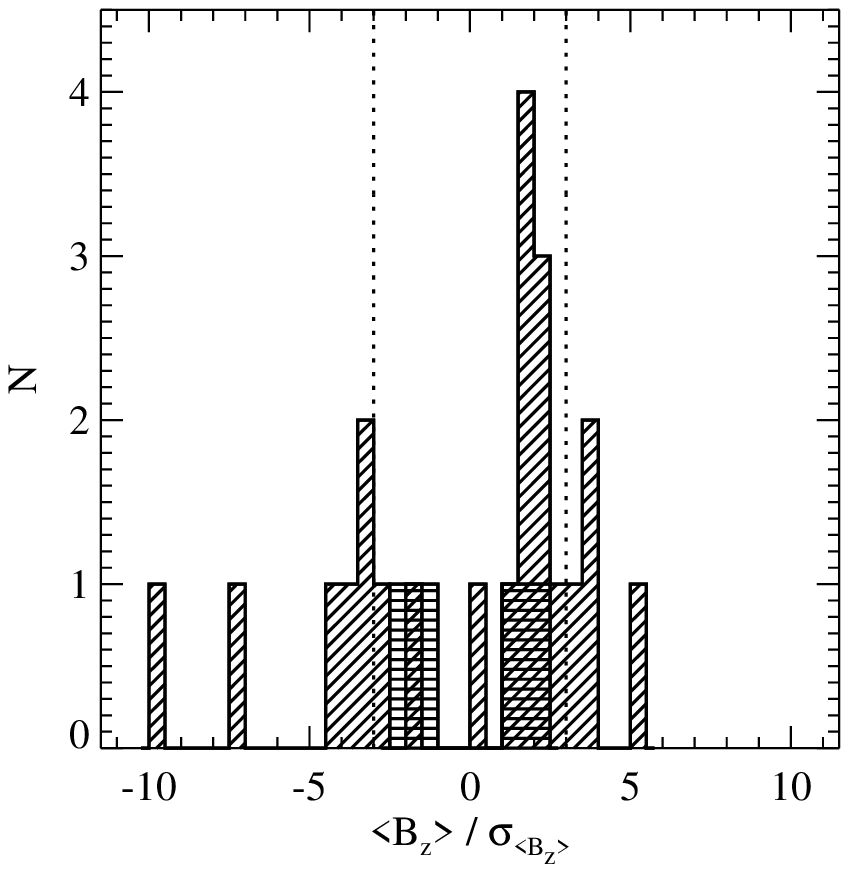}
\includegraphics[width=0.45\textwidth,angle=0,clip=]{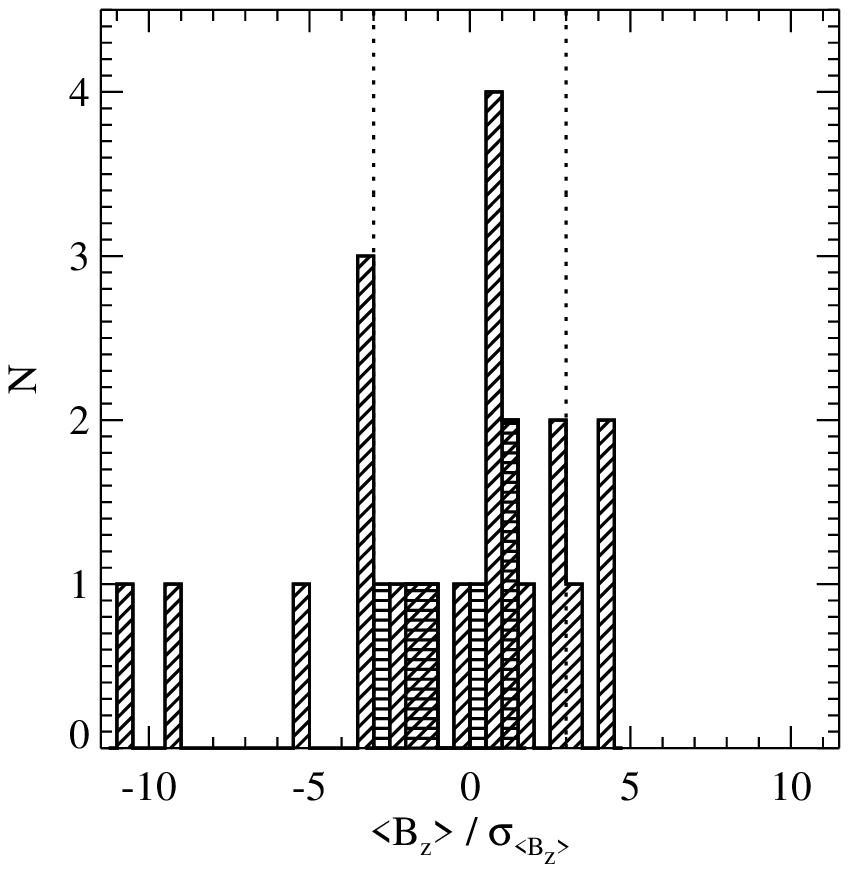}
\end{center}
\caption{
Distribution of the detection significance $\left<B_{\rm z}\right>$/$\sigma_{\left<B_{\rm z}\right>}$ for
our programme stars.
Diagonal lines denote the Herbig Ae/Be stars, while horizontal lines denote the debris disk stars.
Top: Results using the full spectra (Col.~4 of Table~\ref{tab:fields}).
Bottom: Results using only the hydrogen Balmer lines (Col.~5 of Table~\ref{tab:fields}).
}
\label{fig:histos}
\end{figure}

%Because of the strong dependence of the longitudinal magnetic field on the rotational 
%aspect, its usefulness to characterize actual magnetic field strength distributions
%depends on the sampling of the various rotation phases,
%hence various aspects of the magnetic field. 
The obtained magnetic field measurements are presented in Table~\ref{tab:fields}.
In the first two columns, we give the object name of the targets and the modified Julian dates of the 
middle of the exposures. 
In the third column we list the signal-to-noise ratio (SNR) calculated in the final one-dimensional
spectrum around 4750\,\AA{}.
The measured mean longitudinal magnetic field $\left<B_{\mathrm z}\right>$ using all absorption lines 
is presented in Col.~4.
The measured mean longitudinal magnetic field $\left<B_{\mathrm z}\right>$ using all hydrogen lines in 
absorption is listed in Col.~5.
All quoted errors are 1$\sigma$ uncertainties.
In Col.~6, we identify new detections by ND and confirmed detections by CD.
We note that all claimed detections have a significance of at
least 3$\sigma$, determined from the formal uncertainties we derive. These measurements are indicated in bold face.
In Figure~\ref{fig:histos} we show the significance distributions of our measurements.
These distributions are different from the distribution shown by Wade et al.\ (\cite{Wade2007})
for their sample of FORS\,1 observations of Herbig Ae/Be stars.
Assuming that only formal uncertainties apply, this suggests that the 
Herbig stars in our sample are probably weakly magnetic.
%This suggests that the Herbig stars in our sample are probably weakly magnetic, with magnetic
%field strengths on the order of 50\,G.

%The uncertainty of the 
%mean longitudinal magnetic field determination is obtained from the formal uncertainty of the linear regression
%of $V/I$
%versus the quantity
%$-\frac{g_{\rm eff}e}{4\pi{}m_ec^2} \lambda^2 \frac{1}{I} \frac{{\mathrm d}I}{{\mathrm d}\lambda} \left<B_z\right> + V_0/I_0$.
 Apart from the confirmed detections of a magnetic field in the stars HD\,101412, HD\,144668, and HD\,190073
(Wade et al.\ \cite{Wade2007}; Hubrig et al.\ \cite{Hubrig2007b}; Catala et al.\ \cite{Catala2007}),
%eight other stars of our sample, PDS\,2, HD\,53179, HD\,85567, HD\,97048, HD\,100546,
six other stars of our sample, PDS\,2, HD\,97048, HD\,100546,
HD\,135344, HD\,150193, and HD\,176386, show evidence for the presence of a weak magnetic 
field. About half of the stars with magnetic field detections possess longitudinal  magnetic 
fields larger than 100\,G.
These stars are the best candidates for future spectropolarimetric studies to analyze the behaviour of 
their magnetic fields over their rotational cycles to disclose the magnetic topology of their surfaces.
For two Herbig Ae stars, HD\,139614 and HD\,144432, with previously detected weak magnetic fields at a 
significance level of 3$\sigma$
(Hubrig et al.\ \cite{Hubrig2004b}; Hubrig et al.\ \cite{Hubrig2007b}), the magnetic field in the present 
study was diagnosed only 
at a level of 1.8$\sigma$ and 1.6$\sigma$, respectively. 
Wade et al. (\cite{Wade2005}) failed to detect a magnetic field in HD\,139614 with quoted 
uncertainties of 25\,G on two consecutive  nights with the high resolution spectropolarimeter
ESPaDOnS.
The marginal detections of magnetic fields in 
these stars during this observing run 
can probably be explained by the strong dependence of the longitudinal magnetic field on the rotational aspect, 
i.e.\ on the rotation phase. 
For the Herbig Ae star HD\,163296, we found no indication for the presence of a photospheric magnetic field,
in agreement with our 
previous studies (Hubrig et al.\ \cite{Hubrig2006b}; Hubrig et al.\ \cite{Hubrig2007b}).

No definite detection at a significance level of 3$\sigma$ was achieved for any of the stars with debris 
disks. The only measurement close to the 3$\sigma$ level was obtained for the F5/F6V star 
HD\,181327, which belongs to the ($\sim$12\,Myr old) $\beta$\,Pictoris moving group, with a measured 
magnetic field $\left<B_{\rm z}\right>$\,=\,$-$75$\pm$26\,G. 
%It has 86 AU debris viewed at 3.17  inclination from  face-on 
%\changea{SH:: Carol, do you want to add something about the debris ring around this star?}

%\begin{figure*}
%\begin{center}
%\includegraphics[width=0.45\textwidth,angle=0,clip=]{hd101412_IVb.eps}
%\includegraphics[width=0.45\textwidth,angle=0,clip=]{hd101412_IVa.eps}
%\end{center}
%\caption{
%Stokes~$I$ and $V$ spectra of the Herbig Ae/Be star HD\,101412, with the largest detected magnetic field.
%Left panel: Zeeman features in H$_8$,  H$_9$, \ion{Ca}{ii} H\&K, and H$\epsilon$ profiles;
%Right panel: Stokes~$I$ and $V$ spectra in the vicinity of the 
%H$\gamma$ line.
%}
%\label{fig:101412}
%\end{figure*}

\begin{figure}
\begin{center}
\includegraphics[width=0.45\textwidth,angle=0,clip=]{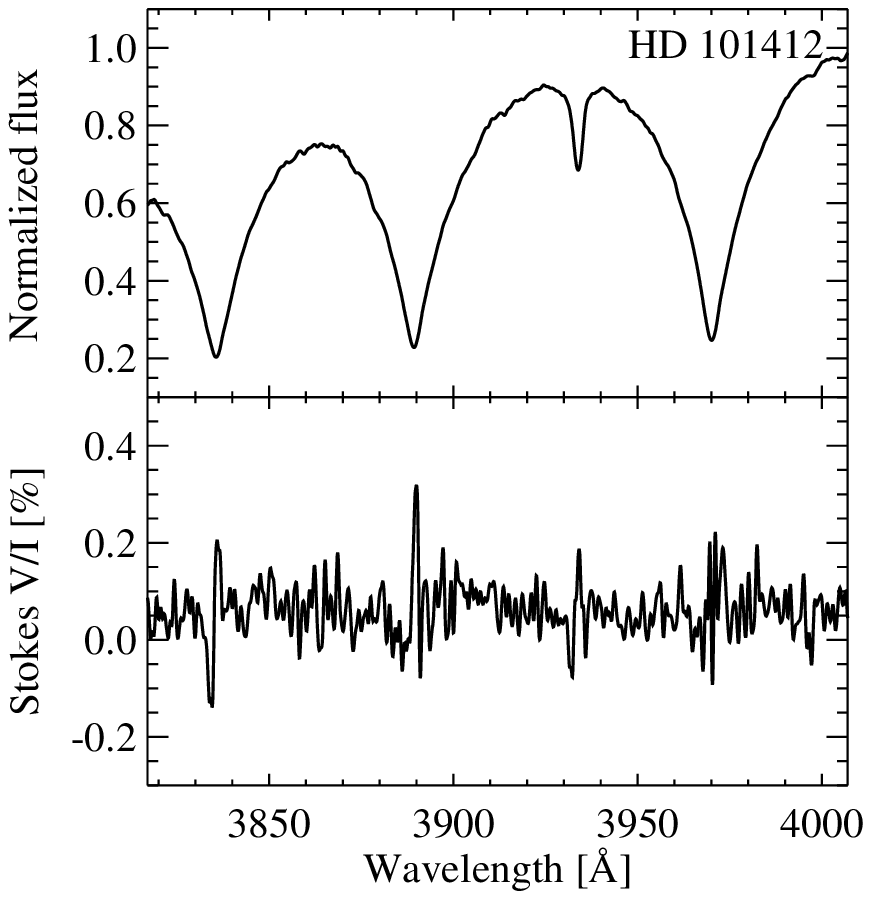}
\includegraphics[width=0.45\textwidth,angle=0,clip=]{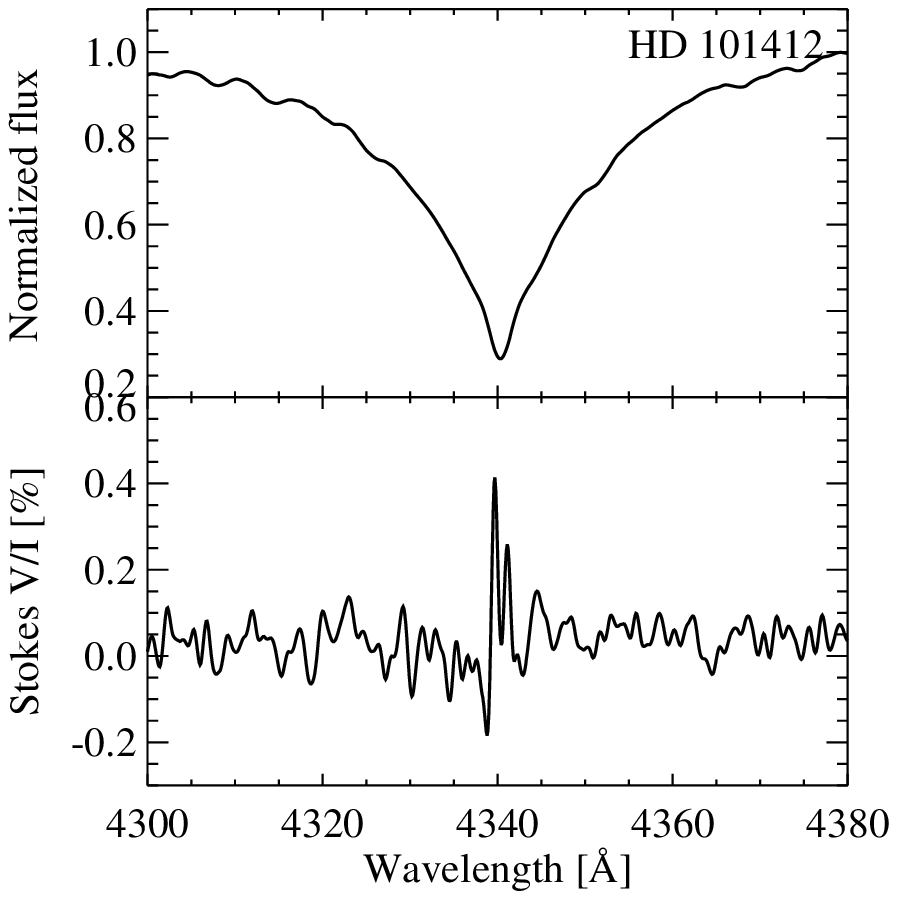}
\end{center}
\caption{
Stokes~$I$ and $V$ spectra of the Herbig Ae/Be star HD\,101412, with the largest detected magnetic field.
Upper panel: Zeeman features in H$_9$,  H$_8$, \ion{Ca}{ii} H\&K, and H$\epsilon$ profiles.
Lower panel: Stokes~$I$ and $V$ spectra in the vicinity of the 
H$\gamma$ line.
}
\label{fig:101412}
\end{figure}

The star HD\,101412, with the largest magnetic field strength measured in our sample stars,
shows a change of the field strength by $\sim$100\,G during two consecutive nights.
In Fig.~\ref{fig:101412} we present distinct Zeeman features detected at the positions of the hydrogen Balmer lines and 
the \ion{Ca}{ii} H\&K lines.
The H$\beta$ line in the Stokes~$I$ spectrum is contaminated by the presence of 
a variable emission in the line core and was not included in our measurements.

%\begin{figure*}
%\begin{center}
%\includegraphics[width=0.45\textwidth,angle=0,clip=]{hd101412_I3_2.eps}
%\includegraphics[width=0.45\textwidth,angle=0,clip=]{hd101412_I3_1.eps}
%\end{center}
%\caption{
%UVES spectra of the Herbig Ae/Be star HD\,101412 in the spectral regions
%around the \ion{Si}{ii} line $\lambda$6371.4 (left) and
%around the two \ion{Fe}{i} lines at $\lambda\lambda$4957.3 and 4957.6 (right).
%Note the strong increase of the line intensities at MJD53871.0.
%}
%\label{fig:101412_UVES}
%\end{figure*}  
%\begin{figure*}
%\begin{center}
%\includegraphics[width=0.45\textwidth,angle=0,clip=]{Halpha_hd101.eps}
%\includegraphics[width=0.45\textwidth,angle=0,clip=]{Hbeta_hd101.eps}
%\end{center}
%\caption{
%UVES spectra of the Herbig Ae/Be star HD\,101412 in the spectral regions
%around the \ion{Si}{ii} line $\lambda$6371.4 (left) and
%around the two \ion{Fe}{i} lines at $\lambda\lambda$4957.3 and 4957.6 (right).
%Note the strong increase of the line intensities at MJD53871.0.
%}
%\label{fig:101H}
%\end{figure*}
%\begin{figure*}
%\begin{center}
%\includegraphics[width=0.45\textwidth,angle=0,clip=]{metal_hd101.eps}
%\includegraphics[width=0.45\textwidth,angle=0,clip=]{silicon_hd101.eps}
%\end{center}
%\caption{
%UVES spectra of the Herbig Ae/Be star HD\,101412 in the spectral regions
%around the \ion{Si}{ii} line $\lambda$6371.4 (left) and
%around the two \ion{Fe}{i} lines at $\lambda\lambda$4957.3 and 4957.6 (right).
%Note the strong increase of the line intensities at MJD53871.0.
%}
%\label{fig:101others}
%\end{figure*}

\begin{figure}
\begin{center}
\includegraphics[width=0.45\textwidth,angle=0,clip=]{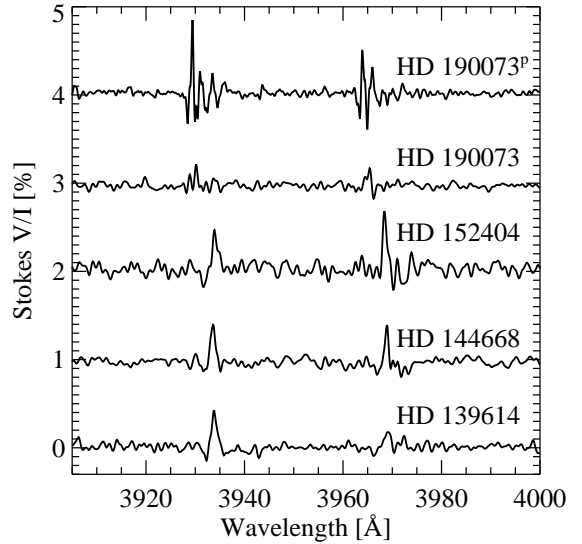}
\end{center}
\caption{
Stokes~$V$ spectra in the vicinity of the \ion{Ca}{ii} H\&K lines of the Herbig Ae/Be stars 
HD\,139614, HD\,144668,  HD\,152404, and  HD\,190073. At the top we present our previous observation of
 HD\,190073, obtained in May 2005. The amplitude of the Zeeman features in the \ion{Ca}{ii} H\&K lines 
observed in our recent measurement has decreased by $\sim$0.5\% compared to the previous 
observations.
}
\label{fig:V5}
\end{figure}

Similar Zeeman features at the positions of the \ion{Ca}{ii} H\&K lines were detected in four other
Herbig Ae/Be stars, HD\,139614, HD\,144668, HD\,152404, and HD\,190073. 
In Fig.~\ref{fig:V5}, we present the Stokes~$V$ spectra for these stars in the region around  the 
\ion{Ca}{ii} doublet, together with our previous observation of HD\,190073, obtained 
with FORS\,1 in 2005. As we already reported in our earlier studies (Hubrig et al.\ \cite{Hubrig2004b}, \cite{Hubrig2006b}, \cite{Hubrig2007b})
these lines are very likely formed at the base of the stellar wind, as well as in 
the accretion gaseous flow, and frequently display multi-component complex structures in both the Stokes~$V$
and Stokes~$I$ spectra.
In two Herbig Ae/Be stars, HD\,31648 and HD\,190073, such a structure was especially noticeable,
and from their study  we concluded that a magnetic field is present in both stars, 
but is most likely of circumstellar origin.
For HD\,31648, we detected a magnetic field $\left<B_{\rm z}\right>$\,=\,87$\pm$22\,G.
Using only the \ion{Ca}{ii} H\&K lines for the measurement of circular polarization in HD\,190073, 
we diagnosed a longitudinal magnetic field at a 2.8$\sigma$ level,  
$\left<B_{\rm z}\right>$\,=\,84$\pm$30\,G.  
Our previous magnetic field measurement for HD\,190073 of the order of 80\,G 
was in full agreement with the high resolution spectropolarimetric measurements obtained for this star
with ESPaDOnS (Catala et al.\ \cite{Catala2007}), who measured a longitudinal magnetic field
 $\left<B_{\rm z}\right>$\,=\,74$\pm$10\,G using metal lines. 
The authors report that they could not see a Stokes~$V$ signal in the \ion{Ca}{ii} H\&K lines, but the 
S/N ratio of their spectra is too low in that spectral region, resulting in a noise level of the order 
of 7$\times$10$^{-3}$ in Stokes~$V$ per spectral bin of 0.0025\,nm, which would not allow them to detect 
as weak a signal as seen in the LSD average, nor the level of signal reported by Hubrig et al.\ (\cite{Hubrig2006b}).
Our new observations reveal a rather large change (by $\sim$0.5\% of circular polarization) in the 
amplitude of the Zeeman features in the \ion{Ca}{ii} H\&K lines compared to our observations from May 2005 
(see the two upper spectra in Fig.~\ref{fig:V5}), indicating a conspicuous variability of the magnetic field. 

%\begin{figure}
%\begin{center}
%\includegraphics[width=0.45\textwidth,angle=0,clip=]{148898_pulkovo2.eps}
%\includegraphics[width=0.45\textwidth,angle=0,clip=]{148898_pulkovo1.eps}
%\end{center}
%\caption{
%The \ion{Ca}{ii} H\&K profiles (lower panel)
%   and H$\beta$ and H$\gamma$ profiles (upper panel) of HD\,148898 and the corresponding
%   Stokes~$V$ spectra.
%}
%\label{fig:he3}
%\end{figure}

Wade et al.\ (\cite{Wade2007}) mentioned the absence of Zeeman features  in the \ion{Ca}{ii} H\&K lines in 
their FORS\,1 spectra of the Herbig stars HD\,31648, HD\,144432, and HD\,144668.
On the other hand, in the same study, they report that their measurements were obtained using 
a variety of slit widths:
  38 of 73 stars were measured with slit widths of 0$\farcs$8 to 1\arcsec{}, 
  and 35 of their 73 stars were measured with a slit width of 
  0$\farcs$5. Using wide slits greatly degrades the resolution of the Stokes 
  V spectra, possibly affecting the deduced longitudinal
  magnetic fields.  Wade et al. (\cite{Wade2007}) looked for this effect in
  their study, generally not finding strong evidence for it.

\section{Discussion of individual stars with magnetic field detections}
\label{sect:discussion}

%Since the discoveries of magnetic fields in Herbig Ae/Be stars are rare, mostly due to their 
%faint magnitudes and the  weakness of detected magnetic 
%fields we decided to introduce this section where 
In the following we discuss the present knowledge of the various observed properties 
of stars with a magnetic 
field detection at a 3$\sigma$ level, placing greater emphasis on the discussion of stars with 
stronger magnetic fields.
 Since a few stars of our sample have been previously 
observed by Wade et al.\ (\cite{Wade2007}), we also compare our results with the results obtained 
by these authors.
% to search for a link between the magnetic nature and the fundamental stellar characteristics.
%Stronger emphasis is put on the characterization of stars with stronger magnetic fields, i.e.\ with 
%longitudinal fields stronger than $\sim$100\,G.

\subsection{Herbig Ae/Be stars with detected magnetic fields}
\label{subsect:discussion_withfield}

{\it PDS\,2:}
This star was observed with FORS\,1 in November 2004 by 
Wade et al.\ (\cite{Wade2007}). A weak magnetic field of the order of 130--140\,G was measured at
a 3$\sigma$ significance level
using metal lines and the full spectrum, but could not be diagnosed from the measurements 
using hydrogen Balmer lines.
We achieve a rather similar result: 
We detect a positive magnetic field $\left<B_{\rm z}\right>$\,=\,103$\pm$29\,G using the full spectrum,
but we are not able to detect the presence of a magnetic field using hydrogen lines.
%PDS\,2 is a pulsating variable with three detected pulsating frequencies (Bernabei et al.\ \cite{Bernabei2007}).
For this star, no spatially resolved observations of circumstellar matter or infrared spectral emission 
features were reported in the literature.

{\it HD\,97048:}
%This object is located in the Chamaeleon I dark cloud.
No detection at a 3$\sigma$ level was achieved by Wade et al.\ (\cite{Wade2007}).
However, in our observations, the magnetic field in this star is rather strong. 
We measure $\left<B_{\rm z}\right>$\,=\,161$\pm$42\,G 
using the full spectrum and $\left<B_{\rm z}\right>$\,=\,188$\pm$47\,G using hydrogen lines.
A large flaring disk with a mean disk inclination of 42.8$^\circ$ was reported by Doucet et al.\ 
(\cite{Doucet2007}) from observations with VISIR. 
No silicate emission band at 10\,$\mu$m was found in the mid-IR spectra of HD\,97048.
The disk has been coronographically imaged by Doering et al.\ (\cite{Doering2007}), and
its inclination angle appears to be closer to $i$=30$^\circ$ than the $i$=42.8$^\circ$ reported by Doucet et al.\ (\cite{Doucet2007}). 
The disk has a radial surface
brightness profile consistent with dust grain growth and settling,  
rather than a highly flared geometry.
No HH knots are visible in the coronographic ACS F606W image presented by Doering et  al.\ (\cite{Doering2007}). 
%\changea{SH: Carol, what is this image - we need to explain this in the text. The next sentence is also 
%not clear.}
This is consistent  with the at best low level of \ion{Mg}{ii}
emission in the archival IUE data.
% and supports the interpretation of HD\,97048 as displaying a transitional disk.
%Recent long slit 
%spectroscopic observations with TIMMI2 show that the aromatic emission features at long wavelengths 
%(i.e., 8.6 and 11.3\,$\mu$m) are extended.
No presence of a binary companion was detected in the study of Corporon \& Lagrange 
(\cite{Corporon1999}). The mid-IR spectrum is dominated by PAH emission,
%(e.g.\ Acke \& van den Ancker \cite{Acke2004}),
which arises mostly from an outer radius at 200--300\,AU (van Boekel 
et al.\ \cite{vanBoekel2004}). Both quiescent H$_2$ emission in the near-IR (Bary et al.\ \cite{Bary2008})
and rotational H$_2$ emission of warm gas in the mid-IR (Martin-Za\"idi et 
al.\ \cite{MartinZaidi2007}) were reported.
HD\,97048 was detected in an {\em XMM-Newton}
survey of Cha\,I. On basis of the negative results of all searches for binarity, Stelzer et al.\ (\cite{Stelzer2004}) 
ascribed the X-ray emission to the Herbig star.

{\it HD\,100546:}
In this star we measure
$\left<B_{\rm z}\right>$\,=\,89$\pm$26\,G using
the full spectrum and $\left<B_{\rm z}\right>$\,=\,87$\pm$28\,G using hydrogen lines.
No detection at a 3$\sigma$ level was achieved by Wade et al.\ (\cite{Wade2007}).
Although no stellar companions have been reported,
there seems to be evidence for a giant planet forming at $\sim$6.5\,AU (Acke \& 
van den Ancker \cite{Acke2006}).
%The star shows silicates in the mid-IR spectra. 
%It seems also to be an X-ray source. van den Ancker \cite{Acke2006}). 
The circumstellar disk was repeatedly resolved at 
optical, near-IR, mid-IR and mm wavelengths. Augereau et al.\ (\cite{Augereau2001})
derive a disk inclination of 51$\pm$3$^\circ$, while the values from 
other authors are consistent within error ranges. There is a circumstellar envelope 
surrounding the disk (e.g.\ Grady et al.\ \cite{Grady2001}). 
%\changea{SH: Carol, is there something else to mention about the envelope?}
Silicate emission 
and PAHs are seen in 10\,$\mu$m spectra (e.g.\ Meeus et al.\ \cite{Meeus2001}). Highly
excited H$_2$ gas was reported by Martin-Za\"idi et al.\ (\cite{MartinZaidi2008}), who estimate 
that the warm H$_2$ is located within about 1.5\,AU from the star.
%The star is detected in the X-ray (Feigelson et al.\ \cite{Feigelson2003},  
%also see Stelzer et al.\ \cite{Stelzer2006}) with L$_X$ typical of accreting Herbig Ae stars. 
The star is detected in X-rays (Feigelson et al.\ \cite{Feigelson2003};  
Stelzer et al.\ \cite{Stelzer2006}) with an $L_{\rm X}$ typical of Herbig Ae stars.
No jet is seen in HST  imagery in the optical
(Grady et al.\ \cite{Grady2001}; Ardila et al.\ \cite{Ardila2007})
or in Ly$\alpha$ (Grady et al.\ \cite{Grady2005a}).

{\it HD\,101412:}
Wade et al.\ (\cite{Wade2007}) measured a positive magnetic field of the order of 500\,G using hydrogen lines.
During the first night we measured
$\left<B_{\rm z}\right>$\,=\,$-$312$\pm$32\,G employing
the full spectrum and $\left<B_{\rm z}\right>$\,=\,$-$454$\pm$42\,G on hydrogen lines.
The magnetic field was also detected on the second night:  $\left<B_{\rm z}\right>$\,=\,$-$207$\pm$28\,G\
using the full spectrum and $\left<B_{\rm z}\right>$\,=\,$-$317$\pm$35\,G on hydrogen lines.
Interestingly, Wade et al.\ (\cite{Wade2007}) found a negative magnetic field of the same order 
when they used metal lines for the measurements.
Contrary to their results, our measurements show rather consistent 
results obtained using the full spectrum, metal lines and hydrogen lines, although the magnetic field 
from metal lines appears somewhat lower, probably due to CS contamination.
From the measurements during the first night, using exclusively metal lines, we obtain
$\left<B_{\rm z}\right>$\,=\,$-$158$\pm$50\,G, 
and $\left<B_{\rm z}\right>$\,=\,$-$95$\pm$46\,G for the second night.
%The star was observed 
%on two consecutive nights, exhibiting 
%a decrease by $\sim$100\,G in the magnetic field strength from the first night to the second.
%As such a noticeable change took place just within two days, we presume that the rotational 
%period should be just a couple of days.  

\begin{figure}
\begin{center}
\includegraphics[width=0.45\textwidth,angle=0,clip=]{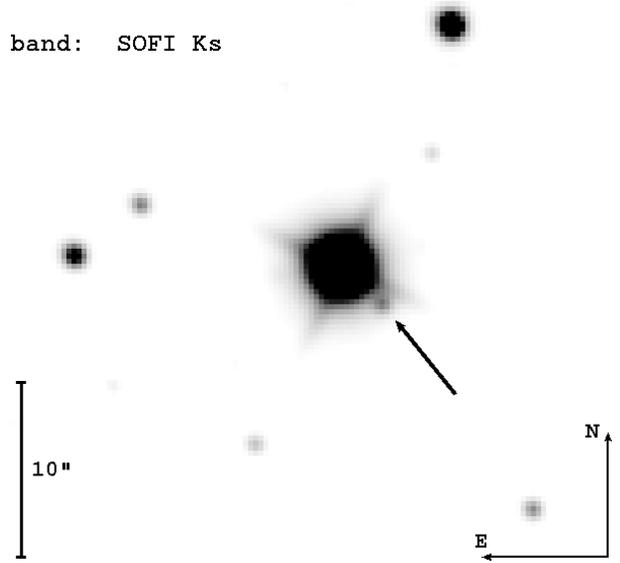}
\end{center}
\caption{
SOFI Ks-band image of a potential faint companion to HD\,101412.
The spatial resolution is 0\farcs{}29/pix and the separation amounts to 3\farcs{}3$\pm$0\farcs2.
}
\label{fig:101412SOFI}
\end{figure}

It is quite possible that HD\,101412 belongs to a visual binary system.
In Fig.~\ref{fig:101412SOFI} we present a Ks-band image of HD\,101412 obtained with SOFI at the NTT
in La Silla. The data were taken in May 2000 in the framework of the ESO programme
65.I-0097(A) and were retrieved from the ESO science data archive. A faint
candidate companion is detected at 3\farcs{}3$\pm$0\farcs{}2 separation to the
southwest. The spatial resolution of the SOFI images was 0\farcs{}29/pix. Whether there is a
real physical association between both stars remains to be confirmed.

%Corporon \& Lagrange (\cite{Corporon1999}) found 
%For this star we found in the ESO archive three high-resolution UVES spectra obtained in the framework of
%ESO programme 077.C-0521(A). An inspection of these spectra, recorded on three different dates,
%indicates variations in line intensities and line profiles of
%\ion{He}{i}, \ion{Si}{ii}, \ion{Fe}{i/ii}, and \ion{Ca}{i} lines.
%A few examples of these variations are presented in Fig.~\ref{fig:101412_UVES}.
%A few examples of these variations are presented in Figs.~\ref{fig:101H} and~\ref{fig:101others}.no spectroscopic binary, but note that the data for this source may have been 
%insufficient to detect radial velocity variations.
%\begin{figure*}
%\begin{center}
%\includegraphics[width=0.45\textwidth,angle=0,clip=]{Halpha_hd101.eps}
%\includegraphics[width=0.45\textwidth,angle=0,clip=]{Hbeta_hd101.eps}
%\end{center}
%\caption{
%UVES spectra obtained on two different MJD dates of the Herbig Ae/Be star HD\,101412 in the spectral regions
%around the H$\alpha$ line (left) and
%around the H$\beta$ line (right).
%%Note the strong increase of the line intensities at MJD53872.540
%}
%\label{fig:101H}
%\end{figure*}

\begin{figure}
\begin{center}
\includegraphics[width=0.45\textwidth,angle=0,clip=]{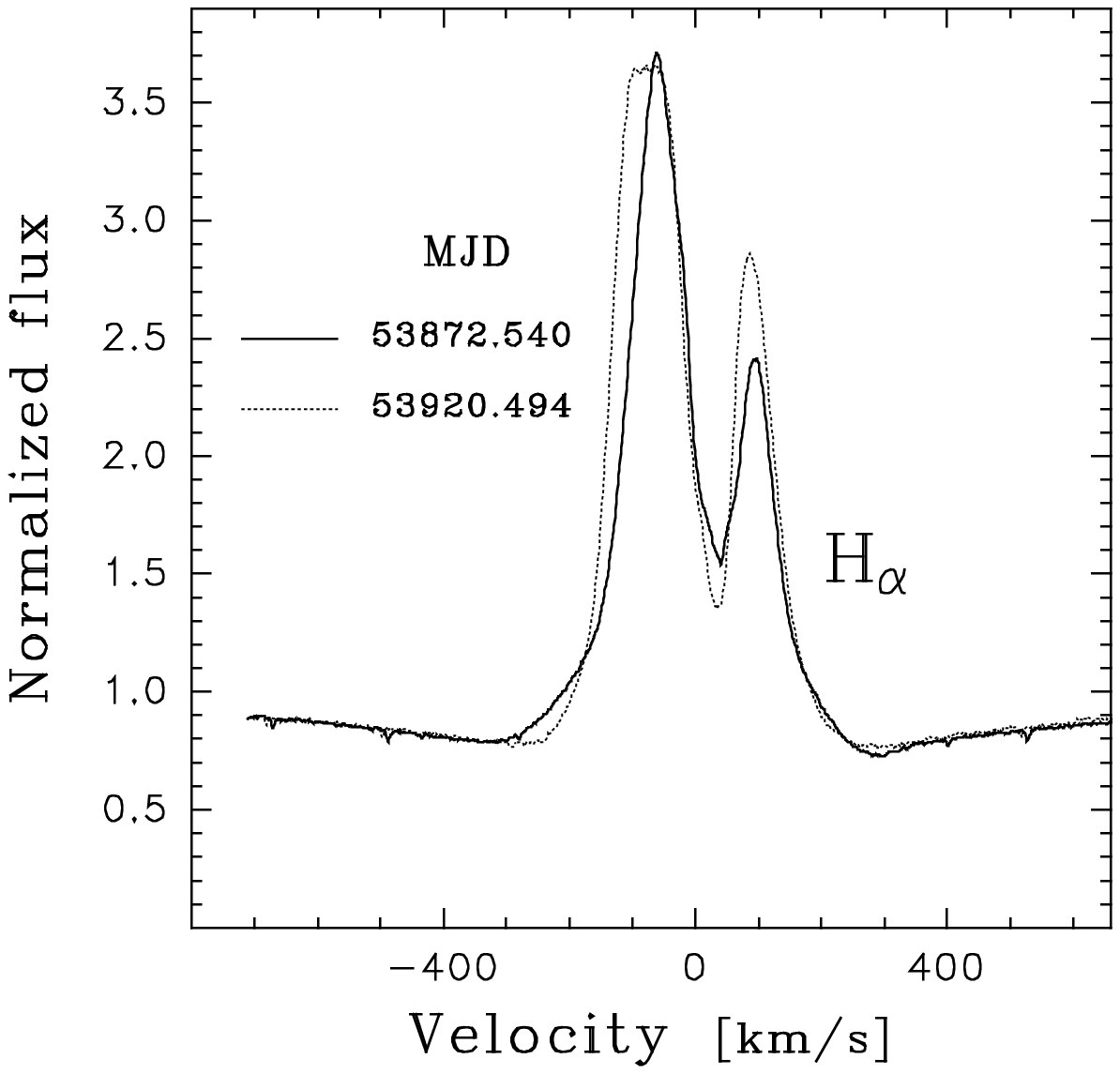}
\includegraphics[width=0.45\textwidth,angle=0,clip=]{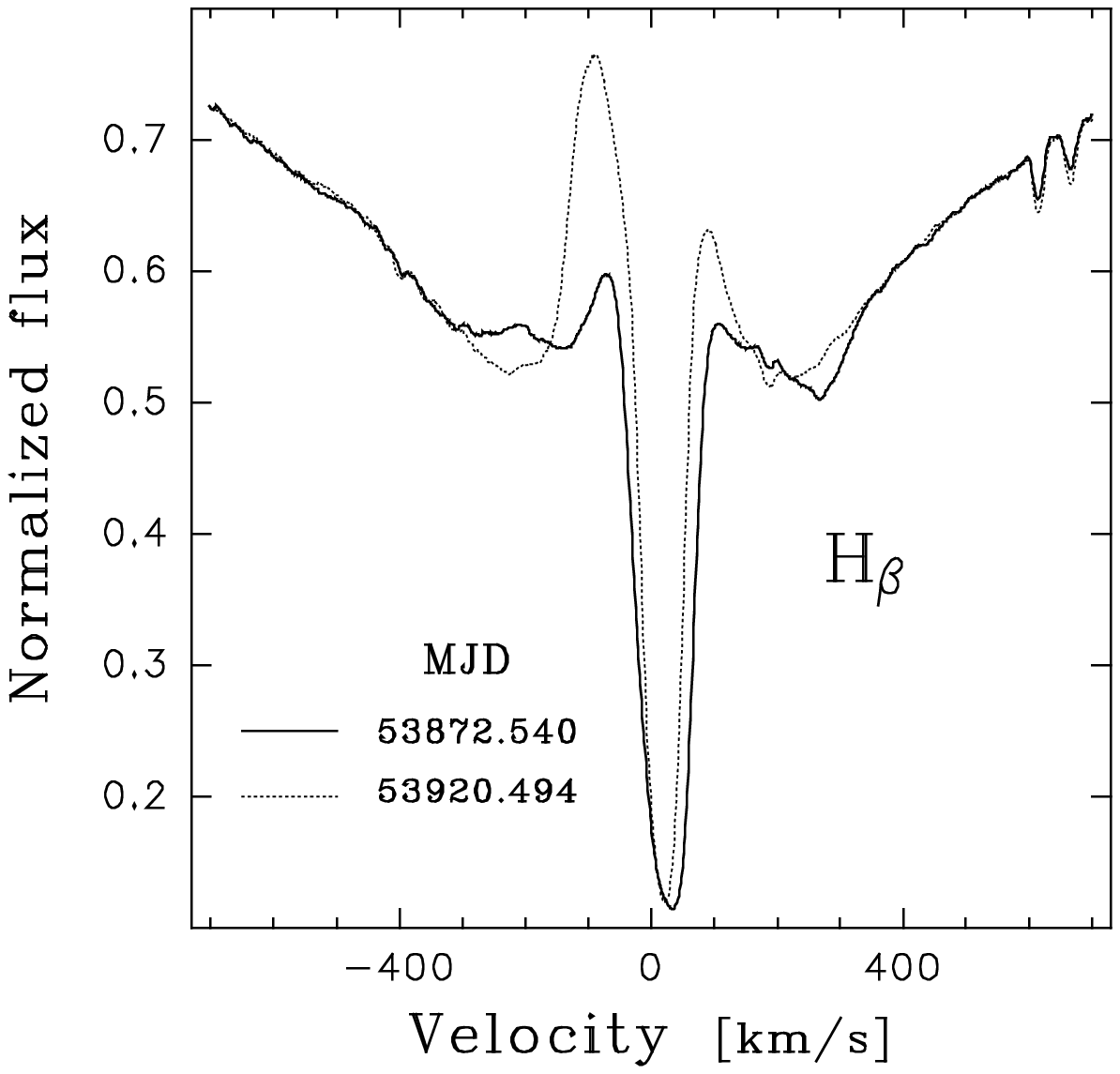}
\end{center}
\caption{
UVES spectra obtained on two different dates of the Herbig Ae/Be star HD\,101412 in the spectral regions
around the H$\alpha$ line (top) and
around the H$\beta$ line (bottom).
%Note the strong increase of the line intensities at MJD53872.540
}
\label{fig:101H}
\end{figure}

%\begin{figure*}
%\begin{center}
%\includegraphics[width=0.45\textwidth,angle=0,clip=]{metal_hd101.eps}
%\includegraphics[width=0.45\textwidth,angle=0,clip=]{silicon_hd101.eps}
%\end{center}
%\caption{
%UVES spectra of the Herbig Ae/Be star HD\,101412 in the spectral regions $\lambda\lambda$5526--5538\,\AA{}
%around the \ion{Cr}{ii}, \ion{Mg}{i}, and \ion{Fe}{ii} lines (left panel) and 
%around two at $\lambda\lambda$4957.3 and 4957.6 \ion{Si}{ii} line $\lambda$6371.4 (left) and
%around the two \ion{Si}{ii} lines at $\lambda$6347 (middle panel) and $\lambda$6371 (right panel).
%%\ion{Fe}{i} lines at $\lambda\lambda$4957.3 and 4957.6 (middle and right).
%Note the strong increase of the line intensities at MJD53920.494.
%}
%\label{fig:101others}
%\end{figure*}

\begin{figure}
\begin{center}
\includegraphics[width=0.45\textwidth,angle=0,clip=]{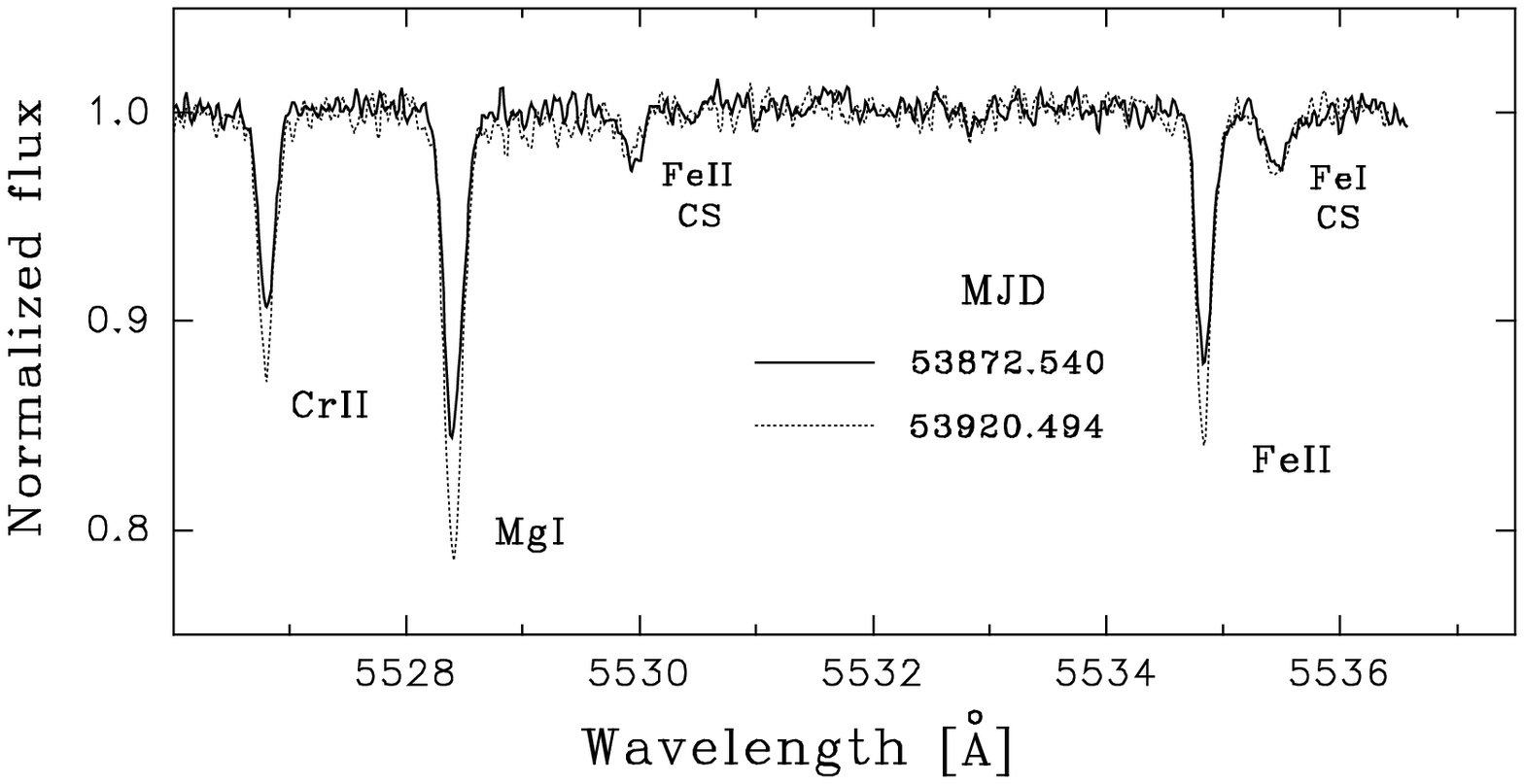}
\includegraphics[width=0.45\textwidth,angle=0,clip=]{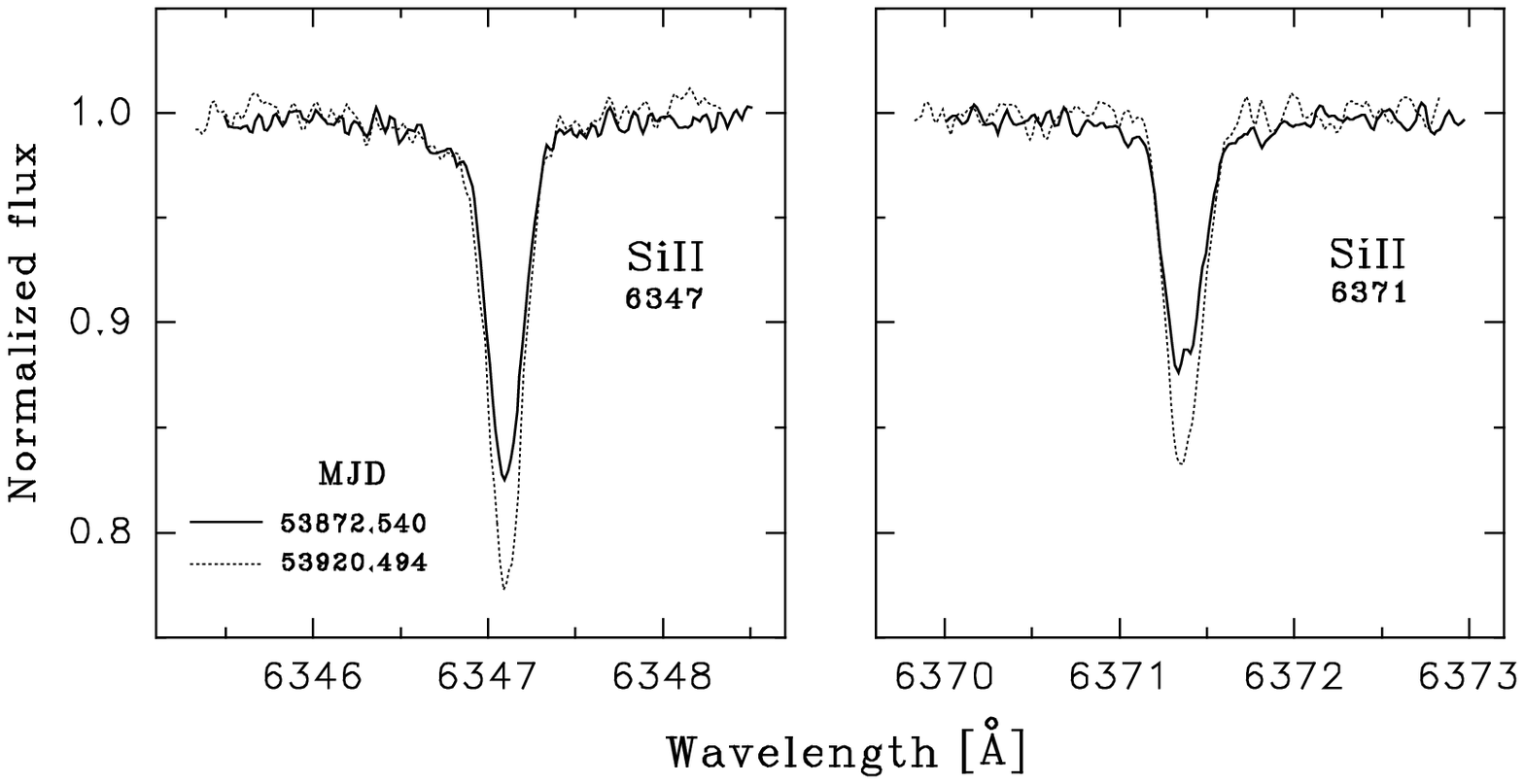}
\end{center}
\caption{
UVES spectra of the Herbig Ae/Be star HD\,101412 in the spectral regions $\lambda\lambda$5526--5538\,\AA{}
around the \ion{Cr}{ii}, \ion{Mg}{i}, and \ion{Fe}{ii} lines (upper panel) and 
%around two at $\lambda\lambda$4957.3 and 4957.6 \ion{Si}{ii} line $\lambda$6371.4 (left) and
around the two \ion{Si}{ii} lines at $\lambda$6347 (lower left panel) and $\lambda$6371 (lower right panel).
CS lines of \ion{Fe}{i} and \ion{Fe}{ii} are clearly visible in the upper spectrum.
%\ion{Fe}{i} lines at $\lambda\lambda$4957.3 and 4957.6 (middle and right).
Note the strong increase of the line intensities at MJD53920.494.
}
\label{fig:101others}
\end{figure}

The very low $v\sin i$=7\,km\,s$^{-1}$ suggests that we may see the disk nearly pole-on.
Since this star exhibits the strongest magnetic field and hence is of special interest,
we retrieved from the ESO archive three high-resolution UVES spectra of HD\,101412 obtained in the framework of
ESO programme 077.C-0521(A). An inspection of these spectra, recorded on three different dates,
indicates variations in line intensities and line profiles.
% and, actually, ee are not able to find any non-variable spectral line in the spectrum.
%\ion{He}{i}, \ion{Si}{ii}, \ion{Fe}{i/ii}, and \ion{Ca}{i} lines. 
All lines seem to be variable, and we were not able to find 
any non-variable spectral line in the spectrum.
The largest variations are observed between 
the dates MJD 53872.540 and MJD 53920.494.
%A few examples of these variations are presented in Fig.~\ref{fig:101412_UVES}.
A few examples of these variations are presented in Figs.~\ref{fig:101H} and~\ref{fig:101others}.
We tested a number of atmospheric models in the range: $T_{\rm eff}$=8000--11\,000\,K and $\log g$=4.0--4.5.
The studied UVES spectra cover both H$\alpha$ and H$\beta$ Balmer lines. However, only the wings 
of the H$\beta$ line can be used to obtain the atmospheric parameters, as
the H$\alpha$ line exhibits strongly variable emission.
% are overlapped by emission up to ± 1000 km/s,
We obtain as a best fit $T_{\rm eff}$=10\,000\,K, $\log g$=4.2--4.3 and a $v \sin i$ value of about 
5\,km\,s$^{-1}$.
This result is in good 
agreement with the study of Guimar\~aes et al.\ (\cite{Guimaraes2006}) who 
published $T_{\rm eff}$=10\,000$\pm$1000\,K, $\log g$=4.1$\pm$0.4, and $v$\,sin\,$i$=7$\pm$1\,km\,s$^{-1}$.
%%V.457, 581 (10000+/-1000, 4.1+/-0.4, $v \sin i$ = 7 ± 1 km/s)
A good fit is also obtained using the model $T_{\rm eff}$=9000\,K and  $\log g$=4.0, but the synthetic spectrum
for these parameters shows many narrow lines which do not appear in the observed UVES
spectra. The spectrum is heavily contaminated by CS lines. Two such lines belonging to \ion{Fe}{i} and 
\ion{Fe}{ii} are clearly visible in Fig.~\ref{fig:101others}.
%We tried to change chemical abundances, but this did not help. So, this model has to be rejected.
%All narrow metal lines can be fitted with models Teff = 8000-9000 K, but with smaller
%depth, having line intensities smaller by a factor of 4 four for Teff = 8000\,K and
%by a factor of two for Teff = 9000\,K. For the model 10000 K we do not 
%see in the synthetic spectrum additional narrow lines visible in the observed spectrum.
%Also the changes of metal abundances give no result.
% because they must be followed by a change of relative intensities of
%metal/non metal lines, but this is not observed.
%By the way, I did not found wide line components of photospheric lines. So, 

As we mentioned above, the low $v \sin i$ value suggests that we observe HD\,101412 close to pole-on, or 
that the star is rotating very slowly.
The inclination angles of disks of Herbig Ae/Be stars (which are expected to be identical with the inclination angle
of the stellar rotation axis) can be reliably derived only for resolved observations of disks.
They are usually determined from millimeter observations, coronographic imagery, or near-IR  
interferometry. 
%No such data exist for HD\,101412.
On the other hand, the orientation of the disk can be constrained using the
emission profile shapes.
Type I P Cygni profiles or single emissions all have $i<40^{\circ}$.
Stars which alternate between type I and type III profiles have $40<i<55^{\circ}$, and stars with double  
emission all have $i>55^{\circ}$. This means that even for
objects for which no good disk imagery exists, it is at least possible   
to constrain their inclination angles from the study of \ion{Mg}{ii} spectral line profiles in the UV or from the line profile
shape of H$\alpha$ (e.g., Finkenzeller \& Mundt \cite{FinkenzellerMundt1984}; Dunkin et al.\ \cite{Dunkin1997}).
The shape of H$\alpha$ in the UVES spectra of HD\,101412 suggests an 
inclination angle $i>55^{\circ}$, i.e.\ the star is viewed far from pole-on, in agreement with the recently 
published value $80^{\circ}\pm7^{\circ}$ by Fedele et al.\ (\cite{Fedele2008}) who used VLTI/MIDI observations.
The slow rotation could in principle be explained by binarity, where the components are synchronized
with $P_{\rm rot}=P_{\rm orb}$. However, we detect neither spectral lines of the companion 
nor any variability of radial velocities in the UVES spectra. It cannot be excluded, though, that 
braking of the star's rotation
is due to its rather strong magnetic field, in analogy with the slow rotation of magnetic Ap and Bp stars.
%We are not able to find any non-variable spectral line in the spectrum. 
Since the intensities of lines of different elements
do not show opposite behaviour, which is usually observed in chemically peculiar Ap and Bp stars,
 we exclude the presence of 
chemical spots on the stellar surface. Also the study of temperature sensitive and insensitive 
lines indicates the absence of temperature spots. Therefore we conclude that the observed spectrum variability
has a CS origin.  
Silicate emission and lower amounts of PAHs are seen in mid-IR spectra 
(e.g., van Boekel et al.\ \cite{vanBoekel2005}; Geers et al.\ \cite{Geers2007}), while the SED is consistent 
with a flat disk type (Geers et al.\ \cite{Geers2007}). 

{\it HD\,135344B=SAO\,206462:}
The detected magnetic field in this star is very weak. 
We observe a change of the polarity between the first and the second observing night, but the measurement on the
first night, using the full spectrum, resulted in a strength of the longitudinal magnetic field
$\left<B_{\rm z}\right>$\,=\,32$\pm$15\,G which is only at a 2.1$\sigma$ significance level. 
On the second night, the magnetic field was detected  at a 3.2$\sigma$ level 
using the full spectrum, $\left<B_{\rm z}\right>$\,=\,$-$38$\pm$12\,G.
As is mentioned in the literature, there has been frequent confusion 
between HD\,135344 and the $\sim$20\arcsec{} close, nearby IR 
source SAO\,206462, a mid-F star (Coulson \& Walther \cite{Coulson1995}), also referred to as HD\,135344B. 
Thus SAO\,206462 was frequently improperly called HD\,135344. 
A further binary pair was found at 
$\sim$5\farcs8 separation from SAO\,206462 by Augereau et al.\ (\cite{Augereau2001}). 
Mid-IR spectra show silicate absorption around 10\,$\mu$m (e.g.\ Acke \& 
van den Ancker \cite{Acke2004}) and PAHs (e.g.\ Sloan et al.\ \cite{Sloan2005}). 
%Doucet et al.\ (\cite{Doucet2006}) 
%derived the disk inclination from mid-IR images to be 46$\pm$5$^\circ$,
Dent et al.\ (\cite{Dent2005}) derived a disk inclination of 11$^\circ$, obtained from measurements of 
CO gas. The presence of warm H$_2$ gas is reported by Thi et al.\ (\cite{Thi2001}), which
should be treated with caution due to the low S/N and low spatial resolution of ISO data.

This star has recently been identified as the host of a transitional or pre-transitional
disk (Brown et al.\ \cite{Brown2007};
Pontoppidan et al.\ \cite{Pontoppidan2008}).
%and is viewed at low inclination (Dent et al.\ 2005).
The disk is coronographically detected at 1.1\,$\mu$m and 1.6\,$\mu$m (Grady et al., in preparation) 
with a radial surface brightness profile consistent with grain growth and settling. No jet  
is seen in HST optical coronographic
imagery (Grady et al.\ \cite{Grady2005b}).
FUV excess light and emission were detected by FUSE, with a mass-accretion
rate reconcilable with the Br$\gamma$ measurements (Garcia Lopez \cite {GarciaLopez2006};  
Grady et al., in preparation).
The bulk of the FUV emission is consistent with stellar activity, rather than with accretion. 
%L$_X$ is estimated to be $\sim$29.6 from a ROSAT HRI detection (Grady et al., in preparation).
%The Doucet et al.\ (2006) inclination is flat out wrong, and should not  be quoted here. I would also be
%cautious about citing the Thi et al.\ (2001) H$_2$ - it hasn't held up for the other sources they claim detections
%for.  The Spitzer data do not suggest silicate absorption:  rather no  
%silicate feature in the presence of PAH emission.  The 5\farcs{}8 candidate binary companion is NOT COMMON  
%PROPER MOTION.

%\begin{figure*}
%\begin{center}
%\includegraphics[width=0.45\textwidth,angle=0,clip=]{hd144668_IVa.eps}
%\includegraphics[width=0.45\textwidth,angle=0,clip=]{hd144668_IVb.eps}
%\end{center}
%\caption{
%Stokes~$I$ and $V$ spectra of the Herbig Ae/Be star HD\,144668. 
%%in the vicinity of the H$\gamma$ line. 
%Left panel: Zeeman feature in the H$\gamma$ line. Right panel: Zeeman feature in the H$\beta$ line.
%}
%\label{fig:144668}
%\end{figure*}

\begin{figure}
\begin{center}
\includegraphics[width=0.45\textwidth,angle=0,clip=]{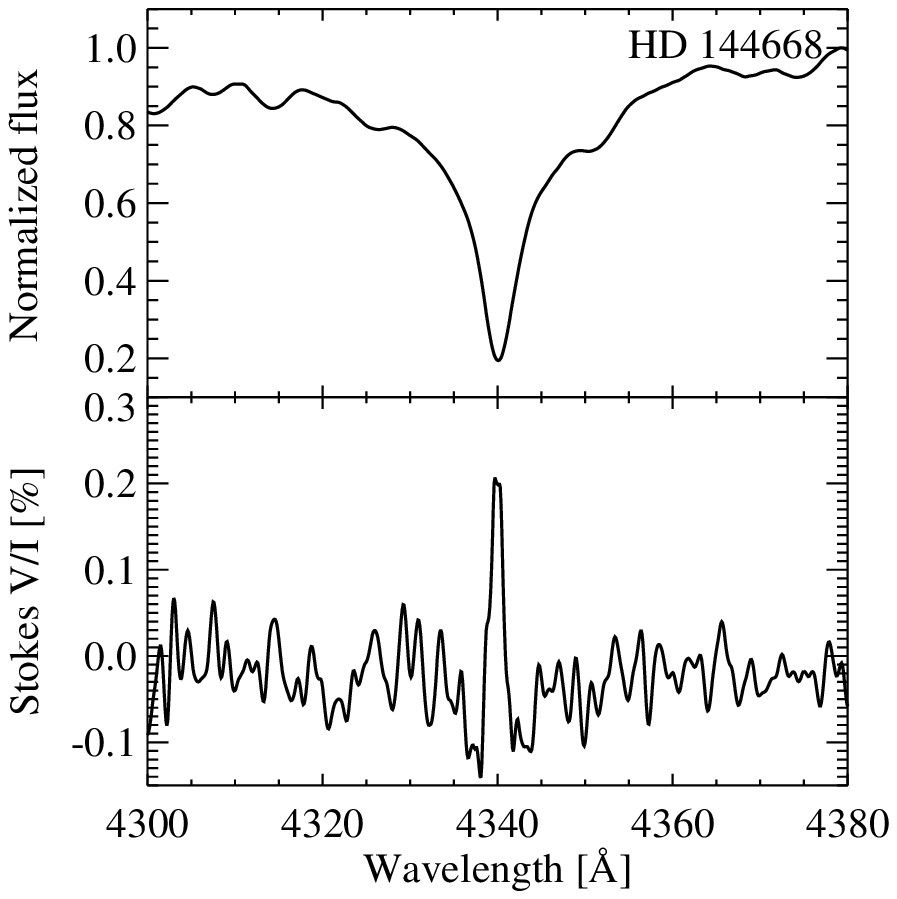}
\includegraphics[width=0.45\textwidth,angle=0,clip=]{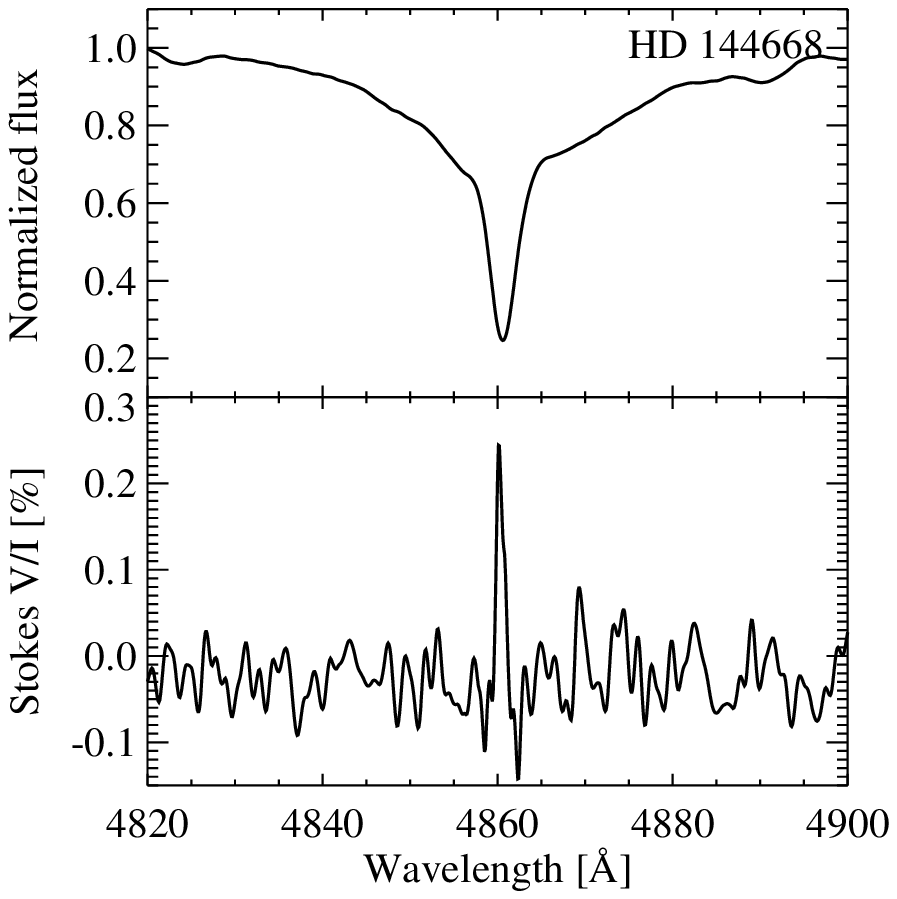}
\end{center}
\caption{
Stokes~$I$ and $V$ spectra of the Herbig Ae/Be star HD\,144668. 
%in the vicinity of the H$\gamma$ line. 
Upper panel: Zeeman feature in the H$\gamma$ line. Lower panel: Zeeman feature in the H$\beta$ line.
}
\label{fig:144668}
\end{figure}

{\it HD\,144668=HR\,5999:}
In our previous study (Hubrig et al.\ \cite{Hubrig2007a}), we showed that
the magnetic field of this star is variable, with the strength 
varying between $-$75\,G and +166\,G. Strong Zeeman features
are clearly visible at the position of the \ion{Ca}{ii} H\&K lines and hydrogen Balmer lines, similar 
to those detected in our previous observations (see Fig.~\ref{fig:V5}). We confirm the presence 
of the magnetic field, measuring $\left<B_{\rm z}\right>$\,=\,$-$62$\pm$18\,G using
the full spectrum and $\left<B_{\rm z}\right>$\,=\,$-$92$\pm$27\,G using hydrogen lines.
In stellar applications, the Stokes~$V$ profiles present disk-integrated observations.
Generally, stellar magnetic fields are not symmetric relative to the rotation axis, so that the polarization 
signal changes with the same period as the stellar rotation. The most simple modeling 
includes a magnetic field approximated by a dipole with an axis inclined to the rotation axis.
Recently, Auri\`ere et al.\ (\cite{Auriere2007}) presented measurements of weak stellar magnetic 
fields using the cross-correlation technique ``Least Squares Deconvolution'', originally introduced 
by Donati et al.\ (\cite{Donati1997}). They showed that  the presence of a weak magnetic field can be diagnosed 
from the detection of Stokes~$V$ Zeeman signatures, even if the longitudinal magnetic field 
is very small, as is usually observed during a cross-over phase.
In our observations of a few Herbig Ae/Be stars, even at the very low resolution of FORS\,1, similar 
Stokes~$V$ Zeeman features are observable in strong 
\ion{Ca}{ii} H\&K and hydrogen Balmer lines, indicating the presence of a cross-over phase.
In Fig.~\ref{fig:144668} we show an example of Stokes~$V$ line profiles of H$\gamma$ and H$\beta$ in HD\,144668,
confirming the presence of a magnetic field in this star at a rotation phase where the longitudinal magnetic field
is rather small.

HD\,144668 belongs to a quadruple system (Dommanget \& Nys \cite{Dommanget1994}), although the 
orbits and physical relations between the companions are not certain. The nearest neighbour at 
a separation of 1\farcs{}4 appears to be a T\,Tauri star (Stecklum et al.\ 
\cite{Stecklum1995}). A broad but rather shallow 10\,$\mu$m silicate emission feature 
indicates grain growth (van Boekel et al.\ \cite{vanBoekel2005}). Preibisch 
et al.\ (\cite{Preibisch2006}) derived an inclination of 58$^\circ$ from MIDI/VLTI observations 
and found that the disk is very compact with an outer edge of $\sim$2.6\,AU, 
potentially caused by a further, even closer, yet undetected companion. 
Recently, HD\,144668 was for the first time resolved from its T\,Tauri companion
in X-rays (Stelzer et al.\ \cite{Stelzer2008}). This observation showed that the bulk of
the X-ray emission comes from the companion, but a weak source is associated with HD\,144668 (or the 
hypothesized sub-arcsecond companion).
%An accretion rate of $10.0 \times 10^{-7}$ M$_{\sun}$/yr is given by Brittain et 
%al. (\cite{Brittain2007}), which constitutes the upper range in their sample of 14 PMS stars.
%This star is the second pulsating Herbig Ae/Be star in our sample. 

{\it HD\,150193:}
No attempt had been made to measure the magnetic field in this object.
The detected magnetic field is comparatively strong: we measure 
$\left<B_{\rm z}\right>$\,=\,$-$144$\pm$32\,G using
the full spectrum and $\left<B_{\rm z}\right>$\,=\,$-$252$\pm$48\,G using hydrogen lines.
A binary companion at 1\farcs{}1 distance (Reipurth \& Zinnecker \cite{Reipurth1993}) 
is a T\,Tauri star (e.g.\ Carmona et al.\ \cite{Carmona2007}). Fukagawa et 
al.\ (\cite{Fukagawa2003}) resolved the circumstellar disk in the near-IR and derived 
a disk inclination of 38$\pm$9$^\circ$. Broad 10\,$\mu$m silicate emission is 
seen in mid-IR spectra (e.g.\ Acke \& van den Ancker \cite{Acke2004}). Warm H$_2$ gas 
is reported by Thi et al.\ (\cite{Thi2001}).
Similar to the case of HD\,144668, resolved {\em Chandra} images show that 
most of the X-ray emission comes from the T\,Tauri companion, but weak emission is 
attributed to the Herbig star (Stelzer et al.\  \cite{Stelzer2006}).

{\it HD\,176386:}
A magnetic field is definitely present: we measure 
$\left<B_{\rm z}\right>$\,=\,$-$119$\pm$33\,G using
the full spectrum and $\left<B_{\rm z}\right>$\,=\,$-$121$\pm$35\,G using hydrogen lines.
This star is part of a triple system (Dommanget \& Nys \cite{Dommanget1994}). 
%[Comment: There is a problem in SIMBAD, which one of these is actually HD\,176386.
Grady et al.\ (\cite{Grady1993}) reported evidence for accretion of circumstellar gas onto the star.
Martin-Za\"idi et al.\ (\cite{MartinZaidi2008}) detected
warm H$_2$ gas.
This source shows the highest H$_2$ column density in their entire sample.
Both IR (Siebenmorgen et al.\ \cite{Siebenmorgen2000}) and far-UV 
spectroscopy (Martin-Za\"idi et al.\ \cite{MartinZaidi2008}) suggest that the circumstellar 
matter has the form of an envelope rather than the form of a disk. PAHs are detected in 
mid-IR spectra (e.g.\ Siebenmorgen et al.\ \cite{Siebenmorgen2000}).
Surprisingly, HD\,176386 is one of the few Herbig Ae/Be stars confirmed to be X-ray dark
($\log{L_{\rm X}}{\rm [erg/s]} < 28.6$).  
Stelzer et al.\ (\cite{Stelzer2006}) showed on basis of {\em Chandra} imaging 
that apparent X-ray detections of this star in previous lower-resolution
observations are to be attributed to the companion at $4^{\prime\prime}$ separation.

{\it HD\,190073:}
A magnetic field of $\left<B_{\rm z}\right>$\,=\,74$\pm$10\,G was detected using photospheric metal lines 
with ESPaDOnS (Catala et al.\ \cite{Catala2007}).
Our old measurements (Hubrig et al.\ \cite{Hubrig2006a}) revealed a magnetic field of the order of 80\,G at the
2.8$\sigma$ level, in full agreement with the high resolution ESPaDOnS spectropolarimetric data, whereas 
Wade et al.\ (\cite{Wade2007}) failed to detect the magnetic field in both previous measurements 
carried out with FORS\,1 in November 2004.
Our new measurements $\left<B_{\rm z}\right>$\,=\,104$\pm$19\,G using the full spectrum and
$\left<B_{\rm z}\right>$\,=\,120$\pm$32\,G using hydrogen lines confirm the
results presented by Catala et al.\ (\cite{Catala2007}).
%This star has been discussed in the literature since more than 70 
%years with various classifications. Being previously often considered as a 
%post-main sequence giant, van den Ancker et al.\ (\cite{vandenAncker1998}) described 
%HD\,190073 as a Herbig Ae/Be star, which is now generally accepted. 
%Considering its large distance, for which only a lower limit is known due 
%to an imprecise Hipparcos parallax (van den Ancker et al.\ \cite{vandenAncker1998}), this 
%unusual object must be much more luminous than other Herbig stars of similar spectral type. 
Baines et al.\ (\cite{Baines2006}) reported the presence of a potential 
companion. Silicates and PAHs are seen in the mid-IR (e.g.\ Boersma et al.\ 
\cite{Boersma2008}). Eisner et al.\ (\cite{Eisner2004}) showed that interferometric visibilities are 
consistent with a disk close to face-on, although a significant nonzero 
inclination cannot be ruled out.

%{\it HD\,31648:}
%This target is a 7 Myr old Herbig Ae star with an accretion rate of 1E-8 solar masses/yr, 
%intermediate between younger, accreting stars and older stars with transitional disks.
%Can we test the hypothesis that the Lx is correlated with accretion signatures, rather than 
%originating in coronal emission associated with a decaying dynamo?

%\subsection{Stars with debris disks}
%{\it HD\,39060:}
%{\it HD\,181327:}

\subsection{Stars with measurements suggestive of the presence of magnetic fields}

{\it HD\,47839=15\,Mon:}
This is a massive O star classified as a pre-main sequence star in SIMBAD.
According to Kaper (\cite{Kaper1996}) and Walborn (\cite{Walborn2006})
it shows distinct peculiarities in the spectra,
which could be typical for stars possessing magnetic fields.
The measured magnetic field is of positive polarity at a 2.6$\sigma$ significance 
level, using the full spectrum. The star should be considered as an important target for follow-up studies 
of magnetic fields in massive stars.

{\it HD\,97300:}
Even though the measured field is only at the $\sim$2$\sigma$ level, we detect a few 
distinct Zeeman features in the Stokes~$V$ spectrum, supporting the evidence of a magnetic field.
This star is a good candidate for future magnetic field measurements.
%This object is located in the Cha~I star forming region. 
%It was also detected as an X-ray source 
%in ROSAT and CHANDRA observations (Hamaguchi et al.\ \cite{Hamaguchi2005}; Stelzer et al.\ \cite{Stelzer2006}).
No quiescent H$_2$ gas was found (Bary et al.\ \cite{Bary2008}). Siebenmorgen et 
al.\ (\cite{Siebenmorgen1998}) reported the presence of a large circumstellar elliptical ring of PAH emission with 
thousands of AU in diameter and an indication for the possible existence of a candidate companion.
It was also detected as an X-ray source 
in {\em ROSAT} and {\em Chandra} observations (Hamaguchi et al.\ \cite{Hamaguchi2005}; Stelzer et al.\ \cite{Stelzer2006})

{\it HD\,139614:}
Our previous measurements of HD\,139614 revealed the presence of a weak magnetic field
in the range from $-$116\,G to $-$450\,G.
(Hubrig et al.\ \cite{Hubrig2004b}; \cite{Hubrig2006b}; \cite{Hubrig2007b}).
The current diagnosis of magnetic fields in this star yielded no detection at a 3$\sigma$ level, neither in 
Balmer nor in metal lines, with error bars as low as 25\,G.  
In analogy with our previous detections, the measured magnetic field is of negative polarity and 
\ion{Ca}{ii} H\&K lines in HD\,139614 show distinct Zeeman features. 
%similar to those detected in our previous studies. 
The nearly featureless mid-IR continuum of HD\,139614 shows small amounts of silicates,
but no PAH (Keller et al.\ \cite{Keller2008}). 

{\it HD\,144432:}
For this object we measured in the past a magnetic field in the range from $-$94\,G to $-$119\,G
(Hubrig et al.\ \cite{Hubrig2004b}; \cite{Hubrig2006b}; \cite{Hubrig2007b}).
Dent et al.\ (\cite{Dent2005}) found CO gas 
and derived a nearly face-on disk orientation. HD\,144432 is a binary 
with $\sim$1\farcs{}4 separation (Dommanget \& Nys \cite{Dommanget1994}). Carmona et al.\ (\cite{Carmona2007}) 
reported the detection of H${\alpha}$ emission and Li absorption in the K5Ve 
T\,Tauri companion. 
The companion is an X-ray source, and weaker X-ray emission is
also detected from the position of the Herbig Ae/Be star (Stelzer et al.\ \cite{Stelzer2008}).
The 10\,$\mu$m silicate emission band shows rather pristine dust, 
while the SED points towards a flat disk, as reported by Meeus et al.\ (\cite{Meeus2001}).

{\it HD\,152404=AK\,Sco:}
The magnetic field $\left<B_{\rm z}\right>$\,=\,107$\pm$37\,G is detected at the 2.9$\sigma$ level using 
the full spectrum. The Stokes~$V$ spectrum exhibits distinct Zeeman features 
at the position of higher number Balmer lines and of the \ion{Ca}{ii} H\&K lines 
(see for example Fig.~\ref{fig:V5}). 
This star is one of the best candidates for future spectropolarimetric observations. 
The {\em Chandra} image presents one faint X-ray source associated with 
HD\,152404 (Stelzer et al.\ \cite{Stelzer2006}).
Some authors refer to this target as a T\,Tauri star, as its spectral 
type may be at the borderline. The 10\,$\mu$m silicate emission feature indicates
very pristine dust (e.g.\ Przygodda et al.\ \cite{Przygodda2003}).

{\it VV\,Ser:}
%This is the third pulsating star in our sample (Ripepi et al.\ \cite{Ripepi2007}). 
The magnetic field is diagnosed 
only at a significance level of 2.7$\sigma$, but additional measurements are desirable.
Alonso-Albi et al.\ (\cite{AlonsoAlbi2008}) 
carried out observations at millimeter and centimeter wavelengths towards VV\,Ser using the Plateau de Bure 
Interferometer and the Very Large Array to compute the SED from the near infrared to centimeter wavelengths. 
The modeling of the full SED has provided insights into the dust properties and a more accurate value for 
the disk mass. The mass of dust in the disk around VV\,Ser was found to be about 
4$\times$10$^{-5}$\,M$_{\sun}$, i.e.\ 
400 times larger than previous estimates. Due to the faintness of the star ($m_V$=11.6), only two series of two 
sub-exposures were taken, each one with an exposure time of 15~min. 
Pontoppidan et al.\ (\cite{Pontoppidan2007b}) compared Spitzer infrared data with a disk model 
and found that the disk is nearly edge-on and self-shadowed by a puffed-up 
inner disk rim. The 10\,$\mu$m silicate emission feature is rather weak.

\subsection{The observed circumstellar properties of Herbig Ae/Be stars with magnetic field 
detections discussed in previous studies}
\label{sect:prev_detect}

{\it HD\,31648:}
Based on interferometric observations, Eisner et al.\ (\cite{Eisner2004}) determined a disk 
inclination of $\sim$30$^\circ$ and reported that a binary model can be ruled 
out with a high degree of confidence. Further, Eisner (\cite{Eisner2007}) found 
a visibility increase in the Br$\gamma$ line for this star, as expected for magnetospheric accretion.
Pi\`etu et al.\ (\cite{Pietu2007}) had obtained
an inclination angle of $\sim$35$^\circ$ from mm measurements. 
%Various further resolved detections at mm wavelengths exist (e.g.\ Hamidouche et al.\ \cite{Hamidouche2006}).
Silicate emission was detected in the mid-IR around 10\,$\mu$m (e.g.\ Acke \& 
van den Ancker \cite{Acke2004}), while the presence of warm H$_2$ gas is reported 
by Thi et al.\ (\cite{Thi2001}).
%silicate emission is seen in the mid-IR around 10\,$\mu$m (e.g.\ Acke \& 
%van den Ancker \cite{Acke2004}). Warm H$_2$ gas is reported by Thi et al.\ (\cite{Thi2001}).

%{\it HD\,36112:}
%Thomas et al.\ (\cite{Thomas2007}) found a binary at 2\farcs{}3 distance by adaptive optics
%imaging. Martin-Za\"idi et al.\ (\cite{MartinZaidi2008}) give a disk inclination of 53-57$^\circ$,
%derived from far-UV spectra, which is very different from the inclination 
%of $\leq$10$^\circ$ given by Dent et al.\ (\cite{Dent2005}), derived from CO emission.
%Silicates and PAHs are present in mid-IR spectra (Boersma et al.\ \cite{Boersma2008}).
%Warm H$_2$ gas is detected by Thi et al.\ (\cite{Thi2001}). Isella et al.\ (\cite{Isella2008}) 
%explain the H-band excess as emission from a hot innermost gaseous 
%accretion disk, reaching down to 0.1\,AU towards the star.

{\it HD\,104237:}
This star is at least a quintuple system (Grady et al.\ \cite{Grady2004}), where some
components are T\,Tauri stars. A further 
spectroscopic companion is suggested by observations of 
B\"ohm et al.\ (\cite{Boehm2004}) and Baines et al.\ (\cite{Baines2006}). 
The Herbig  star itself is the brightest X-ray source of the group (Stelzer et al.\ \cite{Stelzer2006}).
The disk is inclined by
$\sim$18$^\circ$ (Grady et al.\ \cite{Grady2004}). Silicate emission is seen in mid-IR 
spectra (e.g.\ Acke \& van den Ancker \cite{Acke2004}). Martin-Za\"idi et al.\ (\cite{MartinZaidi2008}) 
found excited and hot H$_2$ gas from far-UV spectra, while Tatulli et 
al.\ (\cite{Tatulli2007}) explain the origin of the Br$\gamma$ emission by an inner disk 
wind, originating at about 0.5\,AU from the star.

{\it HD\,200775:}
This object has a companion detected at 2\farcs{}25 (Pirzkal et al.\ \cite{Pirzkal1997}), while spectroscopic
observations suggested further binarity of the main component (e.g.\
Miroshnichenko et al.\ \cite{Miroshnichenko1998}), which was repeatedly discussed in the literature
and later confirmed by Monnier et al.\ (\cite{Monnier2006}) using interferometry.

{\it V380\,Ori:}
This source is actually a triple system with separations of 3\farcs{}0 and 
5\farcs{}0 (Dommanget \& Nys \cite{Dommanget1994}), while the primary itself has a companion 
at 0\farcs{}15 (Leinert et al.\ \cite{Leinert1997}) which is probably another Herbig Ae/Be star. 
Even the superior spatial resolution of {\em Chandra} is not able to resolve this binary,
such that the origin of the X-ray source remains obscure (Stelzer et al.\ \cite{Stelzer2006}).
The broad but shallow 10\,$\mu$m silicate 
feature indicates grain growth, as confirmed by van Boekel et al.\ (\cite{vanBoekel2005}).

{\it BF\,Ori:}
As Mora et al.\ (\cite{Mora2004}) showed, the spectra exhibit circumstellar line absorptions with 
remarkable variations in their strength and dynamical properties.
High velocity gas is observed simultaneously in the Balmer and metallic lines and the gaseous 
circumstellar environment of this star is very complex and active. 
%(Mora et al.\ \cite{Mora2004}).
Silicate emission is detected in 10\,$\mu$m spectra (e.g.\ Acke \& van den Ancker \cite{Acke2004}).

%\begin{figure}
%\begin{center}
%\includegraphics[width=0.45\textwidth,angle=0,clip=]{vsini_H.eps}
%\end{center}
%\caption{
%.
%}
%\label{fig:vsini_H}
%\end{figure}

\section{Searching for a link between the presence of a magnetic field and fundamental stellar characteristics}
\label{sect:link}

Vink et al.\ (\cite{Vink2002}) presented their results of H$\alpha$ spectropolarimetric observations of a sample 
of 23 Herbig Ae/Be stars, pointing out the possibility of the existence of a physical transition 
region in the H-R diagram from magnetospheric accretion, similar to that of cTTS, 
at spectral type A to disk accretion at spectral type B.
The main difference between these scenarios is that in the former case the stellar magnetic field 
truncates the accretion disk at a few stellar 
radii and gas accretes along magnetic channels from the protoplanetary disk to the star,
while in the latter case the accretion flow is not disrupted by the field.
%The Herbig Ae stars might experience magnetically channeled accretion resembling that associated 
%with T\,Tauri stars, whereas the Herbig Be star accretion flow may not be disrupted at a magnetospheric 
%radius. The current understanding of the star-disk interaction region is 
%that this region is controlled by the stellar magnetosphere which truncates the accretion disk at a few stellar 
%radii and channels accreting gas from the protoplanetary disk to the star. 
%Our sample consists of 21 Herbig Ae/Be stars of spectral classification B9 and later spectral types,
%six debris disk stars and of only two hotter Herbig Be stars, HD\,53179 and HD\,85567. 
Our sample consists of 21 Herbig Ae/Be stars of spectral classification B9 and later spectral types and
six debris disk stars.
Since the observations of the 
disk properties of intermediate mass Herbig stars suggest a close parallel to cTTS, 
%revealing the same size range of the disks, similar optical surface brightness and similar structure 
%consisting of inner dark disk and a bright ring, 
%\changea{SH: Carol, Do you know what is it this bright ring? Should we mention it?}
it is quite possible that magnetic fields 
play a crucial role in controlling accretion onto and winds from Herbig Ae stars, similar to the 
case of the lower-mass cTTS. Evidence for disk accretion in Herbig stars from optical 
emission line profiles was presented
by Muzerolle et al.\ (\cite{Muzerolle2004}).
%that especially the magnetic fields 
%play a crucial role in controlling accretion onto, and winds from, Herbig Ae stars, similar to the 
%magnetospheric accretion observed in T\,Tauri stars.
However, contrary to the advances achieved in magnetic studies of cTTS, there is still 
no substantial observational evidence demonstrating the strength, extent, and geometry of magnetic fields in 
Herbig Ae stars. We are aware of the fact that our observations do not present a systematic 
monitoring of the magnetic fields of Herbig stars over 
the rotation period and are just snapshot observations of $\left<B_{\rm z}\right>$ values over two visitor nights.
However, the magnetic field measurements in these stars are rare due to the very small number of spectropolarimetric 
facilities on large telescopes, and presently no other magnetic field data are available.
Still, as we show in the next sub-sections, a few hints and trends can be established 
with the obtained data.
The search for a link with other stellar properties is important to put preliminary constraints on the 
mechanism responsible for magnetospheric activity. As we show in the next sub-sections, we establish 
for the first time preliminary trends.
%they allow us to put several preliminary constraints on the 
%mechanism responsible for the magnetospheric activity. 

\begin{figure}
\begin{center}
\includegraphics[width=9cm]{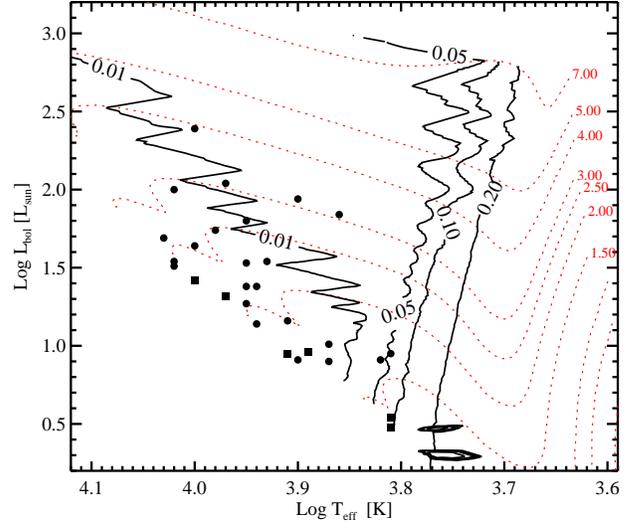}
\caption{H-R diagram for all Herbig Ae (filled circles)
and debris disk (filled squares) stars from Table~\ref{tab:targetvals}
%with known stellar parameters
on the pre-main sequence models
from Siess et al.\ (\cite{Siess00.1}). 
Dotted curves show isomass evolutionary tracks
%solid curves show isochrones,
and solid jagged contours indicate the size of the convective envelopes 
labeled in units of the stellar radius. 
}
%\caption{H-R diagram for all stars from Tables~1 and~2 with known stellar parameters on the evolutionary
%tracks by \citep{Palla1999}. Solid lines are tracks labeled
%by the corresponding mass in solar units, dotted lines are $1, 5, 30$\,Myr isochrones and dashed lines
%are the birthlines for accretion rates of $\dot{M} = 10^{-4}\,{\rm M_\odot/yr}$ and
%$10^{-5}\,{\rm M_\odot/yr}$, respectively.}
\label{fig:hrd}
\end{center}
\end{figure}

The H-R diagram for all Herbig Ae and debris disk stars from Table~\ref{tab:targetvals}
%with known effective temperature and bolometric luminosity 
is shown in Fig.~\ref{fig:hrd}. 
In the following, we study the disk properties, binarity, age, and X-ray emission
for the Herbig Ae stars and debris disk stars. 
We include in the definition of `Herbig Ae' stars targets with spectral type B9 to mid-F.
%such that effectively only two targets, HD\,53179 and HD\,85567, are omitted). 
We have estimated the individual stellar masses from an interpolation 
of the evolutionary tracks from Siess et al.\ (\cite{Siess00.1}), and list the derived values in
Col.~8 of Table~\ref{tab:targetvals}. 
As can be seen from Fig.~\ref{fig:hrd}, the bulk of our sample has masses
between $2-3\,{\rm M_\odot}$.
%and we will use this result in our evaluation
%of magnetospheric accretion and dynamo models (in Sect.~\ref{subsect:accr_rate} and~\ref{subsect:x}).  
A comparison with evolutionary calculations shows that 
the majority of the stars in our sample can be considered fully radiative: the depth of the convection
zone is less than $1$\,\% (cf. contours in Fig.~\ref{fig:hrd}).

%In the following we study the disk properties, binarity, age, and X-ray emission 
%exclusively for the Herbig Ae stars (i.e.\ omitting HD\,53179 and HD\,85567) and debris disk stars
%(for which the presence of magnetospheric accretion is expected)
%with and without magnetic field detections 
%and try to find a connection between the observed 
%longitudinal magnetic field strength and other observed properties.  
%We refer to the notes given in the previous section as well as to Tables~\ref{tab:targetlist} 

\begin{table}
\begin{center}
\caption{Results from statistical analysis of parameter pairs.}
\label{tab:correl}
\begin{tabular}{ccccc}
\hline
\hline
Param 1 & Param 2 & $N$ &   $\tau_{\rm Kend}$ & $\rho_{\rm Spear}$ \\
\hline
\rule{0pt}{2.6ex} $\log{\dot{M}_{acc}}$ & Age                             & 16 & 0.05 & 0.08 \\ 
$B_z$                 & \rule{0pt}{2.6ex} $\log{\dot{M}_{acc}}$           & 16 & 0.68 & 0.57 \\
$\log{L_{\rm x}}$     & Age                             & 19 & 0.13 & 0.28 \\
$B_z$                 & $\log{L_{\rm x}}$               & 19 & 0.09 & 0.07 \\
$B_z$                 & $\log{(L_{\rm x}/L_{\rm bol})}$ & 19 & 0.14 & 0.19 \\
$B_z$                 & $P_{\rm rot}$                   & 20 & 0.50 & 0.42 \\
$B_z$                 & Age                             & 27 & 0.03 & 0.05 \\ 
\hline
\end{tabular}
\begin{flushleft}
Notes:
Parameter pairs are given in Cols.~1 and~2.
$N$ is the number of data points,
$\tau_{\rm Kend}$ and $\rho_{\rm Spear}$ denote the 
probability for no correlation according to Kendall's
and Spearman's rank order correlation test, i.e.\ small
numbers indicate that a correlation is present.
\end{flushleft}
\end{center}
\end{table}

In the remainder of this section, we discuss our search
for dependencies between the magnetic field and other characteristics
of our targets. We have examined various pairs of relevant parameters 
with correlation tests implemented in ASURV (Astronomy Survival Analysis Package;
Lavalley et al.\ \cite{Lavalley1992}).
The probabilities for a correlation between two given parameters
are summarized in Table~\ref{tab:correl}.

\subsection{Accretion rate}
\label{subsect:accr_rate}

The recent results of low resolution linear 
spectropolarimetric observations of Herbig stars in H$\alpha$, H$\beta$, and H$\gamma$ by Mottram et al.\
(\cite{Mottram2007}) support the presence of magnetospheric accretion in Herbig Ae stars.
The authors detected intrinsic line-polarization signatures suggesting
that the magnetic accretion scenario generally considered for cTTS
may be extended to Herbig Ae stars, but that it may not be extended to early Herbig Be stars, 
for which the available data are consistent with disk accretion.

\begin{figure}
\begin{center}
\includegraphics[width=0.45\textwidth,angle=0,clip=]{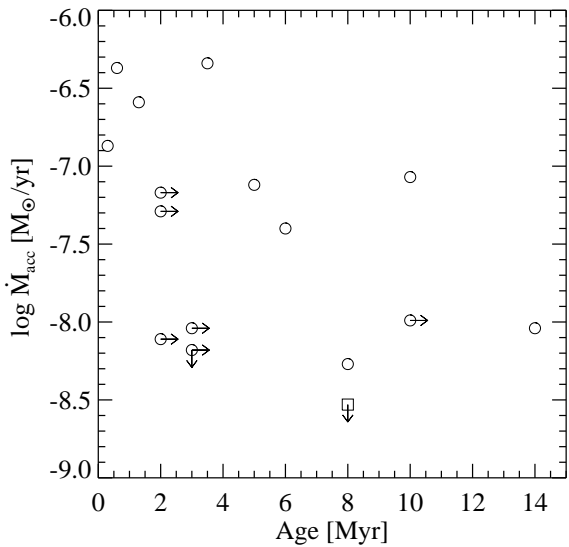}
\end{center}
\caption{
Mass-accretion rate 
versus age.
Open circles denote Herbig Ae stars and open squares indicate debris disk stars.
}
\label{fig:age_accr}
\end{figure}

The number of Herbig Ae
stars with measured mass-accretion rates is still rather low (see Col.~3 in Table~\ref{tab:targetvals}).
In Fig.~\ref{fig:age_accr}, we show the dependence of mass-accretion rate on age for the stars in our sample.
Obviously, the highest mass-accretion rates tend to be found in younger stars. 

%In this figure, also
In the following figures we use for each studied Herbig star 
the maximum measured value of the longitudinal 
field measured from hydrogen lines without taking into account the polarity of the field. 
In addition, we increased the sample of magnetic Herbig Ae stars by adding to our sample four stars with 
previous detections of photospheric magnetic fields, namely HD\,139614, HD\,144432, V380\,Ori, and BF\,Ori. 
As  mentioned in Sect.~\ref{sect:intro}, our previous observations of HD\,139614 and HD\,144432 revealed 
the presence of 
a photospheric magnetic field of the order of $\sim$100\,G (Hubrig et al.\ \cite{Hubrig2007b}). 
The longitudinal magnetic field strengths of 
HD\,139614, HD\,144432, V380\,Ori, and BF\,Ori are listed in Table~\ref{tab:fields_old}. 

\begin{table}
\caption{
Longitudinal magnetic fields of four previously studied Herbig Ae stars with detections at the 3$\sigma$
significance level.
}
\label{tab:fields_old}
\centering
\begin{tabular}{rr @{$\pm$} l}
\hline
\hline
\multicolumn{1}{c}{Object} &
\multicolumn{2}{c}{$\left< B_z\right>_{\rm all}$} \\
\multicolumn{1}{c}{name} &
\multicolumn{2}{c}{[G]} \\
\hline
HD\,139614 & $-$93  & 14 \\
HD\,144432 & $-$111 & 16 \\
V380\,Ori  & 460    & 70 \\
BF\,Ori    & $-$144 & 21 \\
\hline
\end{tabular}
\end{table}

\begin{figure}
\begin{center}
%\resizebox{9cm}{!}{\includegraphics{./B_Mdot.eps}}
\includegraphics[width=0.45\textwidth,angle=0,clip=]{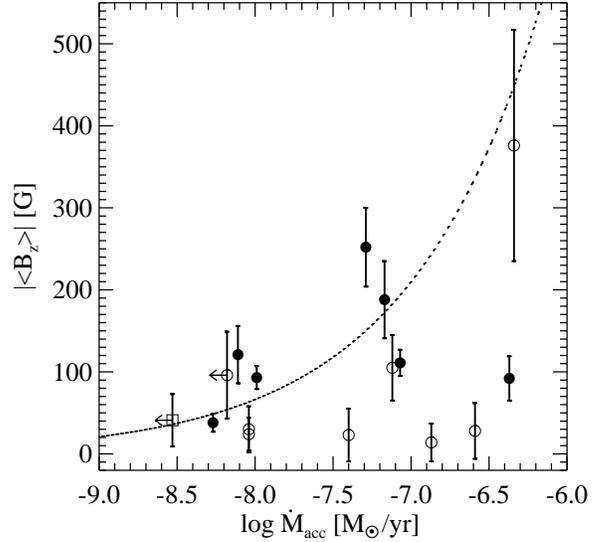}
\caption{The strength of the longitudinal magnetic field as a function
of accretion rate. Overplotted is the relation predicted by 
magnetospheric accretion models (see Eq.~\ref{eqn:JK}) for assumed stellar
parameters of $M_* = 2.5\,{\rm M_\odot}$, 
$R_* = 2.5\,{\rm R_\odot}$, and $P_* = 0.5$\,d.
%Open circles denote Herbig Ae stars and open squares indicate the debris disk stars.
Filled circles denote Herbig Ae stars with a 3$\sigma$ magnetic field detection, while
open circles denote Herbig Ae stars with non-detections.
%a lower $\sigma$.
Squares denote stars with debris disks, of which none has a 3$\sigma$
magnetic field detection.}
\label{fig:accr_H}
\end{center}
\end{figure}

%In Fig.~\ref{fig:accr_H} we present the strength of the longitudinal magnetic field as a function of the 
%mass-accretion rate. 
The values of $\dot M_{\rm acc}$ were derived by Garcia Lopez et al.\ (\cite{GarciaLopez2006}) from the 
measured luminosity of the Br$\gamma$ emission line, using the correlation between $L$(Br$\gamma$) and 
the accretion luminosity $L_{\rm acc}$, established by Muzerolle et al.\ (\cite{Muzerolle1998})
and Calvet et al.\ (\cite{Calvet2004}).
The correlation used is empirical, which makes no assumptions on the 
origin of Br$\gamma$. 
An important result found recently by Kraus et al.\ (\cite{Kraus2008}) is that the Br$\gamma$ line can trace both 
mass infall and outflow, implying that Br$\gamma$ is probably only an indirect tracer of 
the mass-accretion rate.
The authors
used the VLTI/AMBER instrument to spatially and spectrally  resolve the inner ($<$5 AU) 
environment of five Herbig Ae/Be stars (HD\,163296, HD\,104237, HD\,98922, MWC\,297, and V921\,Sco) in the Br$\gamma$ 
emission line as well as in the adjacent continuum. 
The quantitative analysis for HD\,98922 reveals that the line-emitting region is compact enough to be 
consistent with the magnetospheric accretion scenario, and for the stars HD\,163296, HD\,104237, 
MWC\,297, and V921\,Sco 
the authors identify an extended stellar wind or a disk wind as the most likely line-emitting mechanism.
We have not yet searched for a magnetic field in the star HD\,98922, but we observed the 
star HD\,163296 twice.
Both observations were non-detections, supporting the results of Kraus et al.\ (\cite{Kraus2008}) that the magnetospheric 
accretion scenario does not work for this star. No significant magnetic field detection was 
achieved for HD\,104237 by other authors (e.g.\ Wade et al.\ \cite{Wade2007}). 
%indirectly linked to not a primary tracer
%of accretion, but only indirectly linked to the accretion rate.
% e.g.\ via accretion-driven mass loss.
%On the other hand, the magnetospheric accretion theory has been the 
%most successful to date in explaining the origin of atomic hydrogen emission lines. 
In Fig.~\ref{fig:accr_H}, we show the correlation between mass accretion rate and measured
longitudinal magnetic fields for our Herbig Ae sample. While we do not see
a simple correlation between the magnetic field strength and the mass accretion
rate in the data, the observed values are in the range of predictions from
magnetospheric accretion models. 

Magnetospheric accretion models describe the interaction between
a dipolar stellar magnetic field and a surrounding accretion disk
assuming pressure equilibrium.
The analytical approach yields equations that relate the magnetic
field strength to the system parameters (mass, radius, accretion
rate, and rotation period; see e.g., Koenigl \cite{Koenigl1991}; 
Shu et al.\ \cite{Shu1994}). 
Johns-Krull (\cite{JohnsKrull2007}) has shown that for the case of cTTS, the 
models developed by different investigators make consistent
predictions on the magnetic field strength. 

We examine the magnetospheric accretion scenario for our sample
of Herbig Ae stars using the expression for the surface equilibrium field
given by Koenigl (\cite{Koenigl1991}). We present this relation in the form
adopted by Johns-Krull et al.\ (\cite{JohnsKrull1999}): 

\begin{eqnarray}
%\lefteqn{B_{\rm eq}\,{\rm [kG]} = 3.43 \cdot \left(\frac{\epsilon}{0.35}\right)^{7/6} \cdot \left(\frac{\beta}{0.5}\right)^{-7/4} \cdot \left(\frac{M_*}{M_\odot}\right)^{5/6}} \nonumber \\ 
%& \displaystyle ~~~~~~~~~~~~~~~~~~ \cdot \left(\frac{\dot{M}}{10^{-7}M_\odot/{\rm yr}}\right)^{1/2} \cdot \left(\frac{R_*}{R_\odot}\right)^{-3} \cdot \left(\frac{P_*}{\rm d}\right)^{7/6}
B_{\rm eq}\,{\rm [kG]} =& 3.43 \cdot \left(\frac{\epsilon}{0.35}\right)^{7/6} \cdot \left(\frac{\beta}{0.5}\right)^{-7/4} \cdot \left(\frac{M_*}{M_\odot}\right)^{5/6} \nonumber \\ 
 & \cdot \left(\frac{\dot{M}}{10^{-7}M_\odot/{\rm yr}}\right)^{1/2} \cdot \left(\frac{R_*}{R_\odot}\right)^{-3} \cdot \left(\frac{P_*}{\rm d}\right)^{7/6}
\label{eqn:JK}
\end{eqnarray}

\noindent
where $B_{\rm eq}$ denotes the strength of the equatorial magnetic field in kG,
the parameter $\epsilon$ expresses the stellar rotation rate in terms of rotation at the inner boundary of a truncated accretion disk,
the parameter $\beta$ describes how effectively the stellar magnetic field couples to the inner regions of the disk,
$M_*$ denotes the stellar mass in solar masses,
$\dot{M}$ the accretion rate in 10$^{-7}$ solar masses per year,
$R_*$ the stellar radius in solar radii, and
$P_*$ the stellar rotation period in days.

%The field strengths derived with the stellar parameters summarized 
%in Tables~1 and~4 are listed in the last column of Table~3 where they
%can be compared directly to the observed values. 
In Fig.~\ref{fig:accr_H}, the dotted line represents the
prediction from Eq.~\ref{eqn:JK} for a canonical star with $M_* = 2.5\,{\rm M_\odot}$,
$R_* = 2.5\,{\rm R_\odot}$, and $P_* = 0.5$\,d. 
We find that the observed field strengths qualitatively 
support the magnetospheric accretion model for Herbig Ae stars,
although the correlation analysis does not yield a positive result for
$\left<B_{\rm z}\right>$ vs.\ $\dot{M}_{\rm acc}$ (Table~\ref{tab:correl}). 

It must be kept in mind that Eq.~\ref{eqn:JK}
refers to the equatorial dipole field, while our measured values represent
an average of the longitudinal component over the stellar surface and they
also depend on the viewing angle.
%We do not see any simple tracking between magnetic field strength and 
%the mass-accretion rate in Fig.~\ref{fig:accr_H}. We note, however, that  
%a number of magnetic field detections correspond to lower accretion rates. 

%\begin{figure}
%\begin{center}
%\includegraphics[width=0.45\textwidth,angle=0,clip=]{accr_H.eps}
%\end{center}
%\caption{
%The strength of the longitudinal magnetic field as a function of accretion
%rate. The symbols have the same meaning as in Fig.~\ref{fig:incl_H}.
%}
%\label{fig:accr_H}
%\end{figure}
%

%We do not see any simple tracking between magnetic field strength and 
%the mass-accretion rate in Fig.~\ref{fig:accr_H}. We note, however, that  
%a number of magnetic field detections correspond to lower accretion rates. 
Future studies of the magnetic field topology and an improvement of indicators of mass-accretion 
rates are important to understand the role of magnetic fields in the dynamics of 
the accretion processes in Herbig Ae stars.

%\begin{figure}
%\begin{center}
%\includegraphics[width=0.45\textwidth,angle=0,clip=]{accr_H.eps}
%\end{center}
%\caption{
%\The strength of the longitudinal magnetic field as a function of accretion
%\rate. The symbols have the same meaning as in Fig.~\ref{fig:incl_H}.
%}
%\label{fig:accr_H}
%\end{figure}

\subsection{X-ray emission}
\label{subsect:x}

%\begin{table*}
\begin{table}
\centering
\caption{
X-ray emission observed in  Herbig Ae stars and debris disk stars.
}
\label{tab:lxlbol}
\centering
%\begin{tabular}{lrr@{}lcr}
\begin{tabular}{lrr@{}lr}
\hline
\hline
%HD                 log Lx       Instr. + Reference     LogLbol      Reference  Log Lx/Lbol
\multicolumn{1}{c}{Object \rule{0pt}{2.6ex}} &
\multicolumn{1}{c}{log $L_{\rm X}$} &
\multicolumn{2}{c}{Instrument} &
%\multicolumn{1}{c}{log $L_{\rm bol}$} &
\multicolumn{1}{c}{log ($L_{\rm X}/L_{\rm bol}$)} \\
\multicolumn{1}{c}{name} &
\multicolumn{1}{c}{[erg/s]} &
&
&
%\multicolumn{1}{c}{[erg/s]} &
\\
\hline
\hline
HD\,97048    &       29.58 & {\em XMM}     & $^{a}$   & $-$5.64 \\
HD\,97300    &       29.31 & {\em Chandra} & $^{b}$   & $-$5.81 \\
HD\,100453   &       28.82 & {\em Chandra} & $^{c}$   & $-$5.66 \\
HD\,100546   &       28.93 & {\em Chandra} & $^{b}$   & $-$6.16 \\
HD\,135344B  & $\sim$29.6  & {\em ROSAT}   & $^{d}$   & $\sim$$-$4.89 \\
HD\,144432   &       28.9  & {\em Chandra} & $^{e}$   & $-$5.69 \\
HD\,144668    &       28.3 & {\em Chandra} & $^{e}$   & $-$7.22 \\
HD\,150193   &       29.32 & {\em Chandra} & $^{b}$   & $-$5.64 \\
HD\,152404   &       29.09 & {\em Chandra} & $^{b}$   & $-$5.44 \\
HD\,158643   &    $<$28.98 & {\em ROSAT}   & $^{f}$   & $<$$-$6.99 \\
HD\,163296   &       29.6  & {\em Chandra} & $^{b,g}$ & $-$5.36 \\
HD\,169142   &       29.08 & {\em Chandra} & $^{h}$   & $-$5.66 \\
HD\,176386   &    $<$28.56 & {\em Chandra} & $^{b}$   & $<$$-$6.71 \\
\hline
HD\,39060   &       26.5  & {\em Chandra} & $^{i}$    & $-$8.03 \\
HR\,109573    &       29.4  & {\em ROSAT}   & $^{j}$  & $-$5.50 \\
HD\,172555   &    $<$26.9  & {\em Chandra} & $^{k}$   & $<$$-$7.64 \\
HD\,181327   &    $<$29.4  & {\em ROSAT}   & $^{j}$   & $<$$-$4.72 \\
\hline 
V380\,Ori    &       30.96 & {\em Chandra} & $^{b}$   & $-$4.66 \\
BF\,Ori      &    $<$30.2  & {\em ROSAT}   & $^{l}$   & $<$$-$4.91 \\
%HD\,97048    &       29.58 & XMM     & $^{a}$   & 35.22 & $-$5.64 \\
%HD\,97300    &       29.31 & Chandra & $^{b}$   & 35.12 & $-$5.81 \\
%HD\,100453   &       28.82 & Chandra & $^{c}$   & 34.48 & $-$5.66 \\
%HD\,100546   &       28.93 & Chandra & $^{b}$   & 35.09 & $-$6.16 \\
%HD\,135344B  & $\sim$29.6  & ROSAT   & $^{d}$   & 34.49 & $\sim$$-$4.89 \\
%HD\,144432   &       28.9  & Chandra & $^{e}$   & 34.59 & $-$5.69 \\
%HD\,144668    &       28.3 & Chandra & $^{e}$   & 35.52 & $-$7.22 \\
%HD\,150193   &       29.32 & Chandra & $^{b}$   & 34.96 & $-$5.64 \\
%HD\,152404   &       29.09 & Chandra & $^{b}$   & 34.53 & $-$5.44 \\
%HD\,158643   &    $<$28.98 & ROSAT   & $^{f}$   & 35.97 & $<$$-$6.99 \\
%HD\,163296   &       29.6  & Chandra & $^{b,g}$ & 34.96 & $-$5.36 \\
%HD\,169142   &       29.08 & Chandra & $^{h}$   & 34.74 & $-$5.66 \\
%HD\,176386   &    $<$28.56 & Chandra & $^{b}$   & 35.27 & $<$$-$6.71 \\
%\hline
%HD\,39060   &       26.5  & Chandra & $^{i}$    & 34.53 & $-$8.03 \\
%HR\,109573    &       29.4  & ROSAT   & $^{j}$  & 34.90 & $-$5.50 \\
%HD\,172555   &    $<$26.9  & Chandra & $^{k}$   & 34.54 & $<$$-$7.64 \\
%HD\,181327   &    $<$29.4  & ROSAT   & $^{j}$   & 34.12 & $<$$-$4.72 \\
%\hline 
%V380\,Ori    &       30.96 & Chandra & $^{b}$   & 35.62 & $-$4.66 \\
%BF\,Ori      &    $<$30.2  & ROSAT   & $^{l}$   & 35.11 & $<$$-$4.91 \\
\hline
\end{tabular}
\begin{flushleft} 
Notes:
The log~$L_{\rm X}$ values have been collected from the literature, while
the log~$L_{\rm bol}$ are from Table~\ref{tab:targetvals}.\\
References:
%$_X$XMM,
%$_C$Chandra,
%$_R$ROSAT,
$^a$Stelzer et al.\ (\cite{Stelzer2004}),
$^b$Stelzer et al.\ (\cite{Stelzer2006}),
$^c$Collins et al.\ (\cite{Collins2009}),
$^d$Grady et al.\ ApJ, {\sl submitted},
$^e$Stelzer et al.\ (\cite{Stelzer2008}),
$^f$Berghoefer et al.\ (\cite{Berghoefer1996}),
$^g$Swartz et al.\ (\cite{Swartz2005}),
$^h$Grady et al.\ (\cite{Grady2007}),
$^i$Robrade, priv.\ comm.
%Hempel et al.\ (\cite{Hempel2005}),				%(=Robrade,priv.comm)
$^j$Stelzer \& Neuh\"auser (\cite{StelzerNeuhaeuser2000}),
$^k$Feigelson et al.\ (\cite{Feigelson2006}), and
$^l$Hamaguchi et al.\ (\cite{Hamaguchi2005}).
%$^k$Schroeder et al.\ (\cite{Schroeder2008}),
%$^m$van den Ancker et al.\ (\cite{vandenAncker1997}),
%$^n$Carmona et al.\ (\cite{Carmona2008}),
%$^o$van der Plas et al.\ (\cite{vanderPlas2008}),
%$^p$Dominik et al.\ (\cite{Dominik2003}),
%$^q$Tatulli et al.\ (\cite{Tatulli2008}),
%$^r$Di Folco et al.\ (\cite{diFolco2004}),
%$^s$Debes et al.\ (\cite{Debes2008}),
%$^t$Chen et al.\ (\cite{Chen2006b}), and
%$^u$Hillenbrand et al.\ (\cite{Hillenbrand1992}).
\end{flushleft}
\end{table}
%\end{table*}

In Fig.~\ref{fig:hrd} it is demonstrated that our targets have very shallow or completely absent
convection zones. Standard dynamo theory does not predict magnetic field generation for fully
radiative stars. Our targets are also not hot enough to drive strong radiative winds. 
Therefore, any X-ray activity is expected to decay with the dissipation of the primordial field.
Nevertheless, many Herbig Ae/Be stars are known to be X-ray sources (Zinnecker \& Preibisch \cite{Zinnecker1994};
Hamaguchi et al.\ \cite{Hamaguchi2005}; Stelzer et al.\ \cite{Stelzer2006}, \cite{Stelzer2008},
and references therein).
%despite of the fact that neither dynamo action nor strong radiative wind shocks are
%expected for these stars.
In the absence of a theoretical model for X-ray production
in intermediate-mass stars, the detections have often been ascribed to
known or assumed late-type T\,Tauri star companions. However, recent {\em Chandra} imaging
studies have resolved Herbig Ae/Be stars from most known companions and still 
came up with very high detection rates (Stelzer et al.\ \cite{Stelzer2006}, \cite{Stelzer2008}).

We collected X-ray luminosities from the literature, adopting
for each star the value from the instrument with the highest
spatial resolution available. 
The $\log L_{\rm X}$ and $\log L_{\rm bol}$ values and the corresponding 
references are presented in Table~\ref{tab:lxlbol}.
Many of the stars of our sample were included
in dedicated X-ray imaging studies with {\em Chandra} aimed at resolving
them from their visual late-type companion stars (see references
in Table~\ref{tab:lxlbol}). Consequently, only a few of our targets have known unresolved
sub-arcsecond or spectroscopic companions that might be responsible
for the observed X-ray emission. In particular, AK\,Sco (=HD\,152404) is a
spectroscopic binary (Andersen et al.\ \cite{Andersen1989}), V380\,Ori is both a
visual (0\farcs{}154) and a spectroscopic binary (Leinert et al.\ \cite{Leinert1997};
Corporon \& Lagrange \cite{Corporon1999}), and HR\,4796 (=HD\,109573) is 
a 7\farcs{}7 binary (Jayawardhana et al.\ \cite{Jayawardhana1998}) that could not be resolved with {\em ROSAT}.

%\begin{figure}
%\begin{center}
%\includegraphics[width=0.45\textwidth,angle=0,clip=]{age_loglx.eps}
%\end{center}
%\caption{
%X-ray luminosity 
%versus age.
%Open circles denote Herbig Ae stars and open squares indicate the debris disk stars.
%}
%\label{fig:age_loglx}
%\end{figure}
%
%\begin{figure}
%\begin{center}
%\includegraphics[width=0.45\textwidth,angle=0,clip=]{age_loglratio.eps}
%\end{center}
%\caption{
%Ratio of X-ray luminosity over bolometric luminosity
%versus age.
%Open circles denote Herbig Ae stars and open squares indicate the debris disk stars.
%}
%\label{fig:age_loglratio}
%\end{figure}

\begin{figure}
\begin{center}
%\resizebox{9cm}{!}{\includegraphics{./logLx_Age.eps}}
\includegraphics[width=0.45\textwidth,angle=0,clip=]{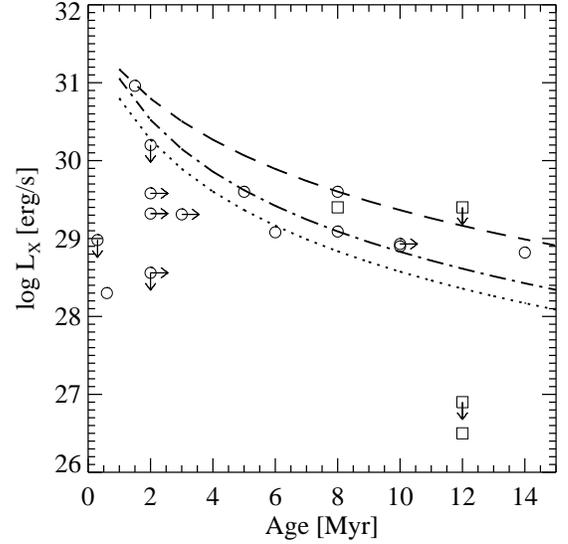}
\caption{
X-ray luminosity versus age.
Symbols are identical to Fig.~\ref{fig:age_accr}.
The decay of $L_{\rm X}$ predicted by the shear dynamo of Tout \& Pringle (\cite{Tout1995})
is shown for different values of the initial parameters $t_{\rm0}$ 
and $L_{\rm X,0}$.
The dotted line represents $t_{\rm0}$=1\,Myr
and $L_{\rm X,0}$=$5\times10^{31}$,
the dashed line represents $t_{\rm0}$=2\,Myr
and $L_{\rm X,0}$=$5\times10^{31}$, and
the dash-dotted line $t_{\rm0}$=1\,Myr 
and $L_{\rm X,0}$=$9\times10^{31}$.
}
\label{fig:age_loglx}
\end{center}
\end{figure}

In Fig.~\ref{fig:age_loglx} 
%and~\ref{fig:age_loglratio} 
we plot $\log L_{\rm X}$ 
%and $\log (L_{\rm X}/L_{\rm bol})$ 
over the age for our sample stars. 
%No clear dependence of the X-ray emission level
%on the age is noticeable, although both figures give some hint that stronger X-ray 
%emission tends to be found in younger Herbig Ae stars.
It is impossible to exclude 
categorically that there are further, as yet  unknown, companion stars
responsible for the X-ray emission. However, given the high detection
rate and high spatial resolution of the X-ray studies, this is unlikely to be the
case for all stars in this sample. 

Tout \& Pringle (\cite{Tout1995}) have suggested a mechanism that can give rise to intrinsic
X-ray emission from Herbig Ae/Be stars. In their model, a dynamo can be
sustained in a radiative star by rotational shear. As a result of 
the decrease of the available rotational energy with time, the 
X-ray luminosity behaves as 

\begin{equation}
L_{\rm X} = L_{\rm X,0} \cdot \left(1 + \frac{t}{t_{\rm 0}}\right)^{-3}
\label{eqn:lx}
\end{equation}

\noindent
i.e.\ it decays from an initial value of $L_{\rm X,0}$ given by the shear
energy. Tout \& Pringle (\cite{Tout1995}) showed that an X-ray luminosity of about 1\% of
the bolometric luminosity can be maintained for a duration of 
$\sim$1\,Myr, and
thereafter decays rapidly according to the law presented in Eq.~\ref{eqn:lx}.
The curves plotted in Fig.~\ref{fig:age_loglx} represent the predicted decay of the X-ray 
emission for $t_{\rm 0} = 1$\,Myr and $2$\,Myr. The initial
X-ray luminosity $L_{\rm X,0}$ was estimated from Eqn.~4.4 of Tout \& Pringle (\cite{Tout1995}) 
for an assumed $M_* = 2.5\,{\rm M_\odot}$, $R_* = 2.5\,{\rm M_\odot}$, and 
$L_{\rm bol} = 50\,{\rm L_\odot}$ (dash-dotted line). Two other solutions
with a factor of two lower $L_{\rm X,0}$ are also shown.

No clear dependence of the X-ray emission level on the age is
noticeable in the data but the absence of stars with high X-ray luminosities at an
advanced age is in agreement with the expected decay of the shear dynamo.

\begin{figure}
\begin{center}
\includegraphics[width=0.45\textwidth,angle=0,clip=]{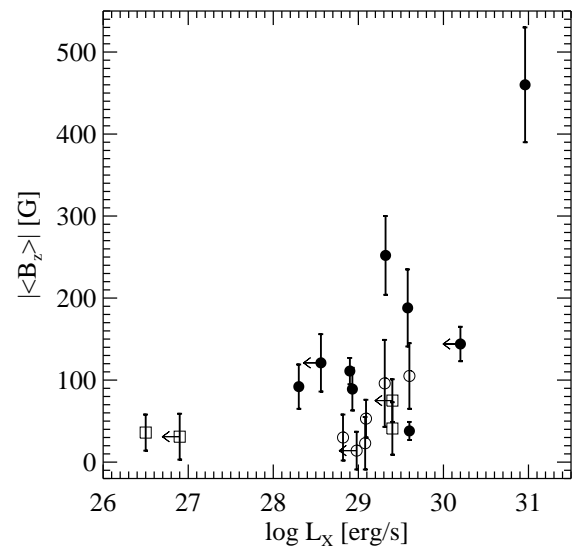}
\end{center}
\caption{
The strength of the longitudinal magnetic field plotted 
over the X-ray luminosity.
Symbols are identical to Fig.~\ref{fig:accr_H}.
}
\label{fig:loglx_H}
\end{figure}

\begin{figure}
\begin{center}
\includegraphics[width=0.45\textwidth,angle=0,clip=]{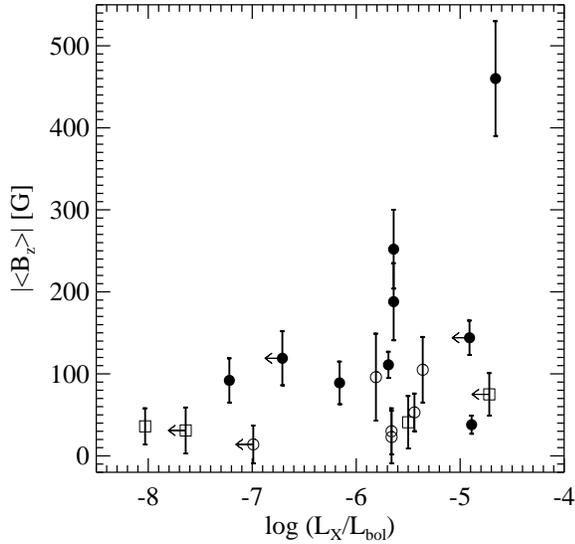}
\end{center}
\caption{
The strength of the longitudinal magnetic field plotted 
over the ratio between X-ray luminosity and bolometric luminosity.
Symbols are identical to Fig.~\ref{fig:accr_H}.
}
\label{fig:loglratio_H}
\end{figure}

In Figs.~\ref{fig:loglx_H} and~\ref{fig:loglratio_H} we present the strength of the magnetic field 
plotted versus $\log L_{\rm X}$ and $\log (L_{\rm X}/L_{\rm bol})$.
In both figures we find a hint for an increase of the magnetic
field strength with the level of the X-ray emission, also supported by
the correlation analysis which yields $>$90\,\% probability that a correlation
is present between $\left<B_{\rm z}\right>$ and $\log{L_{\rm X}}$ 
(Table~\ref{tab:correl}). This could suggest
a dynamo mechanism responsible for the coronal activity in Herbig Ae stars. 
On the other hand, we should keep in mind that the star with the strongest magnetic field in 
both figures is the spectroscopic binary V380\,Ori with $\left<B_{\rm z}\right>$\,=\,460$\pm$70\,G,
and it is not clear yet whether the X-ray emission originates from the primary or from the companion.
The filled circle with the lowest strength of the longitudinal magnetic field,
$\left<B_{\rm z}\right>$\,=\,37$\pm$12\,G, belongs in both figures to the Herbig Ae star HD\,135344B.
%{\bf HD\,135344A was not detected with the ROSAT HRI.}
%As we mentioned above, a binary pair was detected at 
%$\sim$5\farcs{}8 separation. This binary is a background object, not co-moving
%with HD\,135344AB (Grady et al.\ 2009, in preparation).}
%While HD\,135344B was not detected with the ROSAT HRI, the binary pair 
%is not resolved and could be a culprit for the detected X-ray emission (Grady et al.\ 2009, in preparation).
%\changea{SH: Carol, could you resolve with the ROSAT HRI. HD\,135334B and the binary at a distance of 5\farcs{}8?}
%No clear dependence of the X-ray emission level
%on the age is noticeable if we plot $\log L_{\rm X}$ and 
%$\log (L_{\rm X}/L_{\rm bol})$ over the age of our sample stars, although we see some hint that stronger X-ray 
%emission tends to be found in younger Herbig Ae stars.
%Clearly, the presence of the X-ray emission in Herbig Ae stars still remains to be 
%explained and more future spatially resolved observations are necessary, especially for stars with 
%detected magnetic fields.
 
\begin{figure}
\begin{center}
%\resizebox{9cm}{!}{\includegraphics{./logLx_logMagnflux.eps}}
\includegraphics[width=0.45\textwidth,angle=0,clip=]{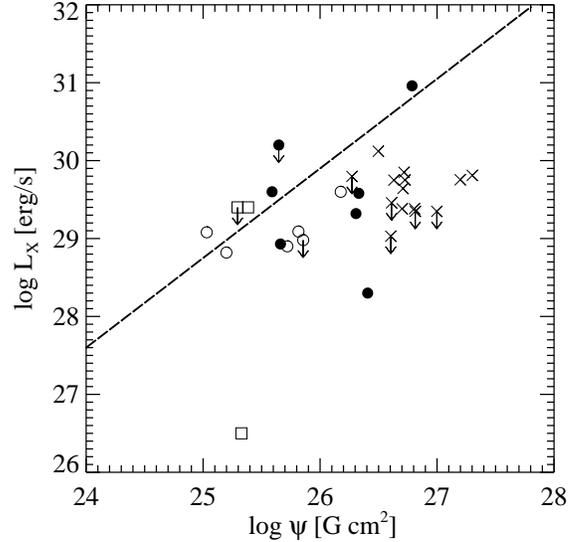}
\caption{X-ray luminosity versus magnetic flux for our sample
of Herbig Ae stars,
and the cTTS sample from Johns-Krull (\cite{JohnsKrull2007}),
compared to the power-law relation derived for the Sun and
active field stars by Pevtsov et al.\ (\cite{Pevtsov2003}).
cTTS are represented by crosses.
All other symbols are identical to Fig.~\ref{fig:accr_H}.
} 
\label{fig:logmagflux_loglx}
\end{center}
\end{figure}

Pevtsov et al.\ (\cite{Pevtsov2003}) have derived a universal relation between
magnetic flux and X-ray luminosity that holds over $10$ orders of
magnitudes from the quiet Sun, over solar active regions to active
late-type field dwarfs. The unique power-law relation 
$L_{\rm X} \sim \Psi^{1.13}$ was taken as evidence that
the coronae of the Sun and the stars are heated by the same kind
of structures. We test whether Herbig Ae stars obey the same relation.
The magnetic fluxes of our stars are approximated as 
%$\Phi_{\rm mag} = B_{\rm z} \cdot \pi R_*^2$, and plotted in
$\Psi$ $= 12 \pi R^2 \overline{\left< B_{\rm z} \right>}$,
assuming a surface magnetic field $B_s \approx 3 \overline{\left< B_{\rm z} \right>}$.
The magnetic fluxes  are plotted in Fig.~\ref{fig:logmagflux_loglx}
versus $L_{\rm X}$ together with the Pevtsov-relation
and a sample of cTTS from Johns-Krull (\cite{JohnsKrull2007}).
The tendency of the cTTS to show lower than expected X-ray luminosities
might be related to the effects of the disk, such as e.g.\ reduced coronal
heating of mass-loaded magnetic field lines (Preibisch et al.\ \cite{Preibisch2005})
or reduced height of coronal loops (Jardine et al.\ \cite{Jardine2006}). 
%Surprisingly, we observe the opposite trend in the Herbig Ae stars; most of them are overluminous
%in X-rays with respect to the expectation from their magnetic flux. 
Apart from HD\,144668 (the lowest filled circle in Fig.~\ref{fig:logmagflux_loglx}),
the Herbig Ae stars cluster around the line
which follows the power law relation derived by Pevtsov.
The lowest symbol corresponds to the debris disk star HD\,172555.

Clearly, the presence of the X-ray emission in Herbig Ae stars still remains to be
explained and more spatially resolved observations are necessary, especially for stars with
detected magnetic fields.
 
%\begin{figure}
%\begin{center}
%%\includegraphics[width=0.45\textwidth,angle=0,clip=]{Bz_logLx.eps}
%\includegraphics[width=0.45\textwidth,angle=0,clip=]{loglx_H.eps}
%\end{center}
%\caption{
%The strength of the longitudinal magnetic field plotted 
%over the log~$L_{\rm X}$. Symbols have the same meaning as in Fig.~\ref{fig:incl_H}.
%}
%\label{fig:loglx_H}
%\end{figure}
%
%\begin{figure}
%\begin{center}
%\includegraphics[width=0.45\textwidth,angle=0,clip=]{loglratio_H.eps}
%\end{center}
%\caption{
%The strength of the longitudinal magnetic field plotted 
%over the log(L$_{\rm X}$/L$_{\rm bol}$). Symbols have the same meaning as in Fig.~\ref{fig:incl_H}.
%}
%\label{fig:loglratio_H}
%\end{figure}
%

\subsection{Rotation}

Any kind of correlation between the magnetic field strength and the rotation period is of particular 
importance in view of the unknown origin of magnetic fields in Herbig Ae stars.
For example, a  trend of the magnetic field strength being lower in more slowly rotating stars 
would be consistent with the usual prediction for the dynamo theory for stars of a given mass
(Mestel \cite{Mestel1975}).
For no Herbig Ae star or debris disk star is the rotation period known.
For stars with known disk inclinations and $v \sin i$  values we can estimate $v_{\rm eq}$ values,
and from the knowledge of $v_{\rm eq}$ and stellar radius we can estimate the rotation period.
The $v \sin i$ values collected from the literature are listed in Col.~4 of Table~\ref{tab:inclinations}
followed by $v_{\rm eq}$ values
in Col.~5. Also most of the radii were collected from the literature. Only for 
three Herbig Ae stars, HD\,101412, HD\,135344B, and HD\,179218,
and two debris disk stars, HD\,109573 and HD\,181327, have the radii not been estimated in the past. 
For these stars, we estimated the radii using the
Stefan--Boltzmann law. The information on bolometric luminosity and effective temperature of the three Herbig stars 
was found in van der Plas et al.\ (\cite{vanderPlas2008}), while
%HD\,101412: Lum.=25 Lo (logL/Lo =1.4), M=2.3 +/-0.2, logTeff=4.02 van der Plas
%HD\,135344  logL/Lo =1.01 M=1.7 +/-0.2 logTeff= 3.82 ---
%HD\,179218 logL/Lo = 1.88 M=2.7 +/-0.3 logTeff= 4.02 
for HD\,109573 we used the values presented by Debes et al.\ (\cite{Debes2008})
and for HD\,181327 we used the values presented by Chen et al.\ (\cite{Chen2006b}).
%HD\,109573 L= 21.1, Teff= 9250
%HD\,181327 Chen 2006, Teff= 6560, L=2.5 
Whenever the literature sources permitted, the uncertainties associated with these parameters were included. 
Using $v_{\rm eq}$ and radii listed in Cols.~5 and 6 of Table~\ref{tab:inclinations}, respectively, we estimated 
the rotation periods, which are listed in Col.~7 of Table~\ref{tab:inclinations}.
% of Table~\ref{tab:inclinations}, respectively.
%In Figs.~\ref{fig:veq_H} and~\ref{fig:periods_H} we present the strength of the 
%longitudinal magnetic field plotted 
%over the $v_{\rm eq}$ and over the rotation period.
%The only rotational studies for Herbig Ae stars were done at Mg  II with IUE
%data for AB Aur and to a lesser extent for HD\,163296 (ref), as  they do not hold  
%up epoch to epoch. 
Magnetic fields are detected in stars with a large range of rotation velocities, from 6\,km s$^{-1}$ up 
to 300\,km s$^{-1}$.
No obvious trend of the strength of the longitudinal magnetic field with the rotation velocity or 
rotation period is identifiable in the statistical tests.
%in these plots.
The star HD\,101412 with the strongest longitudinal magnetic field 
($\left<B_{\rm z}\right>$\,=\,454$\pm$42\,G) clearly stands out in both distributions as it is the most slowly 
rotating Herbig Ae star, with the longest rotation period of more than 17\,days. 
It is possible that the slow rotation is caused by magnetic braking.
%Leaving this star aside, the distribution of the longitudinal magnetic field strength 
%over the rotation period gives a hint that magnetic fields tend to be found in stars with shorter rotation periods.

%\begin{figure}
%\begin{center}
%\includegraphics[width=0.45\textwidth,angle=0,clip=]{veq_H.eps}
%\end{center}
%\caption{
%The strength of the longitudinal magnetic field plotted 
%over the equatorial velocity. Symbols have the same meaning as in Fig.~\ref{fig:accr_H}.
%}
%\label{fig:veq_H}
%\end{figure}
%
%\begin{figure}
%\begin{center}
%\includegraphics[width=0.45\textwidth,angle=0,clip=]{periods_H.eps}
%\end{center}
%\caption{
%The strength of the longitudinal magnetic field plotted 
%over the rotation period. Symbols have the same meaning as in Fig.~\ref{fig:accr_H}.
%}
%\label{fig:periods_H}
%\end{figure}

\subsection{Age}

\begin{figure}
\begin{center}
\includegraphics[width=0.45\textwidth,angle=0,clip=]{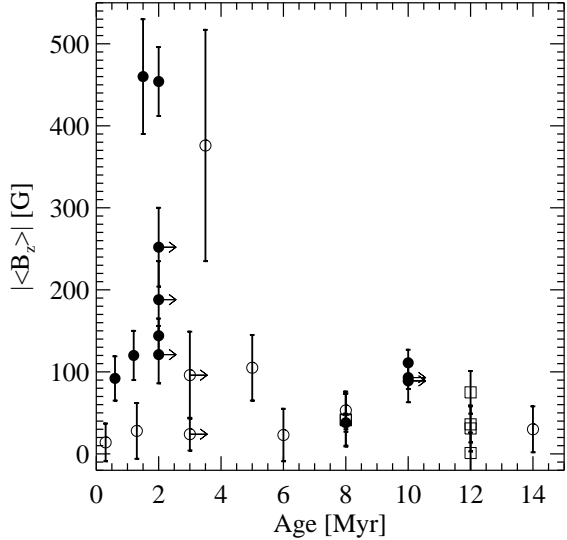}
\end{center}
\caption{
The strength of the longitudinal magnetic field as a function of age.
The symbols are identical to Fig.~\ref{fig:accr_H}.}
\label{fig:age_H}
\end{figure}

There are clear indications for a trend towards stronger magnetic fields in younger
Herbig Ae stars (Fig.~\ref{fig:age_H}), confirmed by statistical tests.
%There is obviously a trend showing that stronger magnetic fields tend to be found in younger Herbig 
%Ae stars (Fig.~\ref{fig:age_H}).
%but age is not the dominant factor, as shows the $>$10 Myr old HD\,100546. 
%The search for magnetic fields in Herbig Ae stars and the study of their structure in the pre-main 
%sequence counterparts to the magnetic Ap stars is a crucial step towards understanding the origin of 
%the magnetic fields in stars of intermediate mass.
The observations of magnetic fields of Herbig Ae stars, their strength and geometry  are of a particular
importance to understand the origin of magnetic fields in Ap stars.
It has been frequently mentioned in the literature that magnetic Herbig Ae stars are potential progenitors of 
the magnetic Ap stars 
(e.g., Stepien \& Landstreet \cite{StepienLandstreet2002}; Catala \cite{Catala2003}; Wade et al.\ \cite{Wade2005}).
%On the other hand, all our previous studies of evolution of magnetic fields in stars across the upper 
%main sequence
%showed that magnetic Ap stars of mass below 3\,M$_\odot$ are significantly evolved and concentrated towards 
%the centre of the main-sequence band (Hubrig et al.\ \cite{Hubrig2000}, \cite{Hubrig2005b}, \cite{Hubrig2007a}).
%and practically no magnetic star of mass below 3\,M$_\odot$ can be 
%found close to the zero-age main sequence (ZAMS) (Hubrig et al.\ \cite{Hubrig2000}, \cite{Hubrig2005b}, 
%\cite{Hubrig2007a}).
%In contrast, magnetic Bp stars with masses $M$$>$3\,M$_\odot$ seem to be concentrated closer to the ZAMS. 
On the other hand, from Fig.~\ref{fig:age_H} it is obvious that stronger magnetic fields appear in very young 
Herbig Ae stars, and the magnetic fields become very weak at the end of their PMS life.
These results clearly 
confirm the conclusions of Hubrig et al.\ (\cite{Hubrig2000}, \cite{Hubrig2005b}, \cite{Hubrig2007a})
that magnetic fields in stars with masses less than 3\,M$_{\sun}$ are rarely found close to the 
ZAMS and that kG magnetic fields appear in A stars already evolved from the ZAMS.
In contrast, magnetic Bp stars with masses $M$$>$3\,M$_\odot$ seem to be concentrated closer to the ZAMS. 
%appear in A-type stars after they completed $\sim$30\% of
%that magnetic fields with masses less than 3\,M$_{\sun}$ are only rarely found close to the 
%ZAMS and that kG magnetic fields 
%appear in A-type stars after they completed $\sim$30\% of their main sequence life.

\subsection{Disk properties}

In the following, we compare the magnetic field and accretion rate to some relevant disk properties.

\subsubsection{Disk inclination}
%{\it Disk inclination:}
%\subsection{Search for a relation between magnetic fields and the 
%star/disk parameters}

\begin{table*}
\centering
\caption{
Rotation periods of stars with known inclinations.
}
\label{tab:inclinations}
\centering
\begin{tabular}{lcccccc}
\hline
\hline
\multicolumn{1}{c}{Object} &
\multicolumn{1}{c}{Inclination} &
\multicolumn{1}{c}{Adopted} &
\multicolumn{1}{c}{$v$\,sin\,$i$} &
\multicolumn{1}{c}{$v_{\rm eq}$} &
\multicolumn{1}{c}{$R$} &
\multicolumn{1}{c}{$P$} \\
\multicolumn{1}{c}{name} &
\multicolumn{1}{c}{Measurements} &
\multicolumn{1}{c}{Inclination} &
\multicolumn{1}{c}{[km s$^{-1}$]} &
\multicolumn{1}{c}{[km s$^{-1}$]} &
\multicolumn{1}{c}{[$R_{\sun}$]} &
\multicolumn{1}{c}{[d]} \\
 &
\multicolumn{1}{c}{[$^\circ{}$]} &
\multicolumn{1}{c}{[$^\circ{}$]} &
 &
 &
 &
 \\
\hline
%HD 53179          & 60                   & 60          & 100            & 115         & 14.4 & 6.3 \\
HD\,95881         & 40$-$55$^1$          & 40$-$55     & 50$^q$         & 61$-$78     & 3.1$^{dd}$ & 2.0$-$2.6 \\
HD\,97048         & 30$-$40$^a$          & 30$-$40     & 140$^r$        & 218$-$280   & 2.5$^{ee}$ & 0.5$-$0.6 \\
                  & 42.8$^b$             &             &                &             &     & \\
HD\,100453        & $>$55$^2$            & $>$55       & 39$^s$         & $<$48       & 1.7$^{ff}$ & $>$1.8 \\
HD\,100546        & 42$^c$               & 47$\pm$5    & 65$^t$         & 82$-$97     & 1.7$^{ff}$ & 0.9$-$1.0 \\
                  & 51$^d$               &             &                &             &     & \\
HD\,101412        & $>$55$^3$            & $>$55       & 5$^u$          & $<$6.1      & 2.1$^{gg}$ & $>$17.4 \\
HD\,135344B       & 11$^e$               & 11          & 69$^v$         & 362         & 2.4$^{gg}$ & 0.3 \\
                  & 14$^f$               &             &                &             &     & \\
HD\,139614        & $<$40$^g$            & $<$40       & 15$^{w}$        & $>$23       & 1.6$^{dd}$ & $<$3.5 \\
HD\,144432        & 48$\pm$10$^e$        & 48$\pm$10   & 70$^{w}$        & 83$-$114    & 3.0$^{ff}$ & 1.3$-$1.8 \\
HD\,144668        & 58$^h$               & 58          & 100$^{w}$       & 118         & 3.9$^{ee}$ & 1.7 \\
HD\,150193        & $\sim$38$^i$         & 30$^5$      & 100$^t$        & 200         & 2.1$^{ee}$ & 0.5 \\
HD\,152404        & 65$-$70$^j$          & 65$-$70     & 18.5$\pm$1$^j$ & 18.5$-$21.5 & 2.6$^{dd}$ & 6.1$-$7.1 \\
HD\,158643        & $\sim$90$^k$         & 90          & 267$\pm$5$^v$  & 267$\pm$5   & 5.3$^{hh}$ & 1.0 \\
                  &                      &             & 228$^{x}$     & 228         &     & \\
HD\,163296        & 51$^{+11,}_{-9}$$^l$ & 51          & 130$^{w}$     & 167         & 2.8$^{ee}$ & 0.8 \\
HD\,169142        & 13$\pm$2$^e$         & 13$\pm$2    & 66$\pm$2$^v$   & 247$-$356   & 1.6$^{dd}$ & 0.2$-$0.3 \\
VV\,Ser           & 72$\pm$5$^f$         & 72          & 142$^{y}$     & 149         & 2.4$^{ee}$ & 0.8 \\
                  &                      &             & 229$^{z}$     & 241         &     & \\
HD\,179218        & 40$\pm$10$^e$        & 40$\pm$10   & 60$^{aa}$      & 78$-$120    & 4.8$^{gg}$ & 2.0$-$3.1\\
HD\,190073        & $<$40$^4$            & $<$40       & 12$^{bb}$      & $>$18.7     & 3.3$^{ff}$ & $<$8.9 \\
\hline 
HD\,9672          & 90$^m$               & 90          & 196$^{x}$     & 196         & 1.7$^{ii}$ & 0.4 \\
HD\,39060         & 87$^n$               & 87          & 130$^k$        & 130         & 1.8$^{jj}$ & 0.7 \\
HD\,109573        & 73$^o$               & 73          & 152$^{x}$     & 159         & 1.8$^{gg}$ & 0.6 \\
HD\,181327        & 31.7$^p$             & 31.7        & 16$^{cc}$      & 30          & 1.2$^{gg}$ & 2.0 \\
%\hline 
%HD\,31648         & 38$^q$               & 38          & 90$^{hh}$      & 146         & 2.5$^{ccc}$ & 0.9 \\
%                  & 36$\pm$1$^r$         &             &                &             &     & \\
%HD\,104237        & 18$^s$               & 18          & 10$^v$         & 32          & 2.7$^{ccc}$ & 4.3 \\
\hline
\end{tabular}
\begin{flushleft}
Notes:
Inclination angles derived for resolved disk detections from the literature and
from the study of \ion{Mg}{ii} spectral line profiles in UV spectra
or the study of H$\alpha$ line profiles in optical spectra and
the inclination values we finally adopt for further use are listed in Cols.~2 and 3.
$v$\,sin\,$i$ values and radii presented in Cols.~4
and 6, respectively, are for the most part gathered from the literature.
In Col.~7 we present the equatorial velocity values which are obtained from the inclination angles 
and the $v$\,sin\,$i$ values. $v_{\rm eq}$ and radii are used to estimate the rotation periods 
listed in the last column.
Radii marked as {\it this study} were calculated using \teff{} and bolometric luminosity, both taken from the literature.\\
References:
$^a$Doering et al.\ (\cite{Doering2007}),
$^b$Doucet et al.\ (\cite{Doucet2007}),
$^c$Ardila et al.\ (\cite{Ardila2007}),
$^d$Augereau et al.\ (\cite{Augereau2001}),
$^e$Dent et al.\ (\cite{Dent2005}),
$^f$Pontoppidan et al.\ (\cite{Pontoppidan2008}),
$^{g}$Meeus et al.\ \cite{Meeus1998},
$^h$Preibisch et al.\ (\cite{Preibisch2006}),
$^i$Fukagawa et al.\ (\cite{Fukagawa2003}),
$^j$Alencar et al.\ (\cite{Alencar2003}),
$^k$Slettebak (\cite{Slettebak1982}),
$^l$Wassell et al.\ (\cite{Wassell2006}),
$^m$Hughes et al.\ (\cite{Hughes2008}),
$^n$Heap et al.\ (\cite{Heap2000}),
$^o$Schneider et al.\ (\cite{Schneider1999}),
$^p$Schneider et al.\ (\cite{Schneider2006}),
$^q$Grady et al.\ (\cite{Grady1996}),
$^r$van den Ancker (\cite{vandenAncker1998}),
$^s$Acke \& Waelkens (\cite{AckeWaelkens2004}),
$^t$Hamidouche et al.\ (\cite{Hamidouche2008}),
$^{u}$this study (see Subsect.~\ref{subsect:discussion_withfield}),
$^v$Dunkin et al.\ (\cite{Dunkin1997}),
$^{w}$Hubrig et al.\ (\cite{Hubrig2007b}),
$^{x}$Royer et al.\ (\cite{Royer2007}),
$^{y}$Vieira et al.\ (\cite{Vieira2003}),
$^{z}$Mora et al.\ (\cite{Mora2001}),
$^{aa}$Bernacca \& Perinotto (\cite{BernaccaPerinotto1970}),
$^{bb}$Pogodin et al.\ (\cite{Pogodin2005}),
$^{cc}$de la Reza \& Pinz\'on (\cite{delaRezaPinzon2004}),
$^{dd}$Blondel \& Tjin (\cite{Blondel2006}),
$^{ee}$Hillenbrand et al.\ (\cite{Hillenbrand1992}),
$^{ff}$Wade et al.\ (\cite{Wade2007}), 
$^{gg}$this study (see table caption),
$^{hh}$Tatulli et al.\ (\cite{Tatulli2008}), 
$^{ii}$Hughes et al.\ (\cite{Hughes2008}), and 
$^{jj}$di Folco et al.\ (\cite{diFolco2004}).\\ 
%$^q$Simon et al.\ (\cite{Simon2000}),
%$^r$Pi\`etu et al.\ (\cite{Pietu2006}),
%$^s$Grady et al.\ (\cite{Grady2004}),
%$^z$Th\'e et al.\ (\cite{The1985}),
%$^{hh}$Beskrovnaya et al.\ (\cite{Beskrovnaya2004}),
%$^a$Meeus et al.\ \cite{Meeus2002},
%$^x$this work,
%$^x$Guimar\~aes et al.\ (\cite{Guimaraes2006}),
%Bernasconi \& Maeder (\cite{BernasconiMaeder1996}),
Comments on the inclination angles deduced in this study, mainly from IUE (LWR and LWP cameras) through analysis of line profile types.
Profile shapes are classified according to Beals (\cite{Beals1951}).
See also the discussion in Subsect.~\ref{subsect:discussion_withfield}.
$^1$\ion{Mg}{ii} alternates between P\,Cygni type I (red emission, with blue-shifted absorption, IUE LWP 30772)  and type III (double emission, IUE  LWP 13082),
$^2$the H$\alpha$ profile shown by Meeus et al.\ (\cite{Meeus2002}) shows characteristic double-peaked H$\alpha$ emission,
$^3$see the discussion in Subsect.~\ref{subsect:discussion_withfield},
$^4$\ion{Mg}{ii} has a type I P\,Cygni profile (IUE LWP 25762, LWR 11977, LWR 08996),
$^5$\ion{Mg}{ii} has a type I  profile (IUE LWP 13083).
\end{flushleft}
\end{table*}

\begin{figure}
\begin{center}
\includegraphics[width=0.45\textwidth,angle=0,clip=]{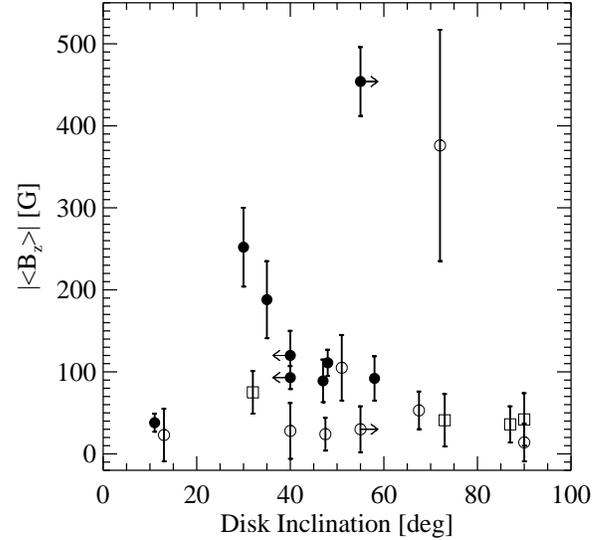}
\end{center}
\caption{
The strength of the longitudinal magnetic field versus disk inclination angles.
Symbols are identical to Fig.~\ref{fig:accr_H}.
}
\label{fig:incl_H}
\end{figure}

%\changea{SH: Carol, can you check the comments on IUE and HST spectra in Table 4?}
Although an expanding sample of Herbig Ae stars have system inclination data from CO and coronographic
imaging surveys,
for a major part of the studied Herbig Ae stars the orientation of the disk still
has to be constrained using 
line profiles as a proxy for the inclination, as described in Sect.~\ref{sect:discussion}.
The emission profile shapes of \ion{Mg}{ii} lines from UV studies 
or the  H$\alpha$ line profiles are frequently used.
We present the inclination angles collected from different papers 
in the second column of Table~\ref{tab:inclinations}. 
Our own estimates of profile types from the study of \ion{Mg}{ii} spectral line profiles in UV spectra are  
listed in the comments with running indices close to the HD numbers. 
In the third column we present the adopted inclinations.
Using inclination angles in combination with $v$\,sin\,$i$ data, we 
derive $v_{\rm eq}$ (Col.~4 of Table~\ref{tab:inclinations}).
However, no correlation between mass-accretion rate and disk inclination is detected.
We also do not find any trend of the measured longitudinal magnetic field with the inclination angle 
(Fig.~\ref{fig:incl_H}).

%HD\,139614 is one such object, where our B field is significantly lower  
%than  measured in the past. It would be interesting to establish whether any such variation  
%is linked to the star or to activity in the disk.
The disk inclinations for our magnetic Herbig Ae stars sample the whole possible range from close 
to face-on to close to edge-on.
%shows a distribution of angles.
% but we note that we find no close to edge-on system among them.
The strongest magnetic fields in our sample are observed for a flared
disk (HD\,97048), for an edge-on disk (HD\,101412) and for close 
to pole-on disks (HD\,150193 and HD\,190073), 
%a disk with moderate inclination (HD\,150193) 
and systems with an envelope rather than a disk (HD\,85567 and HD\,176386). 
These results show that there is no preferred 
disk orientation for the detection of a magnetic field. 
%apart from the edge-on orientation.

\subsubsection{Disk geometry}

We find that magnetic fields are detected in systems both with and without 
strong PAHs, but appear to be more frequent in the first, flared case. This result could
probably indicate an age dependence of the presence of magnetic fields.
Acke et al.\ (\cite{AckeAncker2004}) studied Herbig Ae/Be stars with different IR SEDs
with the conclusion that flaring disks probably evolve into self-shadowed disks.
%generally in the course of disk evolution the flared disks become flat disks. 
The existence of strong PAH emission may point to a flared disk, since the
PAH grains reside further outside in the disk, whereas they would be
destroyed near the hot inner disk rim. 
On the other hand, Keller et al.\ (\cite{Keller2008})
conclude that most Herbig Ae/Be stars have PAH emission at some level, and there is no correlation 
of PAH emission with the disk geometry. Since PAHs originate from the
inner 200--300\,AU of the disk, they are actually not a good measure of  
magnetic properties of the Herbig stars. Silicates usually emerge from $\sim$1\,AU and beyond.
Contrary to the silicates, the PAHs need to be excited by direct stellar irradiation, which may
not work well in a flat disk. In addition, the IR slope of the SED can 
further indicate whether the disk is flat or flared (Meeus et al.\ 
\cite{Meeus2001}). 
We note that magnetic fields are also detected in HD\,176386.
%and the hotter Herbig Be stars 
%HD\,53179 and HD\,85567, which have their 
%circumstellar matter distributed in an envelope rather than in a disk.
%(e.g., HD\,85567, HD\,176386, but also in HD\,53179 the envelope may dominate over the disk).

\subsubsection{Hot disk gas}

%A detection of H$_2$ may point to a young age of the disk, as it is the
%main constituent of the cloud from which the star is formed. We note that
For many of our targets with a detected magnetic field, reports on the detection
of excited, hot inner disk gas exist, which may be attributed to a gaseous
accretion ring close to the star (see especially Tatulli et al.\ \cite{Tatulli2007};
Isella et al.\ \cite{Isella2008}). 

Activity in the inner disk is often 
accompanied by an increased K-band excess. To prove whether there is a 
direct trend, we correlated the K-band excess, derived as the difference
of the measured and expected K-band flux of each star, with the strength
of its magnetic field, but we found no correlation. 
We compared intrinsic V-K colours, as given e.g.\ by Koornneef (\cite{Koornneef1983}),
to the difference V--K in the measured photometry of both passbands, as given
in SIMBAD. Due to a missing luminosity class we assumed a class V for PDS\,2,
HD\,97048, VV\,Ser and HD\,179218. We obtained
excesses between 0.0 and 3.9\,mag, with an uncertainty of 0.1\,mag.
The lack of correlation may be explained 
by the fact that without spectral resolution it is not possible to determine if the $K$-band excess 
is due to gas lines or to a hot inner dust continuum.
%This latter one 
A hot inner dust continuum
will as well produce an increased broad-band $K$
excess, possibly without being related to the magnetic field. 
%hot inner dust will as well produce an increased K-band 
%excess, without possibly being related to the mechanism, which generates
%the magnetic field. 
While there is a clear trend that most of the 
magnetic Herbig Ae stars have reports on hot, inner gas, a selection effect cannot 
be ruled out, as many of the non-magnetic Herbig Ae stars may just not have been 
measured in a similar way.

\subsection{Binarity}

A majority of the Herbig Ae/Be stars are binary or multiple systems. The existence of a
magnetic field, however, does not seem to be related to binarity. 
We note however, that the presence of a close companion contributing to the observed spectropolarimetric 
spectra can cause a non-detection of the magnetic field due to blending of 
spectral lines of the primary. The contribution of the secondary component can be 
disentangled only by means of high resolution spectropolarimetric observations, but 
not with low resolution FORS\,1 spectropolarimetry.
In the present study, some targets, e.g.\
HD\,97048 or HD\,100546, show a magnetic field, but have 
no known companions, while other binary Herbig Ae stars possess no magnetic field.

\section{Summary}
\label{sect:sum}

In the course of our spectropolarimetric study of Herbig Ae/Be stars, we detected
%magnetic fields in eight stars, PDS\,2, HD\,53179, HD\,85567, HD\,97048, HD\,100546,
magnetic fields in six stars for the first time, PDS\,2, HD\,97048, HD\,100546,
HD\,135344B, HD\,150193, and HD\,176386.
The presence of a magnetic field was confirmed in the stars HD\,101412, HD\,144668, and HD\,190073.

We have for the first time examined the relation of the measured field strengths to various
parameters that characterize the star-disk system. 

Among the most important
relations for the interpretation of the fields 
is the one between magnetic field strength and accretion rate. 
We do not find a clear trend between these two parameters in our sample of
Herbig Ae stars 
but the measured field strengths are compatible, in order of magnitude,
with the values expected
from magnetospheric accretion scenarios for a dipole field 
(tens to a few hundreds of Gauss). 
This contrasts with the situation for cTTS. For typical cTTS parameters,
accretion models predict fields that range between $\sim$200--2000\,G.
However, the observed mean magnetic field strengths of cTTS
are not correlated with the predictions (Johns-Krull \cite{JohnsKrull2007}). 
For most cTTS, the observed fields are larger than the expected values,
possibly indicating that magnetic field pressure dominates gas pressure
in these systems. In addition, the dipole approximation is known not
to be valid for the case of cTTS that have complex field geometries
(e.g.\ Gregory et al.\ \cite{Gregory2008}), while our results suggest that 
it may be a reasonable description of Herbig Ae stars.

We find that stronger magnetic fields tend to be found in younger Herbig stars. 
The magnetic fields become very weak or completely disappear in stars when they arrive on the ZAMS.
Similarly, strong X-ray sources are only found at the youngest ages, in qualitative
agreement with the predictions of a shear dynamo that decays within a few Myrs as the
rotational energy of the star decreases (Tout \& Pringle \cite{Tout1995}). 
It is premature, however, to claim a direct
connection between magnetic field and X-ray luminosity.
The Herbig Ae stars seem to follow the power-law between magnetic flux and X-ray luminosity
established for the Sun and main-sequence active stars. 
%They tend to have higher X-ray luminosities than expected for their magnetic flux. 
%We speculate that this difference might indicate non-coronal 
%contributions to the X-ray emission in Herbig Ae stars, such as a binary companion
%or a cooling post-shock region above the equatorial plane such as proposed
%by Babel \& Montmerle (\cite{Babel1997}) for the magnetic Ap star IQ\,Aqr.
%About half of the stars with magnetic field detections possess longitudinal  magnetic 
%fields larger than 100\,G.
%These stars are the best candidates for future spectropolarimetric studies to analyze the behaviour of 
%their magnetic fields over the rotational periods to disclose the magnetic topology on their surfaces 
%and to study the complex interaction between the stellar magnetic field, the disk and the stellar wind. 

We do not find any trend between the presence of a magnetic field and disk inclination angles.
% and  mass-accretion rates although we note that probably the estimation of the mass-accretion rates should be 
%improved by identification of an emission line which can be considered as a primary tracer of the 
%mass-accretion rate.
The membership in binary or multiple systems does not seem to have any impact on the 
presence of a magnetic field, whereas there is a hint that the appearance of magnetic fields is more 
frequent in Herbig stars with flared disks and hot, inner gas.
Since flared disks are the least evolved, 
this is possibly another indication for the decay of magnetic fields with increasing age.
However, no trend of the strength of the magnetic field with rotation velocity and rotation period 
was detected in our study. 
%The stronger magnetic fields tend to be found in younger Herbig stars. The magnetic fields become very 
%weak or completely disappear in stars when they arrive on the ZAMS.
%We also find a hint for an increase of the magnetic field strength of the level of the X-ray emission.
%However, no trend of the strength of the magnetic field with the rotation velocity and rotation period 
%was detected in our study.

While considerable progress has been made with respect to the presence of magnetic fields 
in Herbig Ae stars, a number of questions remain open. The most important question 
is related to the origin of the magnetic fields in these stars. Although our results  provide new clues, 
the observational results presented in this work are still inconclusive as to the magnetic field origin.
Tout \& Pringle (\cite{Tout1995}) proposed a non-solar dynamo that could operate in rapidly 
rotating A-type stars based on rotational shear energy. Their model predicts that the coronal activity 
at the observed rates of log~$L_{\rm X}$ can be sustained for a period of the order of $10^6$\,yr.
Other possible mechanisms causing magnetic activity involve fossil magnetic fields or 
magnetically confined wind shocks (e.g.\ Babel \& Montmerle \cite{Babel1997}).
A more comprehensive survey of the presence of magnetic fields and a detailed study of the magnetic 
field topology in a Herbig star sample of increased size will provide important additional information 
to test the predictions of different theories.
About half of the stars with magnetic field detections possess longitudinal 
magnetic fields larger than 100\,G.
These stars are the best candidates for future spectropolarimetric studies to analyze the behaviour of 
their magnetic fields over the rotational periods to disclose the magnetic topology on their surfaces 
and to study the complex interaction between the stellar magnetic field, the disk and the stellar wind.
%A better statistics based on the more comprehensive survey of magnetic fields in Herbig Ae/Be stars would allow  

{
\acknowledgements
BS acknowledges financial support from ASI/INAF under
contract I/088/06/0, MAP and RVY acknowledge RFBR grant No\,07-02-00535a 
and Sci.Schole No\,6110.2008.2, and
MC acknowledges DIUV grant 08/2007.
This research has made use of the SIMBAD database,
operated at CDS, Strasbourg, France.
}

\end{document}